\documentclass[a4paper, numbers, twoside]{ezthesis}

\usepackage{listings}
\usepackage{multicol}
\usepackage{titlesec} 
\usepackage{multirow}
\usepackage{amsmath} 
\usepackage{enumerate}
\usepackage{graphicx}
\usepackage{float} 
\usepackage[footnotesize]{caption} 
\usepackage{fancyhdr}
\usepackage{multirow}
\usepackage{amssymb}
\usepackage{caption, subcaption} 

\usepackage{amsmath}
\usepackage{amssymb}
\usepackage{amsfonts}
\usepackage{amsthm}
\usepackage[mathscr]{euscript}
\usepackage{bbm}
\usepackage{cancel}
\usepackage{wasysym}
\usepackage{booktabs}
\usepackage{tabularx}
\usepackage{tocloft}
\usepackage{multicol}
\usepackage{nomencl}
\usepackage{slashed}
\pagecolor{white}\usepackage{pdfpages}

\usepackage{setspace}
\makenomenclature

\newcommand{\mc}[1]{\mathcal{#1}}

\newcolumntype{C}{>{\arraybackslash}X}

\newcommand{\blname}{\cellcolor{blue!20}}
\newcommand{\blop}{\cellcolor{blue!10}}
\newcommand{\bhname}{\cellcolor{blue!25}}
\newcommand{\bhop}{\cellcolor{blue!15}}
\newcommand{\bclass}{\cellcolor{blue!40}}
\newcommand{\rlname}{\cellcolor{orange!20}}
\newcommand{\rlop}{\cellcolor{orange!10}}
\newcommand{\rhname}{\cellcolor{orange!25}}
\newcommand{\rhop}{\cellcolor{orange!15}}
\newcommand{\rclass}{\cellcolor{orange!40}}
\newcommand{\glname}{\cellcolor{green!20}}
\newcommand{\glop}{\cellcolor{green!10}}
\newcommand{\ghname}{\cellcolor{green!25}}
\newcommand{\ghop}{\cellcolor{green!15}}
\newcommand{\gclass}{\cellcolor{green!40}}
\newcommand{\olname}{\cellcolor{red!20}}
\newcommand{\olop}{\cellcolor{red!10}}

\newcommand{\oclass}{\cellcolor{red!40}}

\newcommand{\gtick}{\textcolor{green}{\mathbf{\checkmark}}}
\newcommand{\otick}{\textcolor{orange}{\mathbf{\checkmark}}}
\newcommand{\rtick}{\textcolor{red}{\text{\sffamily X}}}

\newcommand{\ii}{\ensuremath{\mathrm{i}}}
\newcommand{\order}[1]{order $\Lambda^{-#1}$ }
\newcommand{\Tr}[1]{\mathrm{Tr}\left[{#1}\right]}
\newcommand{\MSbar}{${\overline{\text{MS}}}$}
\newcommand{\diff}{\ensuremath{{\rm{d}}}}
\newcommand\Tstrut{\rule{0pt}{2.6ex}}         

    \renewcommand*{\nompreamble}{\begin{multicols}{2}}
    \renewcommand*{\nompostamble}{\end{multicols}}
\hyperlinking

\titleformat{\chapter}[display]
{}{\Huge Chapter \thechapter.}{0pt}
{\Huge}

\usepackage{dcolumn}
\newcolumntype{.}{D{.}{\esperiod}{-1}}

\RequirePackage{verbatim}
\usepackage{afterpage}
\usepackage{longtable}

\usepackage{hyperref} 
\usepackage[toc=false,nomain,acronym,automake,nomissingglstext=true]{glossaries-extra}
\setabbreviationstyle[acronym]{long-short}

\newacronym[longplural={Quantum Field Theories}]{qft}{QFT}{Quantum Field Theory}
\newacronym[longplural={Quantum Electrodynamics}]{qed}{QED}{Quantum Electrodynamics}
\newacronym[longplural={Quantum Chromodynamics}]{qcd}{QCD}{Quantum Chromodynamics}
\newacronym{sm}{SM}{Standard Model}
\newacronym{np}{NP}{New Physics}
\newacronym[longplural={Effective Field Theories}]{eft}{EFT}{Effective Field Theory}
\newacronym[longplural={Chiral Perturbation Theories}]{chpt}{$\chi$PT}{Chiral Perturbation Theory}
\newacronym[longplural={Soft Collinear Effective Theories}]{scet}{SCET}{Soft Collinear Effective Theory}
\newacronym[longplural={Heavy Quark Effective Theories}]{hqet}{HQET}{Heavy Quark Effective Theory}
\newacronym{bsm}{BSM}{Beyond the Standard Model}
\newacronym{smeft}{SMEFT}{Standard Model Effective Field Theory}
\newacronym{heft}{HEFT}{Higgs Effective Field Theory}
\newacronym{ew}{EW}{Electroweak}
\newacronym{dimreg}{DimReg}{Dimensional Regularisation}
\newacronym{msbar}{\MSbar}{Modified Minimal Subtraction}
\newacronym{rge}{RGE}{Renormalisation Group Equation}
\newacronym{uv}{UV}{Ultraviolet}
\newacronym{ir}{IR}{Infrared}
\newacronym{ewpo}{EWPO}{Electroweak Precision Observable}
\newacronym{ewpt}{EWPT}{Electroweak Precision Test}
\newacronym{ewpd}{EWPD}{Electroweak Precision Data}
\newacronym{wc}{WC}{Wilson coefficient}
\newacronym[longplural={Equations of Motion}]{eom}{EoM}{Equation of Motion}
\newacronym{rg}{RG}{Renormalisation Group}
\newacronym[longplural={Wavefunction Renormalisation}]{wfr}{WFR}{Wavefunction Renormalisation}
\newacronym{onshell}{OS}{On-shell}
\newacronym{ms}{MS}{Minimal Subtraction}
\newacronym{1pi}{1PI}{One-particle-irreducible}
\newacronym{ngb}{NGB}{Nambu-Goldstone Boson}
\newacronym{ibp}{IBP}{Integration by Parts}
\newacronym[longplural=Anomalous Dimension Matrices]{adm}{ADM}{Anomalous Dimension Matrix}
\newacronym{lnv}{LNV}{Lepton Number Violation}
\newacronym{bnv}{BNV}{Baryon Number Violation}
\newacronym{ewsb}{EWSB}{Electroweak Symmetry Breaking}
\newacronym{nda}{NDA}{Naive Dimensional Analysis}
\newacronym{ftae}{FTAE}{Física Teórica y de Altas Energías}
\newacronym{vev}{vev}{Vacuum Expectation Value}
\newacronym{nnlo}{NNLO}{Next to Next to Leading Order}
\newacronym{nlo}{NLO}{Next to Leading Order}
\newacronym{lo}{LO}{Leading Order}
\newacronym{ll}{LL}{Leading Logarithmic}
\makeglossaries

\newcommand{\myTitle}{Renormalization of the Standard Model Effective Field Theory up to dimension 8 operators\xspace}

\newcommand{\myName}{{\'{A}}lvaro D{\'{i}}az Carmona\xspace}

\hypersetup{
pdfauthor = {\myName (aldiaz (at) ugr (dot) es)},
pdftitle = {\myTitle},
pdfsubject = {},
}

\usepackage{url}
\usepackage{colortbl,longtable}
\usepackage[stable]{footmisc}

\definecolor{gray97}{gray}{.97}
\definecolor{gray75}{gray}{.75}
\definecolor{gray45}{gray}{.45}
\definecolor{gray30}{gray}{.94}

\lstnewenvironment{listing}[1][]
   {\lstset{#1}\pagebreak[0]}{\pagebreak[0]}

\lstdefinestyle{CodigoC}
   {
	basicstyle=\scriptsize,
	frame=single,
	language=C,
	numbers=left
   }
\lstdefinestyle{CodigoC++}
   {
	basicstyle=\small,
	frame=single,
	backgroundcolor=\color{gray30},
	language=C++,
	numbers=left
   }

\lstdefinestyle{Consola}
   {basicstyle=\scriptsize\bf\ttfamily,
    backgroundcolor=\color{gray30},
    frame=single,
    numbers=none
   }

\makeatletter
\def\clearpage{%
  \ifvmode
    \ifnum \@dbltopnum =\m@ne
      \ifdim \pagetotal <\topskip
        \hbox{}
      \fi
    \fi
  \fi
  \newpage
  \thispagestyle{empty}
  \write\m@ne{}
  \vbox{}
  \penalty -\@Mi
}
\makeatother

\usepackage{pdfpages}

\allowdisplaybreaks

\begin{document}

\renewcommand{\tablename}{Table}

\frontmatter

\includepdf[pages=1]{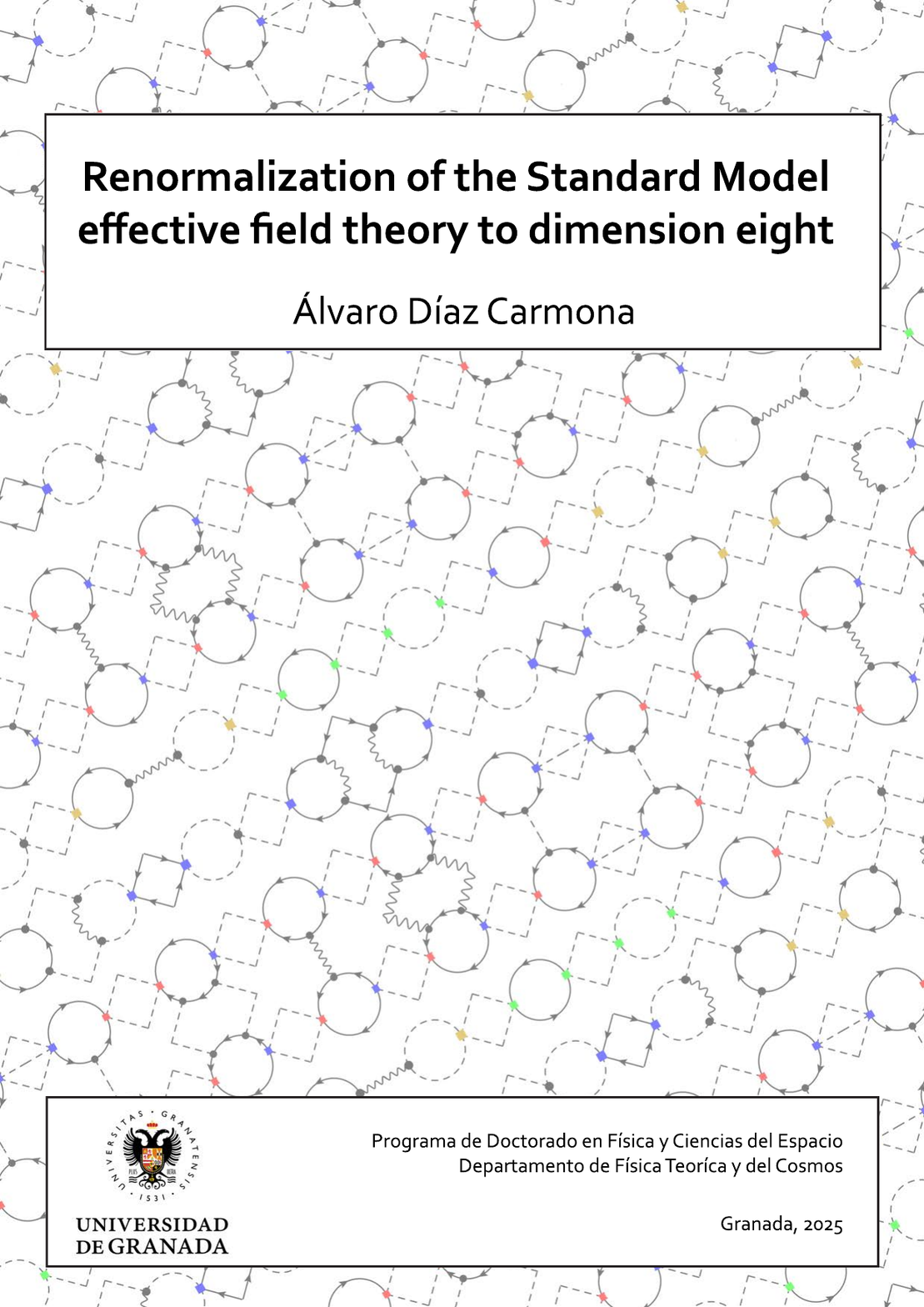}

\begin{titlepage}
		\begin{center}
		\vspace{-3cm}
		{\includegraphics[scale=0.4]{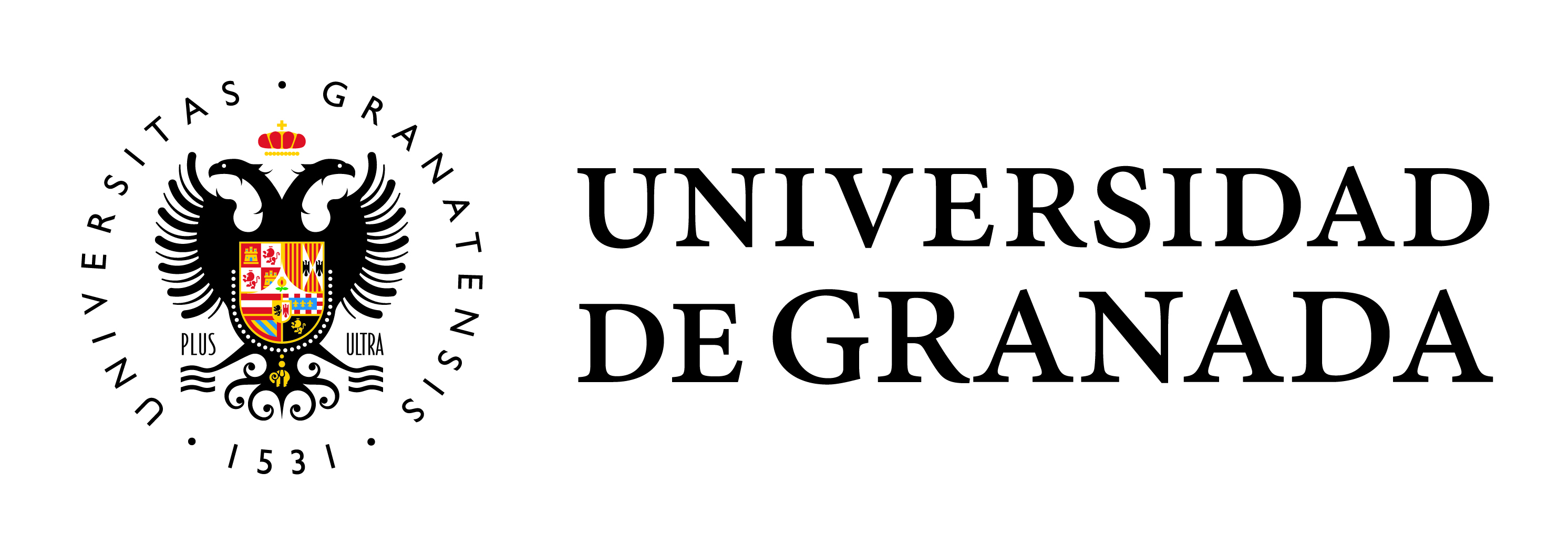} \\}
		
		{\scshape\Large TESIS DOCTORAL \\}
		{\scshape\Large PROGRAMA DE DOCTORADO EN FÍSICA Y CIENCIAS DEL ESPACIO \\}
		\vspace{1cm}
		{\Huge \textbf{Renormalization of the Standard Model effective field theory to dimension eight} \\}
		\vspace{1cm}
		{\normalsize \textbf{Autor} \\}
		{\normalsize Álvaro Díaz Carmona \\}
				{\small Grupo de Física Teórica y de Altas Energías\\
			Departamento de Física Teórica y del Cosmos\\
			Granada, 2025}
		\vfill
		{\normalsize \textbf{Directores} \\}
		{\normalsize Mikael Rodríguez Chala \\}
		{\normalsize Departamento de Física Teórica y del Cosmos\\}
		{\normalsize Adrián Carmona Bermúdez \\}
		{\normalsize Departamento de Física Teórica y del Cosmos\\}
		\vfill

		\vfill
		\end{center}
\end{titlepage}

\clearpage\hbox{}\thispagestyle{empty}\newpage
\vspace{-1cm}\tableofcontents

\clearpage\hbox{}\thispagestyle{empty}\newpage

\addcontentsline{toc}{chapter}{Agradecimientos/Acknowledgements}
\chapter*{Agradecimientos/Acknowledgements}
\markboth{AGRADECIMIENTOS}{AGRADECIMIENTOS}

Se acerca el final de una etapa muy enriquecedora y significativa en mi vida. En los últimos cuatro años, me he desarrollado en el ámbito académico, pero también fuera, 
y tengo muy claro que esto ha sido una consecuencia imprevista de elegir la investigación como profesión. Al matricularme, buscaba el modo de plantear y atacar los problemas de la Física más fundamental, para poder aportar mi granito de arena. En el transcurso de mi corta carrera científica, he empezado a estudiar esta y otras cuestiones, y muy frecuentemente requerí la ayuda de alguien. 

Aunque estoy obsesionado con la Física, también me he involucrado en otros intereses no científicos, que revelaron dimensiones extra en mi mundo. Las lec\-ciones que he aprendido y las experiencias que he vivido se las debo a personas que he conocido a lo largo de mi vida, desde el principio hasta muy recientemente, y a las que quiero agradecer ahora.

Debo empezar dando gracias a mis directores por la paciencia y la con\-fianza que han depositado en mí para organizarme a mi manera, pero a la vez aconsejándome y redirigiéndome cuando me veían perdido o atascado con algún problema. Estoy especialmente agradecido a Miki por la atención y el apoyo que me ha dado, a\-nimándome a defender charlas hasta que perdí el miedo o tratando de transmitirme buenos hábitos y actitudes para la investigación. A Adrián le debo la idea de meterme en este doctorado. De él me llevo el buen rollo en cada conversación y la ilusión por crear proyectos nuevos y luchar por su progreso, a pesar de las dificultades que se presenten.\pagebreak

Uno de los cambios más notables que he vivido ha sido la ocupación exponencial del espacio en los despachos, sobre todo por estudiantes. Antes de llegar, me encontré con un grupo reducido en número pero con gran presencia. Después de mi entrada, se han incorporado muchos investigadores, con mucha ilusión y dedicación, que en mayor o menor medida también están buscando su lugar. He aparecido en el momento justo para conocer varias generaciones de doctorandos y cada una de ellas ha cambiado una pequeña parte de mí.

Maria Ramos, Alejandro Jiménez, Dani, Guilherme y Chema fueron mis re\-ferentes al principio de mi etapa investigadora. Trabajando a su lado adquirí la mentalidad para esforzarme al máximo pero sin sufrir por los problemas rutina\-rios: hay que tomárselo con calma, concentrarse en lo esencial y seguir probando día tras día. 

Pablo Olgoso y Fran fueron mis compañeros de carrera, aunque me sacaron ventaja en la pandemia y, igual que António, empezaron el doctorado antes que yo. Con todos, pero especialmente con ellos, el tiempo pasaba volando en las conversaciones de comedores y los días de trabajo post-pandemia se hicieron más amenos.

Alejandro, Alicia y Luis Pelegrina llenaron el lado experimental de la becaría mientras el bando teórico se vaciaba. Parecían tímidos, hasta que Chicago los cambió y ahora no hay quien los pare. 

Clara, Ramón, Pablo de la Torre, Luis Gil, Adrián Moreno, Javi y Fuen son personas fáciles de querer, sensibles, con gran corazón y muy divertidos. Me transmiten simpatía incluso cuando les da un bajón emocional (cosa no poco frecuente hasta donde he comprobado).

No me olvido de "los nuevos": Rafa, Diogo, Chiara y Cristina; y de los estu\-diantes de máster: Thomas, Sergio, Eddie, Carlos y Víctor. Las nuevas generaciones vienen cargadas de energía positiva y se han integrado rápidamente con el resto. 

También me gustaría agradecer la labor de los equipos de comedores, además de conserjería y limpieza del Edificio Mecenas; a Raúl, instructor de Concept Granada, y a mis compañeros de yoga; a Rosa Pérez, mi fisioterapeuta y osteópata. Sus servicios han facilitado mi día a día en estos cuatro años, y se nota cuando faltan.
\pagebreak

During my travels abroad, I have been introduced to many kind, helpful members of our community. First and foremost, I want to thank Pedro Schwaller and Anke Biekötter for hosting me at JGU for three months. I am also grateful to the nice master's and PhD students I met there, especially to Cristopher, Cristina, Majo and Sascha, for the funny random plans they came up with, including the amazing hyperpop rave. I thank Riccardo Bartocci as well, for literally hosting me during one week.

I would also like to thank the people I met during conferences or travels and who co-starred memorable moments or stories, too long to explain: Sara Leardini, Károly Seller, Ivo de Medeiros, Andrii, Jakub, Hiroki, Viktor and Prisco Lo Chiatto.

Still in the international dimension, I am really thankful to Supratim Das Bakshi for the countless afternoon discussions trying to figure out some puzzling diagram and for refining my taste in Indian cuisine. I would also like to thank my latest office mates: Zhe, Sandra and Clara (again), for the lovely, peaceful environment that room 005 has become.

No puedo cerrar esta sección sin acordarme quienes más me han apoyado: mis amigos y mi familia. Adri y Juaki, que me conocen desde el colegio, han presenciado mi evolución personal desde hace años y aún nos reunimos para ponernos al día y reírnos como en los viejos tiempos. Quique y María Hernández, que se ganaron todo mi cariño desde el día que los conocí y aún siguen haciendo méritos. David y Elena, con los que siempre se puede contar, y que sacan adelante los proyectos más ambiciosos porque nunca se rinden. Alberto, Juanjo, Ana y Carmen, que tienen cada uno su visión del mundo, distinta a la mía. Hablar con ellos siempre me abre nuevas perspectivas.

He tenido la suerte de nacer en una familia muy grande. Por parte de madre, mis tías, tío, primas, primos y correspondientes parejas son gente divertida con la que puedo pasar horas sin aburrirme y que siempre están cuando se los necesita. La familia de mi padre es pequeña pero cercana. De ellos he heredado gran parte de mi carácter, así que siempre me siento como en casa con ellos. Además, quiero incluir a Alba, Lucía y Aleix, que bien podrían ser familia por habernos criado juntos.

Por último, quiero agradecer a mi madre, Gregoria, y mi hermana, Sara, el enorme esfuerzo que han hecho al soportarme y darme apoyo sin tener la más mínima idea de a qué me he dedicado todos estos años (intenté aclararlo, de verdad) y en especial los meses que estuve escribiendo este proyecto. Espero que ver esta tesis imprimida en un libro ayude a calmar su inquietud, y ya de paso que me perdonen por haberme convertido en un espíritu aventurero (pero cauto), obsesionado con explorar y aprender constantemente.

Esta tesis cierra una etapa de mi vida y, a la vez, comienza una nueva. Espero que nuestros caminos no se separen del todo, aunque sé que nos volveremos a cruzar.

\addcontentsline{toc}{chapter}{Financiación/Funding}
\chapter*{\vspace{-1cm}Financiación/Funding}

\markboth{FINANCIACIÓN}{FINANCIACIÓN}




\textbf{Esta tesis es parte de los siguientes proyectos y ayudas:}\par\bigskip

El proyecto PID2022-139466NB-C22, financiado por MICIU/AEI/10.13039/\\
501100011033 y “FEDER/UE”.\par \medskip

El proyecto PID2019-106087GB-C21 financiado por MICIU/AEI/10.13039/\\
501100011033.\par\medskip

La ayuda PRE2020-092144 financiada por MICIU/AEI/10.13039/\\
501100011033 y por “FSE Invierte en tu futuro”.\par\medskip

La ayuda FQM101 de la Junta de Andalucía.\par\medskip
\vspace{1cm}

\textbf{This thesis is part of the following projects and grants:}\par\bigskip

Grant PID2022-139466NB-C22, funded by MICIU/AEI/10.13039/
501100011033 and by “ERDF/EU”.\par\medskip

Grant PID2019-106087GB-C21 funded by MICIU/AEI/10.13039/
501100011033.\par\medskip

Grant PRE2020-092144 funded by MICIU/AEI/10.13039/
501100011033 and by “ESF Investing in your future”.\par\medskip

Grant FQM101 funded by Junta de Andalucía.\par

\clearpage\hbox{}\thispagestyle{empty}\newpage

\addcontentsline{toc}{chapter}{Publicaciones/Publications}
\chapter*{\vspace{-1cm}Publicaciones/Publications}
\markboth{Publicaciones}{Publicaciones}

The following articles were published by the author and collaborators:

\subsection*{Chapter 4}

\begin{enumerate}

\item[(1)]
M.~Chala, \'A.~D\'\i{}az-Carmona and G.~Guedes,
\textit{A Green\textquoteright{}s basis for the bosonic SMEFT to dimension 8},
JHEP \textbf{05} (2022), 138 
doi:10.1007/JHEP05(2022)138
[arXiv:2112.12724 [hep-ph]].

\end{enumerate}

\subsection*{Chapter 5}
\begin{enumerate}

\item[(2)]
S.~Das Bakshi, M.~Chala, \'A.~D\'\i{}az-Carmona and G.~Guedes,
\textit{Towards the renormalisation of the Standard Model effective field theory to dimension eight: bosonic interactions II},
Eur. Phys. J. Plus \textbf{137} (2022) no.8, 973 \\
doi:10.1140/epjp/s13360-022-03194-5 
[arXiv:2205.03301 [hep-ph]].

\item[(3)]
S.~Das Bakshi and \'A.~D\'\i{}az-Carmona,
\textit{Renormalisation of SMEFT bosonic interactions up to dimension eight by LNV operators},
JHEP \textbf{06} (2023), 123 
doi:10.1007/JHEP06(2023)123
[arXiv:2301.07151 [hep-ph]].

\item[(4)]
S.~D.~Bakshi, M.~Chala, \'A.~D\'\i{}az-Carmona, Z.~Ren and F.~Vilches,
\textit{Renormalization of the SMEFT to dimension eight: Fermionic interactions I},
JHEP \textbf{12} (2025), 214 
doi:10.1007/JHEP12(2024)214 
[arXiv:2409.15408~[hep-ph]].

\end{enumerate}

\addcontentsline{toc}{chapter}{Abstract}
\chapter*{\vspace{-1cm}Abstract}
\markboth{ABSTRACT}{ABSTRACT}

The Standard Model Effective Field Theory (SMEFT) provides a powerful, model-independent framework to explore deviations from the Standard Model (SM) by parametrising potential new physics through higher-dimensional operators. This thesis investigates the renormalisation structure of SMEFT, focusing on dimension-eight operators, which are increasingly relevant in precision analyses and in models where dimension-six effects are suppressed.

We review renormalisation in quantum field theory, emphasising dimensional regularisation and the $\overline{\text{MS}}$ scheme, and outline the conceptual foundations of EFTs. One of the central results of this work is the systematic construction and classification of bosonic operators in SMEFT at dimension eight, employing Group Theory techniques and removing redundancies by working in momentum space. Building on this operator basis, we compute the complete one-loop renormalisation group equations (RGEs) involving insertions of dimension-eight-or-lower operators. This includes pure dimension-eight effects, pairs of dimension-six operators and lepton-number-violating sectors. Our calculations use an off-shell Green’s function basis and leverage algebraic simplifications derived from symmetry and gauge invariance.

These results are applied to positivity bounds and oblique parameters, providing essential tools for consistent SMEFT analyses across energy scales. The findings extend SMEFT’s theoretical reach and support its use in high-precision phenomenology.

\clearpage\hbox{}\thispagestyle{empty}\newpage

\addcontentsline{toc}{chapter}{Resumen}
\chapter*{\vspace{-1cm}Resumen}
\markboth{Resumen}{Resumen}

La Teoría de Campos Efectiva del Modelo Estándar (SMEFT, por sus siglas en inglés) proporciona un marco potente e independiente de modelos para explorar desviaciones del Modelo Estándar (SM), parametrizando posibles nuevas físicas mediante ope\-radores de dimensión superior. Esta tesis investiga la estructura de renormalización de la SMEFT, centrándose en operadores de dimensión ocho, los cuales son cada vez más relevantes en análisis de precisión y en modelos donde los efectos de dimensión seis están suprimidos.

Se revisa la renormalización en teoría cuántica de campos, con énfasis en la regularización dimensional y el esquema $\overline{\text{MS}}$, y se presentan los fundamentos conceptuales de las teorías efectivas de campos. Uno de los resultados centrales de este trabajo es la construcción y clasificación sistemática de los operadores bosónicos en SMEFT de dimensión ocho, empleando técnicas de Teoría de Grupos y eliminando redundancias mediante el trabajo en espacio de momentos. A partir de esta base de operadores, se calculan las ecuaciones completas del grupo de renormalización (RGEs) a un lazo que involucran inserciones de operadores de dimensión ocho o inferior. Esto incluye efectos puros de dimensión ocho, pares de operadores de dimensión seis y sectores que violan el número leptónico. Nuestros cálculos utilizan una base de funciones de Green fuera de su capa de masa y aprovechan simplificaciones algebraicas derivadas de la simetría y la invariancia gauge.

Estos resultados se aplican a cotas de positividad y a los `parámetros obli\-cuos', proporcionando herramientas esenciales para análisis coherentes dentro de SMEFT en distintos regímenes de energía. Los hallazgos amplían el alcance teórico de SMEFT y respaldan su uso en estudios fenomenológicos de alta precisión.

\clearpage\hbox{}\thispagestyle{empty}\newpage


\printglossaries\thispagestyle{empty}
\clearpage
\mainmatter

\chapter{Introduction} \label{ch:Intro}
\section{Motivation}
In the latter half of the 20th century, collider experiments revealed that particles once thought to be elementary actually exhibited substructure, leading to the identification of quarks and other subnuclear constituents. 
Such experiments were interpreted within the general framework of \glspl{qft}, which provided the foundation for more specific theories such as \gls{qed}, \gls{qcd}, and ultimately the \gls{sm} of particle physics. The \gls{sm} provided a precise description of the observed particles and their interactions. However, one piece of the puzzle remained missing: a scalar mediator that would give mass to the otherwise massless particles in the model, through the Higgs mechanism.

The Higgs boson was finally detected in 2012 at the LHC~\cite{ATLAS:2012yve,CMS:2012qbp}, decades after its prediction. By that time, other observations from Cosmology, Astrophysics and Particle Physics had already pointed out inconsistencies of the \gls{sm}. Thus, before the \gls{sm} was empirically validated as the most accurate \gls{qft}, there were already new models in development. Supersymmetry~\cite{Wess:1974tw,Wess1974}, Extra Dimensions~\cite{Arkani-Hamed:1998jmv,Randall:1999ee,Randall:1999vf}, Composite Higgs Models~\cite{Kaplan:1983fs,Kaplan:1983sm,Agashe:2004rs}, and String Theory~\cite{Scherk:1974ca,Green:1984sg,Candelas:1985en} are examples of well-established theoretical frameworks developed before the discovery of the Higgs boson and were actively explored in anticipation of LHC-era data. There are also extensions of the \gls{sm} with Axion-Like Particles~\cite{Peccei:1977hh,Peccei:1977ur,Weinberg:1977ma,Wilczek:1977pj,Kim:1979if,Shifman:1979if,Zhitnitsky:1980tq,Dine:1981rt}, Dark Photons~\cite{Holdom:1985ag} or Leptoquarks~\cite{Georgi:1974sy,Pati:1974yy,Buchmuller:1986zs,Dimopoulos:1979es}, which could explain the deviations of experimental data from the \gls{sm}. 

While the detection of the Higgs boson completed the \gls{sm}, it also sharpened the need to address its known limitations, as numerous phenomena remain unexplained. In the search for the Higgs particle, its mass and couplings were constrained, and these restrictions affected models that predicted alternative phenomena in the MeV-GeV region, such as hidden sectors or dark photons. Later on, the focus of experiments moved to the exploration of higher and higher energies. This led to increased interest in theoretical frameworks capable of addressing the limitations of the \gls{sm}. In the absence of clear \gls{np} signals, precision measurements and indirect constraints have become central tools in the search for \gls{bsm} physics. The situation is reminiscent of the pre-Higgs era, in the sense that searches are imposing restrictions on \gls{np} models. But this time, there is no single dominant theory awaiting validation, nor a specific resonance whose discovery would serve as a definitive turning point. The Higgs's expected properties were predicted with precision, and experiments were focused on these specific signals. In the current experimental landscape, there are several open lines of research looking for dark matter, heavy neutrinos or high-energy effects, all of which cover a wide region of the parameter space. The vast range of possibilities contrasts with the limited set of actual results. 

In this context, precision physics gains much more relevance. Data analysis now depends much more on refining the techniques than before, since there is abundant registered data. The emphasis is now on model-independent approaches and robust statistics. It is important to build a consistent, minimal parameter space where most, if not all, experiments can express and compare their results within the same framework. \gls{qft} leaves freedom to build lots of different models, but this variety complicates the comparison of predictions and the interpretation of data. For this reason, the use of \glspl{eft} is crucial. Precision Physics at lower energies already uses form factors and effective interactions in \glspl{eft}, such as \gls{chpt} (applicable to low-energy \gls{qcd}), the \gls{scet} (used in jet physics) or \gls{hqet} (the \gls{eft} of mesons with heavy quarks). In the case of \gls{bsm} physics, it is sensible to set a common framework, for example, with \gls{smeft}, the \gls{eft} parameterising \gls{np} with \gls{sm} fields and symmetries. Other \glspl{eft} are covering similar energy ranges, like \gls{heft}. However, if the traditional Higgs mechanism for the \gls{sm} is assumed and no \gls{np} is discovered below the \gls{ew} scale, then \gls{smeft} is the most reasonable option for a comprehensive analysis, and indeed it is a popular one, although alternative formulations and extensions always have to be considered.

Using all resources from \gls{qft} in general and, in particular, from \gls{eft}, we can exploit \gls{smeft} so that it becomes a thorough tool for analysis. Two key aspects are discussed in this thesis: the calculation of a minimal set of operators that cover all the possible interactions in \gls{smeft} within a limited accuracy range (that we will specify later), and the running of the couplings and parameters across different energy scales. To put these topics into context, Chapter~\ref{ch:Renormalization} reviews renormalisation in \gls{qft}, and specifically, the advantages of \gls{dimreg} and the \gls{msbar} renormalisation scheme. The concept of power counting is introduced here and explained in more detail in Chapter \ref{ch:EFT}, along with another important resource of \gls{eft}: matching. 

Defining a framework to unify theories requires a common parameterisation of the results. In an \gls{eft}, all the allowed interactions can be expressed in terms of a minimal set, known as a \text{basis} of operators. In Chapter~\ref{ch:SMEFT}, we will see the importance of building a basis of operators for \gls{smeft}, although we will allow certain redundancies for our convenience. In the years before the start of the thesis, there were complete bases of operators up to dimension eight~\cite{Grzadkowski:2010es,Liao:2016hru,Li:2020gnx,Murphy:2020rsh,Lehman:2014jma}. We will use a basis extended to include operators that are redundant via the \gls{eom}, which were only known for dimension six~\cite{Gherardi:2020det} before 2021. Constructing such a basis is a non-trivial mathematical task, but over the last few years, there have been key impactful contributions to the problem. Currently, there are already systematic methods to generate sets of independent operators for \gls{smeft}, but it is not easy to compare different bases or to rotate from one to another. We will see how to build such algorithms, including our contribution to the construction of the redundant bosonic operator basis at dimension eight, which plays a key role in the computation of the \glspl{rge}.

The running of \gls{smeft} operators is currently a hot topic, that saw its first peak a decade ago, with the complete one-loop renormalisation up to dimension-six \gls{smeft} operators~\cite{Chankowski:1993tx,Babu:1993qv,Antusch:2001ck,Davidson:2018zuo,Jenkins:2013zja,Jenkins:2013wua,Alonso:2013hga,Alonso:2014zka}, a milestone that enabled systematic \gls{rge} analyses and consistent matching to \gls{uv} models. Recently, attention has been drawn to dimension-eight operators for different reasons:
\begin{enumerate}
\item They are dominant with respect to dimension-six operators in some observables~\cite{Azatov:2016sqh}.
\item Dimension-six operators do not arise in some models, thus, dimension-eight are the leading contribution~\cite{Chala:2018ari,Murphy:2020rsh} (ignoring dimension-seven).
\item If the \gls{np} scale is low, they could be less suppressed by power-counting effects~\cite{Chala:2018ari,Corbett:2021eux}.
\item As the precision of experimental measurements grows, the theoretical precision needs to increase too~\cite{Corbett:2021eux,Panico:2018hal,Ardu:2021koz}.
\end{enumerate}

As dimension-eight operators get involved in calculations, the need for the running of higher-dimensional \glspl{wc} is starting to become apparent among many researchers. In Chapter~\ref{ch:RGE}, we will review the results of the \glspl{rge} at one loop for all operators up to dimension eight. We will explain in detail the chosen method of computation and comment on other methods. This thesis focuses on one-loop results; higher-loop corrections are beyond our current scope.

Finally, we demonstrate how the results of this work, along with those of other groups, are being applied and what the prospects of this line of research are. In particular, we will show applications to positivity bounds~\cite{Adams:2006sv}, restrictions on the signs of scattering amplitudes imposed by the properties of the S-matrix. These constraints can be used either as tests of the unitarity of the S-matrix or as tools to restrict the \glspl{rge}. By their definition, they directly affect operators of dimension eight or greater, so we will explore some of the consequences in Section~\ref{sec:Applications}. Another avenue to be explored is the oblique parameters, where dimension eight operators contribute at \gls{nnlo}. We will not perform fits or very complex studies here, but only remark on some interesting facts directly deduced from the \glspl{rge}.

\section{Notation and conventions}
We assume the reader is familiar with \gls{qft} and, more specifically, the \gls{sm} and its symmetry groups. 
Amplitudes and Feynman diagrams are written in the conventions of \cite{Schwartz2013}, drawn with Jaxodraw~\cite{Binosi:2003yf}, and computed with FeynRules~\cite{Alloul:2013bka}, FeynArts~\cite{Hahn:2000kx} and FormCalc~\cite{Hahn:1998yk}.

In this thesis, we will describe the \gls{sm} with the following Lagrangian:
\begin{align}\nonumber
 \mathcal{L}_\text{SM} = & -\frac{1}{4}G_{\mu\nu}^{A}G^{A\,\mu\nu} -\frac{1}{4}W_{\mu\nu}^{a}W^{a\,\mu\nu} -\frac{1}{4}B_{\mu\nu}B^{\mu\nu}\\\nonumber
 &
+\overline{q^{\alpha}}\ii\slashed{D}q^{\alpha}
+\overline{\ell^{\alpha}}\ii\slashed{D}\ell^{\alpha}
+\overline{u^{\alpha}}\ii\slashed{D}u^{\alpha}
+\overline{d^{\alpha}}\ii\slashed{D}d^{\alpha}
+\overline{e^{\alpha}}\ii\slashed{D}e^{\alpha}
\\\nonumber
& +\left(D_{\mu}H\right)^{\dagger}\left(D^{\mu}H\right)
+m_H^{2}|H|^{2}-\lambda|H|^{4}
\\
&-\left(
y_{\alpha\beta}^{u}\overline{q^{\alpha}}\widetilde{H}u^{\beta}
+y_{\alpha\beta}^{d}\overline{q^{\alpha}}H d^{\beta}
+y_{\alpha\beta}^{e}\overline{l^{\alpha}}H e^{\beta}
+\text{h.c.}\right)~,
\end{align}
where $B$, $W$ and $G$ represent the gauge bosons of $U(1)_Y$, $SU(2)_L$ and $SU(3)_C$. Likewise, $\ell^\top\equiv~(\begin{matrix} \nu_{L} & e_{L} \end{matrix})^\top$ and $q^\top\equiv~(\begin{matrix} u_{L} & d_{L} \end{matrix})^\top$ are the left-handed leptons and quarks, with $e,\,u,\,d$ as their right-handed counterparts. We will always assume there are $n_f=3$ generations unless otherwise stated. We use $$H^\top\equiv~(\begin{matrix} H^+ & H^0 \end{matrix})^\top\equiv~(\begin{matrix} H_1+\ii H_2 & H_3+\ii H_4 \end{matrix})^\top$$ for the Higgs doublet. We also define the dual vector $\widetilde{H}^\top=\ii (H^\dagger) (\sigma^2)$ and the dual field strength tensors: $$\widetilde{F}_{\mu\nu}=\frac{1}{2}F^{\rho\sigma}\epsilon_{\rho\sigma\mu\nu}$$ where $F_{\mu\nu}=B_{\mu\nu},\,W^I_{\mu\nu},\,G^A_{\mu\nu}$ and $\epsilon_{0123}=+1$.

We use the minus-sign covariant derivative:
\begin{equation}
D_\mu = \partial_\mu - \ii g_1 Y B_\mu -\ii g_2\frac{\sigma^I}{2} W_\mu^I -\ii g_3\frac{\lambda^A}{2} G_\mu^A\,,
\end{equation}
with $Y$, $\sigma^I$ and $\lambda^A$ being the hypercharge operator, Pauli and Gell-Mann matrices, respectively; while $g_1$, $g_2$ and $g_3$ stand for the gauge boson couplings.

We will declare consistent conventions for the indices, and follow them strictly unless there is no possible confusion:
\begin{description}
\item[Isospin:] $i$, $j$, $k$, $l$ going from $1$ to $2$
\item[Flavour:] $\alpha$, $\beta$, $\gamma$, $\delta$, $\epsilon$, $\zeta$ going from $1$ to $3$
\item[Colour:] $a$, $b$, $c$ going from $1$ to $3$
\item[$SU(2)$ adjoint generators:] $I$, $J$, $K$, $L$ going from $1$ to $3$
\item[$SU(3)$ adjoint generators:] $A$, $B$, $C$ going from $1$ to $8$
\item[Lorentz:] $\mu$, $\nu$, $\rho$, $\sigma$, $\kappa$, $\lambda$ going from $0$ to $3$
\end{description}

\chapter{Regularisation and renormalisation}\label{ch:Renormalization}

Renormalisation is a foundational concept in \gls{qft}, with deep mathematical roots in the study of self-similar systems and scaling behaviour. Broadly speaking, renormalisation refers to the redefinition of a theory's parameters such that its predictions remain consistent when probed at different energy scales\footnote{This connection between \gls{qft} renormalisation and fractals was inspired by the lecture notes of Prof. McGreevy~\cite{McGreevy2018}.}. A well-known illustration of this idea is found in fractals: mathematical structures that display identical patterns upon successive magnification, a property known as `scale invariance'~\cite{Mandelbrot1999}.

Self-similarity is not exclusive to abstract mathematics; it is also observed in physical systems, such as turbulent flows, biological growth patterns, and critical phenomena in statistical mechanics~\cite{Mandelbrot1999}. In the context of \gls{qft}, it underlies the structure of the Lagrangian formalism. The parameters appearing in a Lagrangian—such as coupling constants and masses—are not directly observable and, when inserted into perturbative calculations, often lead to \gls{uv} divergences in loop integrals \cite{Peskin:1995ev,Schwartz2013}.

To address these divergences, a two-step procedure is implemented. First, \textit{regularisation} is introduced: a mathematical scheme, such as dimensional regularisation or a momentum cutoff, that renders the divergent integrals well-defined. Second, the divergences are absorbed through \textit{renormalisation}, which redefines the theory's bare parameters to cancel out infinities, resulting in finite, physically meaningful predictions \cite{Schwartz2013}. The resulting \textit{renormalised} Lagrangian maintains the same form as the original but with scale-dependent, analytic couplings and field normalizations.

This procedure can be understood conceptually as a type of scale transformation: just as the recursive definition of a fractal preserves essential structural properties, renormalisation in \gls{qft} modifies the parameters while preserving the form of the theory. This idea is formalized in the language of the \gls{rg}, which governs the flow of parameters across energy scales \cite{Wilson:1971bg, Weinberg:1995mt}.

In this chapter, we provide a structured overview of regularisation and renormalisation in \gls{qft}. We begin by examining the origin of divergences in perturbative calculations and survey several regularisation techniques. Special emphasis is placed on \gls{dimreg} and the modified minimal subtraction \gls{msbar} scheme, which are used throughout the remainder of this thesis. We also discuss the distinction between renormalisable and non-renormalisable theories, and set the stage for the use of \glspl{eft}, such as the \gls{smeft}, in handling non-renormalisable interactions in a consistent and predictive framework \cite{Manohar:1996cq, Manohar:2018aog}.

\section{Divergences}\label{sec:Divergences}

\begin{figure}
\begin{center}
\includegraphics[scale=0.13]{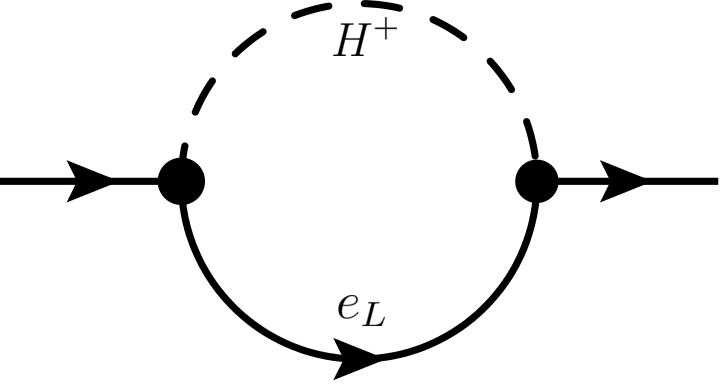}\qquad
\includegraphics[scale=0.13]{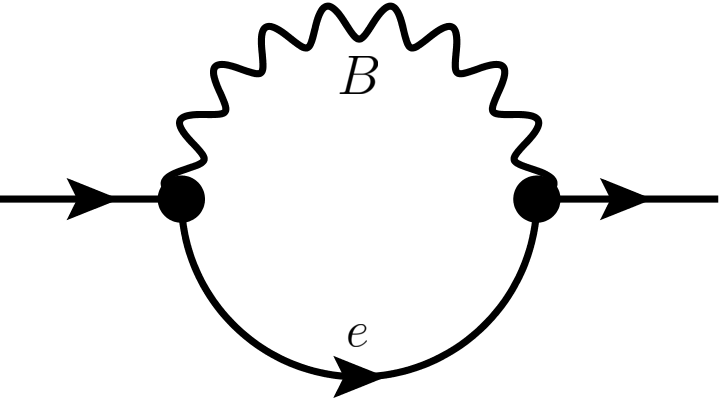}\qquad
\includegraphics[scale=0.13]{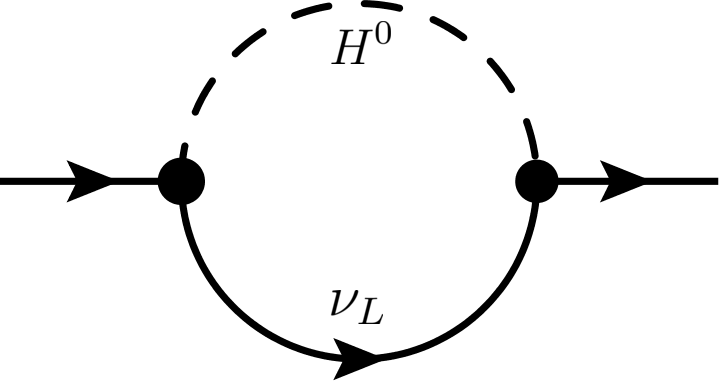}
\end{center}
\caption{\label{fig:WFRexample}One-loop self-energy diagrams contributing to the propagator of the right-handed electron $e$ in the unbroken Standard Model. Each diagram involves a fermion-boson loop and is labelled according to the hypercharge of the internal fermion line: Left: $ \Sigma_{-\frac{1}{2}} $ (left-handed electron and charged Higgs). Center: $ \Sigma_0 $ (right-handed electron and $ B $ gauge boson). Right: $ \Sigma_{\frac{1}{2}} $ (left-handed neutrino and neutral Higgs). These diagrams yield logarithmic divergences and motivate the introduction of regularisation and renormalisation.
}
\end{figure}

In perturbative quantum field theory, observables such as cross-sections and decay rates are computed using Green's functions, which are derived from the theory's Lagrangian. However, the Lagrangian itself is not directly physical: it contains parameters—such as couplings and masses—that serve as inputs to a formalism rather than measurable quantities. These parameters often lead to divergences when inserted into loop-level Feynman diagrams.

The divergences encountered in loop calculations originate from the \gls{uv} region of the momentum integrals, where virtual particles probe arbitrarily high energies. Without a systematic method to regulate and absorb these divergences, the Green's functions become ill-defined and physical predictions are lost. This issue necessitates the introduction of regularisation and renormalisation, which are discussed throughout this chapter. These methods allow us to redefine the parameters in such a way that physical predictions remain finite and consistent with experimental results~\cite{Peskin:1995ev, Schwartz2013}.

To illustrate the nature of such divergences, we begin with a concrete example: the \gls{wfr} at one loop for a right-handed charged lepton in the unbroken \gls{sm}. The relevant two-point Green's function is defined as $ \ii G(p) = \langle e_\alpha e_\alpha \rangle $, which corresponds to the free propagator at the tree level. 

At one loop, three diagrams contribute corrections to this quantity. These corrections can be expressed in the form:
\begin{equation}\label{eq:WFRexample1LampDef}
\ii G_{\text{TL}}(p) \, \ii \Sigma(p) \, \ii G_{\text{TL}}(p),
\end{equation}
where $\ii G_{\text{TL}}(p)$ is the tree-level propagator and $\Sigma(p)$ represents the self-energy insertion. The contributing diagrams, shown in Figure~\ref{fig:WFRexample}, involve various combinations of bosons and fermions circulating in the loop. We label them by the weak isospin of the internal fermion line:
\begin{itemize}
    \item $ \Sigma_{-\frac{1}{2}} $: A charged Higgs loop with a left-handed electron.
    \item $ \Sigma_0 $: A gauge boson loop with a right-handed electron.
    \item $ \Sigma_\frac{1}{2} $: A neutral Higgs loop with a left-handed neutrino.
\end{itemize}

For example, the diagram where a left-handed electron and a charged Higgs circulate in the loop yields:
\begin{align}\label{eq:Sigma_minus1}
\ii[\Sigma_{-\frac{1}{2}}]&\equiv ([y^e]^\top [y^e]^*)_{\alpha\alpha} \mathcal{I} \quad\text{(no sum over }\alpha)\,,\\
\mathcal{I} &=\int \frac{\rm{d}^4 k}{(2\pi)^4}  \frac{ \slashed{k}}{k^2+\ii\varepsilon}\frac{1}{(p-k)^2+m_H^2+ \ii\varepsilon}\,,
\end{align}
where $y^e$ is the Yukawa coupling of the lepton.

Using Feynman parameters, Wick rotation and variable change, we can transform the integral into a standardised expression with a general solution:
\begin{equation}\label{eq:Divergence}
\mathcal{I} = \int \frac{{\rm{d}}^d k}{(2\pi)^d} \frac{1}{(k^2-\Delta+ \ii\varepsilon)^2} = \frac{\ii}{(4\pi)^\frac{d}{2}}\frac{1}{\Delta^{2-\frac{d}{2}}}\Gamma\left(\frac{4-d}{2}\right)\,,
\end{equation} which is divergent as $k\rightarrow \infty$ for $d=4$ dimensions.

These integrals are logarithmically divergent in the \gls{uv} limit and must be regulated. In the sections that follow, we will evaluate them using dimensional regularisation to isolate and cancel the divergent pieces through the renormalisation process.

\section{Regularisation}

To address \gls{uv} divergences in loop integrals, \gls{qft} employs various regularisation methods, each introducing a formal prescription (a `regulator') that modifies the divergent integrals to make them finite and computable. These methods affect the structure of the theory to different extents and vary in their compatibility with symmetries. Some examples are:

\paragraph{Cutoff}

The cutoff method introduces a physical energy scale $\Lambda$ that restricts the integration domain of the loop momentum. This approach is intuitive but breaks Lorentz invariance and is not suited for preserving gauge symmetries. Its simplicity makes it useful for rough or pedagogical calculations, though it lacks theoretical elegance.

\paragraph{Derivative method}

This method involves differentiating the divergent integral with respect to a dimensionful parameter until the result becomes convergent, then integrating it back. While this cancels divergences formally, it introduces arbitrary integration constants $\Lambda_1, \Lambda_2, \dots$, making it ambiguous unless symmetry constraints (like gauge invariance) uniquely determine those constants.

\paragraph{Pauli-Villars}

A more systematic variation of the derivative method, this regulator introduces unphysical `ghost' fields\footnote{Not to be confused with Faddeev-Popov ghosts.} (with modified statistics) to cancel divergences. These fields do not correspond to real particles but are inserted in the Lagrangian. While it works well in Abelian theories, this approach can violate gauge invariance in non-Abelian settings and becomes cumbersome with higher-loop corrections.

\paragraph{Lattice}

Spacetime is discretised into a lattice, effectively regulating integrals by removing the \gls{uv} limit. This method preserves unitarity and is well-suited for non-perturbative problems (e.g., \gls{qcd}), but breaks Lorentz symmetry and is computationally demanding. It is not easily adapted to theories with chiral fermions.

\subsection{(Naive) Dimensional Regularisation} 

\gls{dimreg} stands out as the most widely adopted regularisation scheme available in \gls{qft}. Its popularity stems from its ability to preserve key symmetries of the theory --Lorentz, gauge, and, in the case of massless fermions, chiral symmetry--while retaining the analytic structure of Green's functions. Although somewhat abstract in its formulation, dimensional regularisation has become the standard in loop calculations and is the default approach in the majority of modern perturbative analyses.

Let us return to the loop integral presented earlier in Equation~\eqref{eq:Divergence}, which was shown to be logarithmically divergent. The divergence arose from the high-momentum behaviour of the integrand, which scales as 
$k^3 \frac{\diff k}{k^4}=\frac{\diff k}{k}$. Rather than altering the limits of integration or modifying the particle content of the Lagrangian, dimensional regularisation instead modifies the dimension of the integration measure itself.
This is achieved by analytically continuing the number of spacetime dimensions from four to $d = 4 - 2\epsilon $, with the limit $\epsilon \to 0$ taken at the end of the calculation. While the notion of non-integer dimensions is mathematically formal and lacks direct physical interpretation, the method allows for a controlled expansion of divergent quantities in terms of poles in $\epsilon$.

Using \gls{dimreg} the scalar loop integral appearing in Equation~\eqref{eq:Divergence} becomes:
\begin{equation}\label{eq:dimregInt}
\mathcal{I}=\frac{\ii}{(4\pi)^\frac{d}{2}}\frac{1}{\Delta^{2-\frac{d}{2}}}\Gamma\left(\frac{4-d}{2}\right)=\frac{\ii}{(4\pi)^{2-\epsilon}}\frac{1}{\Delta^{\epsilon}}\Gamma\left(\epsilon\right)\,,
\end{equation}
where the divergence is parametrised by $\epsilon\rightarrow 0$.

\gls{dimreg} also affects the mass dimensions of fields and couplings. For example, the dimension of a Yukawa coupling $y_e$ shifts as:
\begin{equation}\label{eq:DimensionShift}
[y^e]=0\mapsto [y^e]=\epsilon \Rightarrow y^e\mapsto\mu^\epsilon y^e\,,
\end{equation}
where $\mu$ is an arbitrary mass scale introduced to compensate for the change in dimension, ensuring that the interaction terms in the Lagrangian remain $d$-dimensional in $d$ spacetime dimensions.

Including the mass dimension correction $\mu^{2\epsilon}$ by the two Yukawa couplings in the amplitude~\eqref{eq:Sigma_minus1} and expanding~\eqref{eq:dimregInt} around $\epsilon=0$ we get\footnote{We use $\Gamma(\epsilon)=1/\epsilon-\gamma_E+\mathcal{O}(\epsilon)$ and $x^{-\epsilon}=1-\epsilon \ln x + \mathcal{O}(\epsilon^2)$.}:
\begin{equation}\label{eq:DivergenceExpanded}
\mu^{2\epsilon}\frac{\ii}{(4\pi)^{2-\epsilon}}\frac{1}{\Delta^{\epsilon}}\Gamma\left(\epsilon\right)\approx \frac{\ii}{16\pi^2}\left(\frac{1}{\epsilon} + \ln\left(\frac{4\pi e^{-\gamma_E}\mu^2}{\Delta}\right) + \mathcal{O}(\epsilon)\right)\,,
\end{equation}
where $\gamma_E$ is the Euler-Mascheroni constant. Thus, the regulated amplitude~\eqref{eq:Sigma_minus1} is:
\begin{equation}\label{eq:regulatedSigma_minus1}
\ii [\Sigma_{-\frac{1}{2}}]= \ii \frac{([y^e]^\top [y^e]^*)_{\alpha\alpha}}{16\pi^2 \epsilon}\slashed{p}+\text{finite}\,.
\end{equation}

It is important to emphasise that the scale $\mu$ introduced here is not a physical cutoff. It only appears to ensure dimensional consistency, but eventually disappears from physical observables after renormalisation. Nonetheless, intermediate expressions --such as beta functions and anomalous dimensions-- can depend on it explicitly.

\section{Counterterms and renormalisation conditions}

Renormalisation is the process by which divergences in loop amplitudes are absorbed into redefinitions of the theory's bare parameters and fields. These redefinitions introduce counterterms that cancel divergent contributions from loop diagrams~\cite{Peskin:1995ev}. The freedom in choosing the finite parts of these counterterms is what defines a \textit{renormalisation scheme}.

The choice of scheme does not affect physical observables, such as S-matrix elements, but it does alter intermediate quantities like Green's functions and running couplings. Each scheme imposes different conditions on the counterterms and thus results in different expressions for renormalised parameters. 

The formalism applies at all-loops order. In perturbation theory, we expand the counterterms $Z$ in terms with the same loop suppression $Z=1+\delta+\mathcal{O}(\text{2-loop})$, where $\delta$ will also be called \textit{counterterms} without possible confusion in this text, since we always work at one loop.

\subsection{Onshell scheme}

In the \gls{onshell} scheme, renormalisation conditions are imposed so that renormalised quantities match physically measured ones. For example, propagator poles are fixed at physical particle masses, and the residues at those poles are normalized to unity~\cite{Weinberg:1995mt}. 

\begin{figure}
\includegraphics[width=\textwidth]{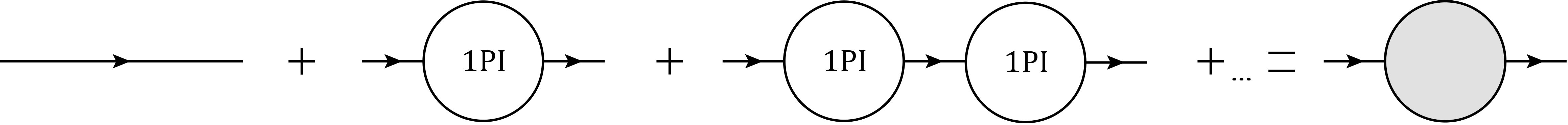}
\caption{\label{fig:DressedPropagator} The resummation of 1PI diagrams leads to a diagram formally similar to the tree-level expression: the dressed propagator.}
\end{figure}

Applying this to the right-handed lepton propagator, we resum the \gls{1pi} 1-loop contributions due to the self-energy $\Sigma(p)$ into the dressed propagator (see Figure~\ref{fig:DressedPropagator}). Then, we fix the renormalised self-energy $\Sigma_R(p)$ such that
\begin{equation}
\left.\frac{\rm{d}}{\rm{d}\slashed{p}}\Sigma_R(\slashed{p})\right|_{\slashed{p}=0}=0.
\end{equation}
 
Although physically motivated, the \gls{onshell} scheme becomes cumbersome in theories with many parameters or in \glspl{eft}, where not all couplings are physical observables.

\subsection{Minimal Subtraction scheme}
The \gls{ms} scheme introduces counterterms that cancel only the divergent parts of loop amplitudes. In dimensional regularisation, divergences appear as poles in $ \epsilon = (4 - d)/2 $. \gls{ms} counterterms $\delta^{\text{MS}}$ subtract these poles without touching finite terms~\cite{tHooft:1973wag}:
\begin{equation}
\delta^{\text{MS}} = \ii \frac{([y^e]^\top [y^e]^*)_{\alpha\alpha}}{16\pi^2 \epsilon}\slashed{p}\,.
\end{equation}

This approach is simple and efficient, especially when dealing with large numbers of parameters, and is well-suited to theories like \gls{smeft}. However, it lacks direct physical interpretation since renormalised masses and couplings do not correspond to physical observables.

\subsection{Modified Minimal Subtraction scheme}\label{sec:MSbar}

The \gls{msbar} scheme~\cite{Bardeen:1978yd} improves on \gls{ms} by removing not only the $1/\epsilon$ poles but also associated constants such as $\ln(4\pi)$ and the Euler–Mascheroni constant $\gamma_E$ (as was the case in equation \eqref{eq:DivergenceExpanded}). This modification improves the behaviour of beta functions and is the standard scheme used in the computation of renormalisation group equations.

For a generic parameter $g$ such that $[g]_{d}=1-\epsilon$ the relation between bare and renormalised forms is written as
\begin{equation}\label{eq:ScaleDependence}
g_0 = \mu^\epsilon Z_g g_R(\mu),
\end{equation}
where $\mu$ is the renormalisation scale\footnote{We will discuss its relevance in Section~\ref{sec:CallanSymanzyk}.}, and $Z_g$ contains only the divergent and $\gamma_E$-dependent pieces in \gls{msbar}. 

\subsection{Renormalised Lagrangian}

Once a renormalisation scheme is chosen, the renormalised Lagrangian is obtained by substituting bare parameters and fields with their renormalised counterparts and associated counterterms. For instance, the kinetic term of a fermion field $\psi$ is written at one-loop order as
\begin{equation}
\mathcal{L} = Z_\psi \overline{\psi} \ii \slashed{\partial} \psi = \overline{\psi}_R i\slashed{\partial} \psi_R + \delta_\psi \overline{\psi}_R i\slashed{\partial} \psi_R.
\end{equation}
This decomposition separates the renormalised term and the counterterm. The same applies to mass and interaction terms. The complete renormalised Lagrangian thus consists of the original Lagrangian (in terms of renormalised parameters) plus a sum of counterterms determined by the chosen scheme.

These renormalisation techniques are foundational for the treatment of loop corrections in \glspl{eft}. In particular, they are applied throughout Chapter \ref{ch:RGE} to study the \gls{smeft}, where the dimension-six and dimension-eight operators require careful handling of counterterms and running couplings across scales.

\section{Renormalised perturbation theory}

Perturbative quantum field theory systematically expands around the free theory using a series of small parameters—usually couplings or inverse mass scales. However, we have seen loop diagrams often lead to divergences that must be handled carefully. Renormalised Perturbation Theory provides a consistent framework to perform these calculations by starting from a Lagrangian expressed in terms of renormalised fields and couplings. Counterterms are introduced from the beginning and are determined order-by-order in perturbation theory~\cite{Peskin:1995ev}.

Instead of adding counterterms after encountering divergences in each process, the approach assumes their presence and uses physical constraints or matching conditions to fix their coefficients. This method is not only more elegant but also essential in theories like the \gls{smeft}, where the structure of the theory at low energies reflects the influence of physics at higher scales.

In the \gls{sm}, renormalised perturbation theory leads to a finite set of counterterms sufficient to absorb all one-loop divergences. These include approximately\footnote {We are not counting gauge fixing or redundant terms that can also be included in the \gls{sm}.} 33 counterterms~\cite{Gross:1973id,Politzer:1973fx,Gross:1973ju}
\begin{itemize}
    \item 19 \gls{wfr} (for gauge bosons, Higgs, and all \gls{sm} fermions),
    \item 13 coupling renormalisations (for $g_1$, $g_2$, $g_3$, all Yukawas, and the Higgs quartic coupling),
    \item 1 mass parameter renormalisation (the Higgs potential parameter $m_H^2$).
\end{itemize}
Once these are fixed by appropriate renormalisation conditions, physical predictions such as cross-sections and decay rates become finite~\cite{Weinberg:1995mt}.

Renormalised Perturbation Theory allows for a structured and symmetry-preserving approach to loop calculations in both renormalisable theories and \glspl{eft}. Its general principles—introduction of counterterms, renormalisation conditions, and gauge-invariant regularisation methods—serve as the core of modern amplitude computations, and will be extensively applied in the remainder of this thesis, particularly in the context of \gls{smeft}. However, we must first clear out the role renormalisation plays in theories where there is no finite number of counterterms.

\section{Renormalisability \label{sec:DefiningRenormalizability}}

A quantum field theory is said to be renormalisable if all \gls{uv} divergences in its amplitudes can be absorbed into a finite number of counterterms. This condition applies not only to Green’s functions but also to S-matrix elements since the latter are obtained from the former via the LSZ reduction formula~\cite{Lehmann1955}, which does not introduce additional divergences. Thus, the divergences of S-matrix elements are controlled by the divergences of the underlying Green’s functions, meaning that a consistent renormalisation of the latter ensures the finiteness of the former.

In practice, the divergences arise from \gls{1pi} diagrams. These are diagrams that cannot be disconnected by cutting a single internal line. Connected (reducible) diagrams are built by combining \gls{1pi} diagrams and do not introduce new divergences beyond those already present in their \gls{1pi} components. This is because any additional line connecting two \gls{1pi} blobs carries no loop momentum and hence does not generate \gls{uv} divergences. Therefore, it suffices to renormalise the \gls{1pi} Green’s functions at a given loop order. 

Of course, not every \gls{1pi} diagram is divergent. Since our interest lies in renormalisation, it is convenient to define a tool to discard finite diagrams. This classification is non-trivial, thus we begin by using a simpler, though only approximate, instrument.

We define the \textit{superficial degree of divergence} $\mathscr{D}$ of a loop diagram as the scaling power of the diagram in the \gls{uv} limit, where the loop momentum becomes large~\cite{Weinberg:1959nj}. For a \gls{1pi} diagram with $n_X$ gauge bosons, $n_\psi$ fermions, $n_\phi$ scalars, and $n_i$ insertions of operator $\mathcal{O}_i$ with mass dimension $r_i=[\mathcal{O}_i]$\footnote{Assuming canonical mass dimension for all fields.}, the superficial degree of divergence is 
\begin{equation}\label{eq:DefSuperfDiv4}
\mathscr{D}=4-3/2 n_\psi-n_\phi-n_X-\sum n_i (4-r_i)
\end{equation}
in a $4$-dimensional theory. This can be deduced by counting loop momenta and noting that the superficial degree of divergence is related to the mass dimension of the matrix element that contains the \gls{1pi} Green's function.

It is a useful diagnostic tool in determining whether a Feynman diagram is divergent:
\begin{itemize}
    \item $\mathscr{D}<0$: the diagram is convergent,
    \item $\mathscr{D}=0$: it is logarithmically divergent,
    \item $\mathscr{D}>0$: it exhibits power-law divergence proportional to $ \Lambda^\mathscr{D} $, where $\Lambda$ is a \gls{uv} cutoff.
\end{itemize}

This estimate, however, is \textit{superficial} — it assumes no cancellations due to symmetries or specific vertex structures. Therefore, diagrams with non-negative $\mathscr{D}$ might still be finite, or even vanish by construction.

\begin{figure}
\begin{center}
\includegraphics[scale=0.13]{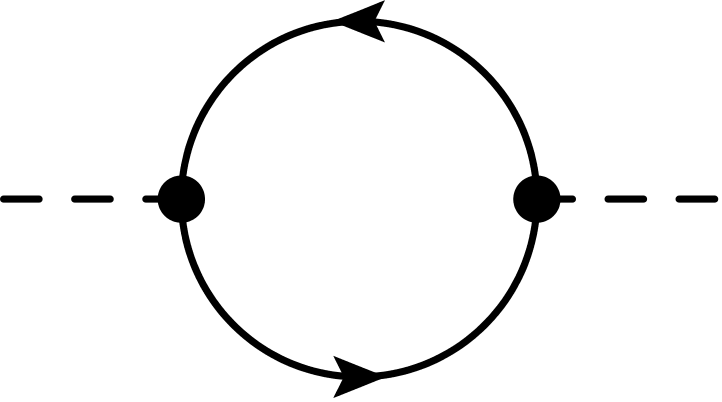}\qquad
\includegraphics[scale=0.13]{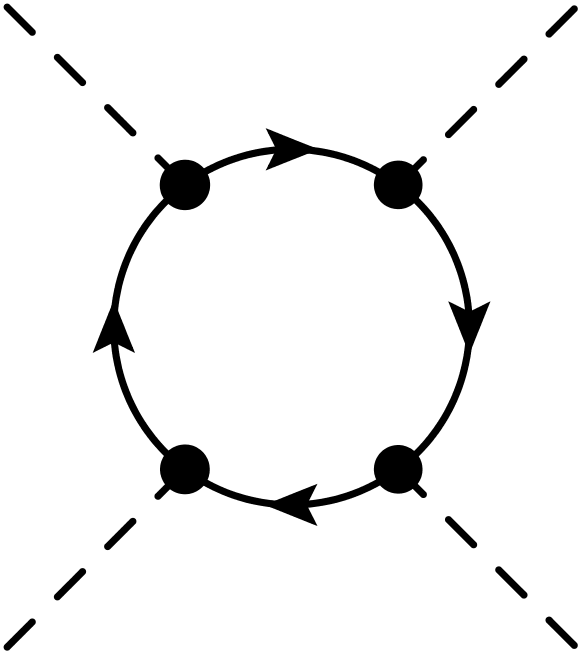}\qquad
\includegraphics[scale=0.13]{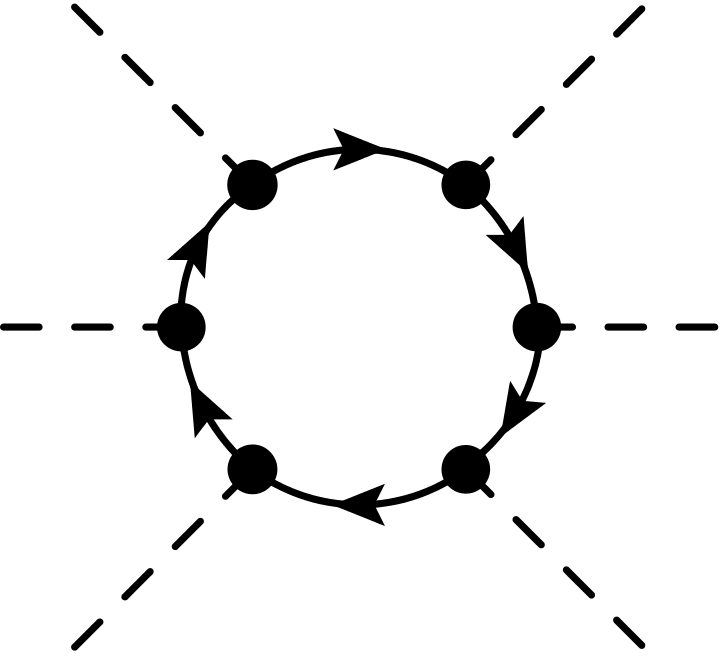}
\end{center}
\caption{\label{fig:SuperfDegreeDivergExample}Left: a Green's Function with a positive superficial degree of divergence. Centre: A Green's Function with zero superficial degree of divergence (in this case, it is divergent). Right: A Green's Function with a negative superficial degree of divergence.}
\end{figure}

To illustrate this, consider the diagrams shown in Figure~\ref{fig:SuperfDegreeDivergExample}. The diagram on the left, which contributes to Higgs wave-function renormalisation, has a positive $\mathscr{D}$ and is divergent. The diagram on the centre, contributing to the quartic Higgs coupling, has $\mathscr{D}=0$ and is logarithmically divergent. In contrast, the diagram on the right has $\mathscr{D}<0$ and yields finite results.

In the \gls{sm}, all operators have $r_i\le 4$, but one can wonder what happens to $\mathcal{D}$~\eqref{eq:DefSuperfDiv4} when the mass dimension is higher than $4$. Each insertion of a higher-dimensional operator increases $\mathcal{D}$ by $r_i-4$, making the superficial degree of divergence more positive with each insertion, and thus the corresponding diagram will be increasingly divergent. The effect of such divergences will need to be renormalised with more higher-dimensional operators and so on, leading to potentially an infinite tower of counterterms.  

In renormalisable theories, power counting implies that only diagrams with a small number of external fields can be divergent. This ensures that only a finite number of counterterms are needed, corresponding to operators of dimension four or less. This is not the case for non-renormalisable theories, like \glspl{eft}, which have higher-dimensional operators. As an anticipation for the following chapter, this issue is addressed by fixing a certain power counting above which all contributions are negligible. Thus, higher-dimensional operators will only be renormalised provided they have a sufficiently low power counting order.

\chapter{EFTs and matching}\label{ch:EFT}

\section{Separation of scales}

The principle of \emph{separation of scales} is foundational in physics, underpinning both theoretical frameworks and experimental methodologies. It refers to the idea that physical phenomena occurring at vastly different energy (or length) scales can often be studied independently. Historically, this approach has enabled the simplification of complex systems by focusing only on the relevant degrees of freedom at a given scale.

A classic example is the description of planetary motion: Kepler's laws and Newtonian gravity successfully predict orbital dynamics without requiring knowledge of planetary topography. Even in modern high-precision measurements, detailed surface features such as mountains and craters have negligible influence on celestial mechanics. Likewise, the Earth's electromagnetic field can be characterised without accounting for local, small-scale sources like refrigerator magnets.

This principle is not specific to gravity or electromagnetism. It is a general feature of many physical theories, including those that govern subatomic interactions. For instance, chemical reactions are typically studied using \gls{qed}, while ignoring weak and strong interactions—these only become relevant at much higher energies or shorter distances. At nuclear scales, the strong interaction becomes significant, yet quark and gluon degrees of freedom remain inaccessible until energies reach the subnuclear regime.

The separation of scales enables simplifications across physical contexts. Classical mechanics is adequate for macroscopic systems, while \gls{qft} is essential at subatomic distances. The key insight is that high-energy (or \gls{uv}) physics becomes irrelevant to low-energy (or \gls{ir}) observables—except through renormalised parameters or suppressed corrections.

\glspl{eft} systematise this concept within the language of Lagrangians. In \glspl{eft}, short-distance physics is encoded through higher-dimensional operators in a local, low-energy theory. These operators are suppressed by powers of a large scale $\Lambda$ and their contributions are organised by a power-counting scheme. This allows theoretical predictions to be systematically improved by including higher-order terms as required by experimental precision.

Importantly, \glspl{eft} can be formulated even without full knowledge of the \gls{uv} theory. This makes them especially valuable in contexts where the high-energy completion is unknown or inaccessible. Calculations are simplified because irrelevant operators—those suppressed by high powers of $1/\Lambda$—can be safely neglected below the matching scale.

Conversely, \glspl{eft} can also serve as a bottom-up tool. By matching high-energy experimental data to the \gls{eft} parameters, one can constrain or even infer properties of potential \gls{uv} theories. This dual role—as a predictive low-energy model and a tool for model-independent \gls{uv} inference—makes \glspl{eft} indispensable in contemporary particle physics.

\section{Motivation and description of EFTs}

In its most general form, an \gls{eft} is a \gls{qft} designed to describe physical processes at energies below a certain cutoff scale $\Lambda$. This framework is motivated by the principle of separation of scales: at low energies, the effects of high-energy degrees of freedom can be encoded through local interactions without requiring an explicit treatment of the full \gls{uv} theory.

Typically, an EFT is constructed by identifying the relevant light fields and their symmetries, while systematically integrating out the heavy degrees of freedom. This process yields an infinite series of higher-dimensional operators, suppressed by inverse powers of the cutoff scale.

From a computational perspective, EFTs offer predictive power because calculations can be organised as a power series expansion in the small parameter $p/\Lambda$ where 
$p$ is the characteristic momentum of the process. If the theory is known to order 
$n$, the resulting theoretical uncertainty is of order 
$\mathcal{O}(p/\Lambda)^{n+1}$. This truncation ensures that only a finite number of operators need to be considered at any desired accuracy.

Historically, \glspl{eft} have been employed long before their formal methodology was established. A notable example is Fermi's theory of weak interactions, which effectively described beta decay well before the discovery of the electroweak gauge bosons. Modern developments, such as the \gls{smeft}, extend this philosophy by incorporating higher-dimensional operators that capture the effects of possible new physics.

The flexibility, precision, and universality of \glspl{eft} have made them an indispensable tool in both theoretical and experimental particle physics. Their formulation allows for systematic improvements and error estimation, even in the absence of a fully known \gls{uv} theory.

\section{Power Counting and mass dimension}\label{sec:PowerCounting}
\glspl{eft}, as quantum field theories, are expressed through Lagrangians that encapsulate all relevant physics below a cutoff scale. A defining feature of \glspl{eft} is that all observables can be expanded as a power series in the ratio of two scales: the typical energy of the process $m$ and the heavy scale $M$, often associated with new physics. This ratio defines the power-counting parameter:
\begin{equation}
\lambda=\frac{m}{M}.
\end{equation}
Typically, \glspl{eft} are constructed to include only a single mass scale $M$, with all heavier degrees of freedom integrated out. Consequently, all quantities in the \gls{eft} can be assigned a scaling behaviour in terms of $\lambda$. To establish this behaviour, we introduce a power counting scheme—a systematic prescription that assigns a scaling dimension to each field, coupling, or operator in the theory:

\begin{equation}
p_\mu \sim m (1,1,1,1) \equiv \lambda M (1,1,1,1),
\end{equation}
which implies
\begin{equation}
x_\mu \sim \frac{1}{m} \sim \frac{1}{\lambda M} \quad \text{and} \quad  \rm{d}^4x=\frac{1}{M^4}\lambda^{-4}\,.
\end{equation}
Thus, momentum scales as $\lambda$ and position scales as $\lambda^{-1}$.

To assign a scaling to fields, we consider their kinetic terms, which dominate at high energy and must be included in the free theory. These terms must be of order $ \lambda^4 $ to match the scaling of the action (which is dimensionless). For example, the kinetic term of a scalar field $\phi$ is:
\begin{equation}
\mathcal{L}_{\text{kin}} = |\partial \phi|^2 \sim \lambda^2 (\lambda_\phi)^2.
\end{equation}
Imposing $\mathcal{L}_{\text{kin}} \sim \lambda^4 $ implies $\phi \sim \lambda$. Analogous arguments apply to gauge and fermion fields, yielding similar results consistent with their canonical mass dimensions.

Since observables are computed by evaluating matrix elements of operators, the power counting of an operator $\mathcal{O} $ scales as:
\begin{equation}
\mathcal{O} \sim \lambda^{[\mathcal{O}]},
\end{equation}
where $ [\mathcal{O}] $ is the canonical mass dimension. This scaling holds for weakly coupled \glspl{eft}, where the kinetic terms dominate. For strongly coupled theories, a different power counting scheme may be required, as interactions can modify the dominant scaling.

In weakly coupled \glspl{eft}, the expansion is often performed simultaneously in powers of $ \lambda$ and the coupling constant $g $. \gls{dimreg} ensures that loop corrections introduce logarithmic dependence on the ratio $ p/\Lambda $, leading to expansions in parameters like $ g \log(p/\Lambda) $. To maintain accuracy, terms with logarithms must be resummed when the logs are large.

This power counting framework enables consistent truncation of the \gls{eft} Lagrangian and allows for reliable estimation of theoretical uncertainties.

\section{Examples}

Before we develop the technical structure of the \gls{smeft}, it is instructive to consider simpler \glspl{eft} that exemplify key features of the framework. These examples, drawn from both particle and atomic physics, serve to highlight conceptual foundations such as separation of scales, power counting, matching, and the emergence of higher-dimensional operators—features that are central to \gls{smeft} but are often obscured by its complexity.

While the concepts of decoupling and low-energy expansion are straightforward in idealised settings, real \gls{uv} theories often involve subtleties such as nontrivial field content or strong interactions. These complications motivate the need for examples that isolate specific technical tools and conceptual strategies. In what follows, we explore a series of \glspl{eft} with distinct structures and purposes, each selected to emphasise a particular principle relevant to the construction and interpretation of \gls{smeft}.

Through this approach, we aim not only to build intuition but also to motivate the technical choices and methods employed later in this thesis. This section was heavily inspired by \cite{Manohar:2018aog,Pich:1998xt}

\subsection{Fermi Theory: A Prototype for Matching in EFTs}

The Fermi theory of weak interactions stands as one of the earliest examples of an \gls{eft}. Before the establishment of \gls{qft} and the \gls{ew} \gls{sm}, Enrico Fermi proposed a contact interaction between four fermions to describe processes like beta decay~\cite{Fermi:1934hr}.

In modern notation, the Lagrangian takes the form:
\begin{equation}
\mathcal{L}_\text{Fermi} = -\frac{G_F}{\sqrt{2}} \left[\bar{\psi}_e \gamma^\mu (1 - \gamma^5) \psi_\nu\right] \left[\bar{\psi}_p \gamma_\mu (1 - \gamma^5) \psi_n\right],
\end{equation}
where $ G_F$ is the Fermi constant, determined experimentally from muon decay\footnote{In the Fermi theory, the interactions responsible for muon decay and beta decay have the same coupling. This can only be understood after the \gls{eft} is \gls{uv}-completed.}. Dimensional analysis reveals that this operator has mass dimension six, making the theory non-renormalisable. Power counting of this Lagrangian term scales as $\lambda^2 \sim G_F E^2 $, which implies a breakdown of perturbative unitarity around energies 
\begin{equation}
E \sim \Lambda_\text{Fermi} \sim 1/\sqrt{G_F} \approx 300 \, \text{GeV}.
\end{equation}

This behaviour is a distinguishing feature of \glspl{eft}: predictive power at low energies, but inconsistencies emerge at energies approaching the cutoff. In this case, the \gls{uv} completion is known—the full electroweak theory, where weak interactions are mediated by massive $ W^\pm $ bosons. The Fermi operator emerges from the tree-level matching of the full Standard Model onto the EFT by integrating out the $W $ boson. The resulting Wilson coefficient is:
\begin{equation}
\frac{G_F}{\sqrt{2}} = \frac{g_2^2}{8 m_W^2},
\end{equation}
where $g_2$ is the $ SU(2)_L $ gauge coupling. This example demonstrates tree-level matching and the identification of Wilson coefficients, a procedure central to \gls{smeft}.

The Fermi theory is also an example of how gauge symmetry can be hidden at low energies. The full electroweak theory has local $SU(2)_L \times U(1)_Y$ symmetry, but in the \gls{eft}, this is effectively replaced by approximate global symmetries. Understanding how symmetry principles constrain operator structure is crucial for constructing consistent \glspl{eft} like \gls{smeft}, which preserve the gauge symmetries of the \gls{sm}.

Thus, Fermi theory serves not just as a historical curiosity, but as a pedagogical prototype for modern \gls{eft} techniques—matching, operator classification, and power counting—that reappear throughout \gls{smeft}.

\subsection{Chiral Perturbation Theory: Constructing EFTs from Symmetry Principles}

\gls{chpt} is the low-energy effective field theory of \gls{qcd} in the presence of light quarks. It provides a canonical example of how an \gls{eft} can be built from symmetry considerations alone, without direct knowledge of the \gls{uv} dynamics — a perspective that is also central to the construction of \gls{smeft}.

The starting point is the observation that \gls{qcd} with $n_q$ massless quarks has a global chiral symmetry:
\begin{equation}
SU(n_q)_L \times SU(n_q)_R \rightarrow SU(n_q)_V,
\end{equation}
which is spontaneously broken by the \gls{qcd} vacuum. The resulting \glspl{ngb} are the light pseudoscalar mesons (pions for $n_q = 2$, or the octet including kaons and the $\eta$ for $n_q = 3$). These \glspl{ngb} are described by a unitary matrix-valued field $ U(x) \in SU(n_q) $, parameterised as:
\begin{equation}
U(x) = \exp\left(\frac{\ii  \sqrt{2} \Phi(x)}{F}\right),
\end{equation}
where $ F $ is the pion decay constant and $\Phi $ collects the meson fields in the adjoint representation.

To construct the effective Lagrangian, \gls{chpt} employs a derivative and mass expansion, with all terms organised by their transformation properties under chiral symmetry. The low-energy theory is built from all operators consistent with:
\begin{itemize}
\item The symmetries of \gls{qcd} (chiral symmetry, parity, Lorentz invariance),
\item The field content (\glspl{ngb} in $U(x)$),
\item A well-defined power-counting in momenta or derivative order.
\end{itemize}

External source fields $v_\mu, a_\mu, s, p$ are coupled to \gls{qcd} via quark bilinears and introduced as background fields transforming under chiral symmetry~\cite{Gasser:1983yg}. These spurions enable a systematic construction of invariant terms in the chiral Lagrangian and allow one to define a generating functional:
\begin{equation}
Z_{\chi\text{PT}}[v,a,s,p],
\end{equation}
which reproduces the same Green’s functions as \gls{qcd} in the low-energy regime:
\begin{equation}
Z_\text{QCD}[v,a,s,p] = Z_{\chi\text{PT}}[v,a,s,p].
\end{equation}

The \gls{lo} Lagrangian, which contains the lowest number of derivatives and quark mass insertions, is:
\begin{equation}
\mathcal{L}_{\text{LO}} = \frac{F^2}{4} \text{Tr}\left[D_\mu U^\dagger D^\mu U + U^\dagger \chi + \chi^\dagger U\right],
\end{equation}
where:
\begin{itemize}
\item $ D_\mu U = \partial_\mu U - \ii r_\mu U + \ii  U l_\mu $, with $ r_\mu = v_\mu + a_\mu $, $ l_\mu = v_\mu - a_\mu $,
\item $ \chi = 2B(s + \ii  p) $, encodes explicit symmetry breaking due to quark masses.
\end{itemize}

This construction mirrors the principles used in \gls{smeft}:
\begin{itemize}
\item Operators are classified by their dimension and symmetry structure~\cite{Grzadkowski:2010es},
\item Background field techniques and spurions are used to build invariant terms~\cite{Henning:2015daa},
\item A power-counting scheme organises the expansion in increasing orders of $ p/\Lambda $, as explained in \ref{sec:PowerCounting} and as was done in~\cite{Buchmuller:1985jz}.
\end{itemize}

In \gls{smeft}, as in \gls{chpt}, we do not require full knowledge of the \gls{uv} theory. Instead, symmetry and field content dictate the allowed operators, which are suppressed by powers of the cutoff scale. While \gls{smeft} is weakly coupled and \gls{chpt} is non-perturbative in its \gls{uv} origin, both serve as examples of systematic, symmetry-based \gls{eft} construction.

\subsection{Soft-Collinear Effective Theory: Factorisation and\\ Mode Separation}

\gls{scet} is a powerful framework developed to describe the interactions of energetic, collimated particles (such as those in jet physics), especially when soft and collinear emissions dominate the dynamics~\cite{Bauer:2000ew,Bauer:2001yt,Bauer:2002aj}. It provides a clear illustration of \gls{eft} techniques adapted to systems with a preferred direction, as is common in collider experiments.

Unlike \gls{chpt}, \gls{scet} does not integrate out entire fields, but instead integrates out energy \textit{modes}. Fields are decomposed into contributions from distinct momentum regions: hard (high-energy), collinear (boosted along a lightlike direction), anti-collinear, and ultrasoft. Each mode is treated as a separate field in the effective theory.

This decomposition gives rise to a novel power-counting scheme based on a small parameter $\lambda $, related to the energy hierarchy between the soft and collinear modes. A generic four-momentum $ p^\mu $ is expressed using light-cone vectors $ n^\mu = (1,0,0,1) $, $ \bar{n}^\mu = (1,0,0,-1) $, and transverse components:
\begin{equation}
p^\mu = n \cdot p \, \frac{\bar{n}^\mu}{2} + \bar{n} \cdot p \, \frac{n^\mu}{2} + p^\mu_\perp.
\end{equation}
Collinear momenta scale as $ p_c \sim (\lambda^2, 1, \lambda) $, anti-collinear as $ ( 1,\lambda^2, \lambda) $, and ultrasoft as $ (\lambda^2, \lambda^2, \lambda^2) $. The \gls{eft} Lagrangian is built to preserve the gauge symmetry and scaling properties of each sector.

This feature distinguishes \gls{scet} from other \glspl{eft}: the power-counting is aniso\-tropic and depends on both the direction and energy of fields. The Lagrangian includes only interactions that respect the scaling laws of the modes involved. For instance, an operator contributing to collinear-quark interactions must scale consistently with $ \lambda $ and conserve gauge symmetry in each sector.

While \gls{smeft} does not distinguish field modes in the same way, both frameworks rely on systematic power counting and matching procedures. The analogy lies in:
\begin{description}
\item[Separation of scales:] \gls{smeft} assumes a hierarchy between the electroweak scale and the new physics scale $ \Lambda $, while \gls{scet} separates hard and collinear scales.
\item[Matching across modes or theories:] Both use diagrammatic and functional \\matching to determine \glspl{wc}.
\item[Power counting:] \gls{scet}’s anisotropic counting has its \gls{smeft} counterpart in operator dimension and loop suppression.
\end{description}

Moreover, the \gls{smeft} analogue of \gls{scet} has been explored for \gls{ew} processes with boosted final states, emphasising the need for \gls{eft} tools that can combine scale hierarchy and directionality in collider phenomenology~\cite{Beneke:2002ni}.

\chapter{SMEFT}\label{ch:SMEFT}

The \gls{smeft} is the \gls{eft} of the \gls{sm} extended with non-renormalisable interactions. Conceptually, this implies the existence of a more fundamental \gls{uv} theory, from which \gls{smeft} can be obtained by integrating out heavy degrees of freedom. While numerous \gls{uv} completions are theoretically possible—each with distinct phenomenological implications—experimental constraints restrict the space of viable models~\cite{Ellis:2018gqa,Ellis:2014jta,Falkowski:2017pss}. Nevertheless, \gls{smeft} remains a valuable framework even without assuming a specific \gls{uv} completion, particularly in the context of weakly coupled extensions of the \gls{sm}.

Classifying operators within an \gls{eft}—and especially within \gls{smeft}—is a nontrivial task that requires careful attention to Lorentz invariance, gauge invariance, and algebraic identities. The subsequent sections develop the formal machinery necessary to implement these constraints and systematically construct a complete operator basis. Although the concepts introduced here are generalisable to any \gls{eft}, our discussion from this point onward will focus specifically on \gls{smeft}.

\section{SMEFT operators and their classification}\label{sec:dependancies}
The field content of the \gls{smeft} is identical to that of the \gls{sm}. We consider \gls{smeft} to be valid up to a cutoff scale $ \Lambda $, beyond which the effective description breaks down. This scale also serves as the matching point in a top-down approach. However, in this thesis, we adopt a bottom-up perspective: we systematically construct all operators consistent with the symmetries of the \gls{sm}, organised by increasing mass dimension and suppressed by powers of $ 1/\Lambda $. The general \gls{smeft} Lagrangian is written as:
\begin{equation}
\mathcal{L}_{\text{SMEFT}} = \mathcal{L}_{\text{SM}} + \sum_{r=5}^{\infty} \sum_q \sum_{p=1}^{n_{\text{ops}}(q)} \frac{c_{r;\,q}^{(p)}}{\Lambda^{r-4}} \mathcal{O}_{r;\,q}^{(p)},
\end{equation}
where $ r$ is the operator dimension, $q $ labels the operator class, and $ p = 1, \dots, n_{\text{ops}}(q) $ indexes all operators within a given class $q $. Unless otherwise noted, all \gls{wc} $ c_{r;\,q}^{(p)} $ are dimensionless.

Operators are grouped into \emph{classes}, defined by their field content and Lorentz structure. For classification purposes, fields are typically grouped into three categories: gauge field strength tensors $ X $, fermions and their conjugates $ \psi $, and Higgs fields and their conjugates $ H $. Covariant derivatives are denoted by $D $. An operator of mass dimension $ r $ belongs to a class $ X^{n_X} \psi^{n_\psi} H^{n_H} D^{n_D} $, satisfying the dimension relation:
$
2n_X + \frac{3}{2}n_\psi + n_H + n_D = r.
$
Some classifications make further distinctions, such as helicity decomposition (e.g., $ X \in \{X_L, X_R\} $, $ \psi \in \{\psi_L, \psi_R\} $) or charge conjugation ($\psi \mapsto \{ \psi, \psi^\dagger \}$, $H \mapsto  \{H, H^\dagger \}$). In this work, we adopt a minimal and generic naming convention for operator classes.

Within each class, operators can be grouped into \emph{subclasses} or \emph{types} that specify the exact field content. For instance, $ \ell^2 H^2 $ is the only subclass of the dimension-five operators $ \psi^2 H^2 $, while the dimension-six class $ X \psi^2 H $ contains multiple subclasses such as $ B \ell^2 H $, $ B e^2 H $, and $ W \ell^2 H $.

When fermions are involved, it is often useful to further group operators into \emph{terms} that reflect flavour indices. Assuming $ n_f $ fermion families, such terms may represent up to $ n_f^{n_\psi} $ operators. Flavour symmetries can reduce this number. For example, the Weinberg operator $ \left[\mathcal{O}_{5;\ell^2H^2}\right]_{\alpha\beta} $ is symmetric under exchange of its flavour indices, yielding $ \frac{n_f(n_f+1)}{2} $ independent operators instead of $ n_f^2 $.

\glspl{eft} generally can contain a large set of higher-dimensional operators at a given dimension. However, many of these operators yield identical contributions to physical observables—such as S-matrix elements—and are therefore considered dependent. From a practical and conceptual standpoint, it is desirable to identify a minimal, non-redundant set of operators that fully capture the dynamics of the theory. To achieve this, one must carefully examine the various mechanisms by which operator dependencies arise, including symmetry constraints, integration by parts, field redefinitions, and equations of motion.
\subsection{Global and gauge symmetries}\label{sec:dependanciesGauge} 
\subsubsection{Bianchi identities}
The Bianchi identities express a geometric constraint on field strength tensors in gauge theories and can be written as a cyclic identity involving covariant derivatives~\cite{Peskin:1995ev}:
 \begin{equation}
D_\rho F_{\mu\nu} + D_\mu F_{\nu\rho} + D_\nu F_{\rho\mu}=0.
\end{equation}

Contracting this identity with the Lorentz-invariant Levi-Civita tensor $\epsilon^{\rho\sigma\mu\nu}$ leads to a condition involving the dual field strength tensor:
\begin{equation}\label{eq:BianchiIdentityForDualFStensor}
D_\sigma \widetilde{F}^{\rho\sigma} =0
\end{equation}
where $\widetilde{F}^{\rho\sigma}=\epsilon^{\rho\sigma\mu\nu}F_{\mu\nu}$. Although this equation resembles an \gls{eom} for the dual field strength, it arises purely from geometric consistency—specifically, from Bianchi identities—and not from the variation of an action.

This distinction becomes conceptually important when constructing operator bases such as Green’s bases, where the goal is to eliminate redundant operators. Since~\eqref{eq:BianchiIdentityForDualFStensor} follows identically from the structure of the gauge field, operators involving this combination are considered redundant and are excluded, even in bases that permit \gls{eom} redundancies. In this sense, dual field strengths are effectively treated as “on-shell” objects to address the Bianchi identities.

\subsubsection{Fierz identities}
In operator constructions involving fermions, spinor algebra often leads to cumbersome expressions involving products of Dirac matrices. These expressions can be systematically simplified using Fierz identities~\cite{Fierz:1937wjm}, which express products of bilinears in terms of alternative contractions. The general identity in a four-dimensional spacetime is given by~\cite{Nishi:2004st}:
\begin{equation}\label{eq:FierzIdentities}
\Gamma_{mn}^P\Gamma_{st}^Q= \frac{1}{4} \text{Tr}[\Gamma^P\Gamma_S\Gamma^Q\Gamma_T]\Gamma_{mt}^T\Gamma_{sn}^S,
\end{equation}
where $P, Q, S, T=1,...16$ and $m,n,s,t=1,...4$ run over spinor components. The set $\Gamma^P$ spans a chiral basis of the  space of $4\times 4$ complex matrices:
\begin{equation}
\Gamma^P\in \left\lbrace R, L, R\gamma^\mu, L\gamma^\mu, \sigma^{\mu\nu}\right\rbrace,
\end{equation}
with projectors $R=\frac{1}{2}(1+\gamma_5)$, $L=\frac{1}{2}(1-\gamma_5)$ and the antisymmetric combination $\sigma^{\mu\nu}=\frac{\ii}{2}\left[\gamma^\mu,\gamma^\nu\right]$. The dual basis is 
\begin{equation}
\Gamma_P\in \left\lbrace R, L, L\gamma_\mu, R\gamma_\mu, \frac{1}{2}\sigma_{\mu\nu}\right\rbrace.
\end{equation} 
These identities are particularly useful for simplifying operators containing at least four fermionic fields. For example, at the tree level in four dimensions, one can relate operators of the form:
\begin{align}
\mathcal{O}_1&=(\overline{\ell} \gamma^\mu \ell)(\overline{e} \gamma_\mu e)=\overline{\ell}_{m}\ell_{n}\overline{e}_{s}e_{t}(R\gamma^\mu)_{mn}(L\gamma_\mu)_{st},\\
\mathcal{O}_2&=(\overline{\ell}_m e_n)(\overline{e}_s \ell_t) \delta_{mn} \delta_{st},
\end{align}
where flavour indices are omitted for brevity. These two operators are related via the Fierz identity~\eqref{eq:FierzIdentities} by:
 \begin{equation}
\mathcal{O}_2=-\frac{1}{2}\mathcal{O}_1. 
 \end{equation}

However, at the loop level—particularly when using dimensional regularisation in $d=4-2\epsilon$ dimensions—this identity no longer holds exactly. The difference between the two forms becomes an operator of order $\mathcal{O}(\epsilon)$, which, when combined with the $1/\epsilon$ poles from \gls{uv} divergences, can generate finite contributions at one loop. At two or more loops, the violation of the Fierz identity can affect divergent structures directly~\cite{Fuentes-Martin:2022vvu}.

To account for this, one formally introduces a new operator defined by:
\begin{equation}
\mathcal{E}=\frac{1}{2}\mathcal{O}_1+\mathcal{O}_2,
\end{equation}
which vanishes in four dimensions but must be retained in $d\neq 4$ calculations. These are known as evanescent operators. Although they contribute only at loop level in \glspl{rge}, they do not affect tree-level amplitudes. In the context of this thesis, we restrict our attention to one-loop renormalisation, where the \glspl{rge} are unaffected by evanescent operators. As such, we will treat Fierz-related redundancies as valid identities, even in \gls{dimreg}.

\subsubsection{Schouten identities}

Another important class of algebraic identities relevant to the simplification of operator bases in \glspl{eft} are the Schouten identities. These identities arise from the linear dependence of vectors in a finite-dimensional space. Specifically, in four-dimensional spacetime, any set of five (or more) four-vectors must be linearly dependent~\cite{Remiddi:2013joa}. This leads to an identity among tensors, involving the metric and the Levi-Civita symbol:
\begin{equation}\label{eq:Schouten}
 g_{\mu\nu}\epsilon_{\alpha\beta\gamma\delta} + g_{\mu\alpha}\epsilon_{\beta\gamma\delta\nu} + g_{\mu\beta}\epsilon_{\gamma\delta\nu\alpha} + g_{\mu\gamma}\epsilon_{\delta\nu\alpha\beta} + g_{\mu\delta}\epsilon_{\nu\alpha\beta\gamma} = 0\,.
\end{equation}

Contracting this identity with a product of five four-vectors $a^\alpha b^\beta c^\gamma d^\delta e^\nu$ demonstrates that a generic linear combination of these vectors must vanish. If the determinant of the matrix formed by these vectors is zero—as enforced by this identity—the vectors are linearly dependent. This relation is fundamental and purely geometric, and it holds only in four dimensions.
  
The Schouten identity leads to nontrivial relations among operator structures. For example, given two field strength tensors $X$ and $F$, and tensors $A_{\mu\nu},\,S_{\mu\nu},\,T_{\mu\nu}$ (antisymmetric, symmetric, and generic rank-2 tensors, respectively), the following identities hold:
\begin{align}\label{eq:4dim1}
  A_{\mu\nu} X^\mu_{\,\,\rho} \widetilde{F}^{\nu\rho} &= A_{\mu\nu} \widetilde{X}^{\mu}_{\,\,\rho} F^{\nu\rho}\,,\\
 S_{\mu\nu} X^\mu_{\,\,\rho} \widetilde{F}^{\nu\rho} &= -S_{\mu\nu} \widetilde{X}^{\mu}_{\,\,\rho} F^{\nu\rho}+\frac{1}{2}S^{\mu}_\mu \widetilde{X}^{\nu\rho} F_{\nu\rho}\,,\\
 T_{\mu\nu} X^{\mu\rho}\widetilde{X}^\nu_{\,\,\rho} &= \frac{1}{4} T_\mu^\mu X^{\nu\rho}\widetilde{X}_{\nu\rho}\,.
\end{align}\pagebreak

These identities are useful in identifying and eliminating redundancies among higher-dimensional operators, particularly those involving dual-field strengths. They effectively allow one to “move the dual” between tensors under contractions. As with Fierz identities, Schouten identities depend on the dimensionality of spacetime. Therefore, in \gls{dimreg}, violations of these relations give rise to evanescent operators, which vanish in four dimensions but can affect loop-level amplitudes when extended to $d=4-2\epsilon$.

\subsection{Integration by parts}\label{sec:ibp} 

Operators containing derivatives can exhibit redundancies arising from total derivatives. Specifically, if an operator takes the form $D_\mu \mathcal{O}^\mu$, it contributes a surface term when inserted into the action. Assuming that all fields vanish at spatial infinity—a common boundary condition that ensures well-defined conserved charges in local, flat \glspl{qft}—such total derivatives integrate to zero in the path integral. This leads to a linear relation among the terms that result from expanding the derivative:
\begin{equation}\label{eq:IBPref}
D_\mu \mathcal{O}^\mu\equiv \mathcal{O}_1 + \mathcal{O}_2 + \dots=0\,.
\end{equation}
Each of the operators $\mathcal{O}_i$ contains the derivative acting on only a subset of the fields in the original expression. This identity follows from applying \gls{ibp} within the path integral of the action.
  
As a concrete example, consider the following scalar operators constructed from the Higgs doublet $H$:

\begin{align}
\mathcal{O}_1 &=\left( H^\dagger H\right)D_\mu\left( H^\dagger H\right)D^\mu\left( H^\dagger H\right)\,,\\
\mathcal{O}_2 &=\left( H^\dagger H\right)^2\left(D^2 H^\dagger H+ H^\dagger D^2 H\right)\,,\\
\mathcal{O}_3 &=\left( H^\dagger H\right)^2\left(D_\mu  H^\dagger  D^\mu H\right).
\end{align}

These operators are related through the following total derivative:\pagebreak
\begin{align}\label{eq:IBPExample}
0 &=D_\mu\left[\left( H^\dagger H\right)^2D^\mu\left( H^\dagger H\right) 
\right]\nonumber\\
& = 2\left( H^\dagger H\right)D_\mu\left( H^\dagger H\right)D^\mu\left( H^\dagger H\right) + \left( H^\dagger H\right)^2\left(D^2 H^\dagger H+ H^\dagger D^2 H\right) \nonumber\\&+ 2\left( H^\dagger H\right)^2\left(D_\mu  H^\dagger  D^\mu H\right) = 2\mathcal{O}_1+\mathcal{O}_2+2\mathcal{O}_3\,.
\end{align}

This relation implies that only two of the three operators are linearly independent. Redundancies of this type, originating from \gls{ibp} identities, represent one of the primary obstacles in constructing a complete, non-redundant operator basis in \glspl{eft}. In practice, distributing derivatives over composite field expressions without generating redundant terms demands either a systematic algebraic method or a very meticulous—and often tedious—manual analysis.

\subsection{Field redefinitions}\label{sec:FieldRedefinitions}

A field redefinition is a local transformation of a quantum field $\phi$ into a new field $ \phi^\prime $, defined by a functional $ \phi = F[\phi^\prime] $, where the functional $ F $ is assumed to be expressible as a finite power series in $\phi^\prime$ and its derivatives. The redefined field $ \phi^\prime $ is of the same type as the original field $ \phi $, though possibly with a different normalization. If $ F[\phi^\prime] $ creates a one-particle state from the vacuum, i.e:
\begin{equation}
\left\langle p|F[\phi^\prime]|0\right\rangle\neq 0\,,
\end{equation}
then the physical S-matrix elements computed from the transformed Lagrangian $ \mathcal{L}[F[\phi^\prime]] $ are identical to those derived from the original Lagrangian $ \mathcal{L}[\phi] $.

This invariance follows from the behaviour of the generating functional under a change of variables in the path integral. The original generating functional with source $J $ is

\begin{equation}
Z[J]=\int {\rm{D}\phi} \exp (\ii \mathcal{S}[\phi] + J\phi) \equiv \int {\rm{D}\phi} \exp \left(\ii \int {\rm{d}x} \mathcal{L}[\phi] + J\phi\right)\,.
\end{equation}

Since the path integral sums over all configurations of $ \phi $, we may treat $ \phi $ as a dummy integration variable. Under the change of variables $ \phi = F[\phi^\prime] $, the measure transforms as:\pagebreak
\begin{align}
\phi&= F[\phi^\prime(x)]\,,\\
{\rm{D}\phi}&= {\rm{D}\phi^\prime} \left|\frac{\delta F}{\delta \phi^\prime}\right|\,,
\end{align} 
where the Jacobian determinant is unity when \gls{dimreg} is employed~\cite{Fujikawa:1979ay}. The generating functional becomes:

\begin{align}\label{eq:GenFunctionalFieldRedef}
Z[J]&=\int {\rm{D}\phi^\prime} \exp \left(\ii \int {\rm{d}x} \mathcal{L}^\prime[\phi^\prime] + JF[\phi^\prime(x)]\right),
\end{align}
where \( \mathcal{L}^\prime[\phi^\prime] \equiv \mathcal{L}[F[\phi^\prime]] \).

In contrast, the generating functional constructed from $ \mathcal{L}^\prime $ with a source coupled directly to $ \phi^\prime $ is:

\begin{equation}\label{eq:GenFunctionalFieldRedefTransformed}
Z^\prime[J]=\int {\rm{D}\phi^\prime} \exp \left(\ii \int {\rm{d}x} \mathcal{L}^\prime[\phi^\prime] + J\phi^\prime\right)\,.
\end{equation}

The difference between $ Z[J] $ and \( Z^\prime[J]$ manifests in their respective Green’s functions, which are defined by functional derivatives with respect to $ J $. However, physical S-matrix elements depend only on the poles of Green’s functions, which remain invariant under such transformations provided $ F[\phi'] $ can also create a one-particle state from the vacuum.

Field redefinitions therefore represent a redundancy at the level of the Lagrangian: two Lagrangians related by such a transformation yield identical S-matrix elements. However, one may choose to build operator bases from off-shell Green’s functions. In this context, operators related by field redefinitions are not considered redundant, as they yield distinct correlators. This perspective is often advantageous for explicit computations, and we will adopt it throughout this thesis. Nonetheless, we remain mindful of field redefinitions, as they play a crucial role in the structure and interpretation of \glspl{eft} operator bases in specific contexts.

\subsection{Equations of motion}\label{sec:EoM}
Given a Lagrangian $\mathcal{L}[\phi,\partial_\mu \phi]$ involving generic fields $\phi$, the classical \glspl{eom} are defined as:

\begin{equation}
E[\phi,\partial_\mu \phi]\equiv\partial_\mu\frac{\delta \mathcal{L}}{\delta (\partial_\mu\phi)}-\frac{\delta \mathcal{L}}{\delta \phi}.
\end{equation}

These \gls{eom} terms can appear as components of higher-dimensional operators. Consider a generic operator $\mathcal{O}=O(\phi)\cdot E(\phi)$ involving such a structure. Now perform the field redefinition $\phi(x)= \phi^\prime(x) - \varepsilon O(\phi^\prime)$, where $\varepsilon\ll 1$ is a small expansion parameter. This transformation induces a shift in the Lagrangian:

\begin{equation}\label{eq:FieldRedefasEoMTaylor}
\mathcal{L}[\phi]=\mathcal{L}[\phi^\prime]-\varepsilon O(\phi^\prime)\left(\partial_\mu\frac{\delta \mathcal{L}}{\delta (\partial_\mu\phi^\prime)}-\frac{\delta \mathcal{L}}{\delta \phi^\prime}\right) + \mathcal{O}(\varepsilon^2)\equiv
\mathcal{L}[\phi^\prime]- \varepsilon\mathcal{O}(\phi^\prime)+ \mathcal{O}(\varepsilon^2).
\end{equation}

The linear term in $ \varepsilon $ reproduces the \gls{eom} operator $\mathcal{O} $, thus demonstrating that such operators can be removed via field redefinitions.

Field redefinitions can be extended to a power series in $ \varepsilon $, and applying such expansions to operators involving \glspl{eom} introduces new higher-dimensional terms into the effective Lagrangian. This makes clear that eliminating an \gls{eom} operator by substitution is not equivalent to simply setting it to zero—it corresponds instead to a controlled reorganisation of the theory under a non-linear change of variables. In practice, \gls{eom} redundancies are often used to replace operators involving derivatives with other operator classes. Notably, this is the only redundancy mechanism that allows operators of distinct field content to be identified as equivalent.

In the context of renormalisation, especially when working off-shell, it is essential to consider these redundancies. They can be systematically eliminated via field redefinitions after the \gls{uv} divergences have been computed. However, to simplify the analysis, we will use the \gls{eom} directly to remove such operators. This approach is valid for one-loop computations involving only a single insertion of a redundant operator, which is the case for all diagrams considered in this work\footnote{Since we intend to renormalise at one-loop, the insertion of two (one-loop-renormalised) redundant operators would be formally a two-loop contribution to the redefined operator.}.

At \gls{lo} in the \gls{smeft} power counting, the Standard Model \glspl{eom} are:
\begin{align}\label{eq:EoMforSMEFT}
D^2 H_j& = m_H^2 H_j-\lambda| H|^{2} H_j-\bar{q}^{\alpha,k}\epsilon_{kj}(y^u)_{\alpha\beta}u^\beta - \bar{d}^\beta(y^d)^*_{\alpha\beta}q^{\alpha}_j -\bar{e}^\beta(y^e)^*_{\alpha\beta}\ell^{\alpha}_j, \nonumber\\
\ii \slashed{D}q^j_\alpha &= (y^u)_{\alpha\beta}u^\beta H^\dagger_k\epsilon^{kj} + (y^d)_{\alpha\beta}d^\beta H^j, \nonumber\\
\ii \slashed{D}\ell^j_\alpha&= (y^e)_{\alpha\beta}e^\beta H^j, \nonumber\\
\ii \slashed{D}d_\beta &= (y^d)^*_{\alpha\beta}q^\alpha H^{\dagger}, \nonumber\\
\ii \slashed{D}u_\beta &= (y^u)^*_{\alpha\beta}q^\alpha\tilde{ H}^{\dagger}, \nonumber\\
\ii \slashed{D}e_\beta&= (y^e)^*_{\alpha\beta}\ell^\alpha H^{\dagger},\nonumber\\
D^\mu B_{\mu \nu} &= - g_1 \sum_{\psi=u,d,q,e,\ell} \bar{\psi}Y\gamma_\nu \psi - \frac{g_1}{2} H^\dagger \ii \overleftrightarrow{D}_\nu  H, \nonumber\\
D^\mu W^I_{\mu \nu} &= - \frac{g_2}{2} \sum_{\psi=q,\ell} \bar{\psi}\sigma^I\gamma_\nu \psi -  \frac{g_2}{2}H^\dagger \ii \overleftrightarrow{D}^I_\nu  H, \nonumber\\
D^\mu G^A_{\mu \nu} &= - \frac{g_3}{2} \sum_{\psi=u,d,q} \bar{\psi}\lambda^A\gamma_\nu \psi, 
\end{align}
where $ H^\dagger \ii \overleftrightarrow{D}^I_\nu  H =  \frac{1}{2} (H^\dagger \sigma^I \ii D_\nu  H - \ii D_\nu H^\dagger \sigma^I   H)$.

These expressions can be extended to include higher-order terms in the power-counting expansion. For example, Refs.~\cite{Chala:2021pll,Bakshi:2024wzz} apply the \glspl{eom} up to $ \Lambda^{-2} $~\cite{Barzinji:2018xvu} to eliminate redundant dimension-six operators and study their contribution to the \glspl{rge} of physical operators at dimension eight.

\subsection{Repeated fields}

If all operators contained only distinct fields, then the total number of terms would be determined by the number of independent Lorentz- and gauge-invariant combinations, which can be computed using Group Theory standard methods. In that case, incorporating flavour simply involves multiplying the count by the appropriate powers of the number of flavours associated with each field.

However, when operators involve repeated fields, this naive estimate typically overcounts the number of independent terms. This reduction arises from additional internal symmetries, including field (anti)commutation properties, flavour symmetry, and gauge index permutations. These symmetries are not independent—they often interplay in nontrivial ways—so they must be treated collectively.

Intuitively, the effect of these symmetries can be understood by analysing the inequivalent permutations of fields and indices. Consider, for instance, operators of the form $\ell^2 H^2 $, which involve two identical scalar fields and two identical fermionic doublets\footnote{We take this example from~\cite{Fonseca:2019yya}.}. The relevant symmetry considerations are as follows:

\begin{itemize}
\item The Higgs fields $H $ are complex scalar doublets under $ SU(2)_L $, commute under field exchange, and are Lorentz scalars without colour charge.
\item The lepton doublets $ \ell $ are also $ SU(2)_L $ doublets, but they are anticommuting Grassmann fields, and hence antisymmetric under exchange.
\item Lorentz invariance requires that the fermion bilinear be contracted into a Lorentz scalar. The antisymmetry of the spinor contraction compensates for the Grassmann sign.
\item Gauge invariance under $ SU(2)_L $ allows contractions of the four doublets either into singlets or triplets. However, a singlet contraction of two identical bosons would vanish due to the symmetry of the fields; thus, the Higgs pair must be contracted as a triplet.
\item Consequently, the lepton doublets must also be contracted as an $SU(2)_L$ triplet, which is symmetric in their flavour indices.
\item Finally, overall gauge invariance requires hypercharge neutrality, which further constrains the allowed combinations.
\end{itemize}

As a result, instead of the naive $ n_f^2$ flavour structures one might expect from two lepton fields, the actual number of independent flavour contractions for this operator class is reduced to $ \frac{n_f(n_f+1)}{2} $, reflecting the symmetry under flavour exchange.

While in this simple example one can construct an explicit operator form by hand, the general task of counting and constructing independent operators is far more involved. Fortunately, the types of redundancies discussed here—arising from Lorentz, gauge, and flavour symmetries—can be addressed systematically using modern algebraic and computational methods, such as those based on Hilbert series techniques or symmetry-group classification algorithms.

\subsection{Notational Choices} 

The counting of effective operators has been automated using several different approaches. As long as the field content and symmetries of a theory are known, these methods can be systematically extended to a wide range of models. In this work, we have primarily adopted an intuitive, traditional algorithm~\cite{Fonseca:2019yya} to characterise operator dependencies and count independent terms. An alternative, fully algebraic approach is provided by the Hilbert series formalism~\cite{Benvenuti:2006qr,Feng:2007ur,Gray:2008yu,Jenkins:2009dy,Hanany:2010vu}, which yields equivalent results~\cite{Lehman:2015via,Lehman:2015coa,Henning:2015daa,Henning:2015alf}. In terms of computational structure, both methods require essentially the same input (field content and symmetries) and produce the same output (the number of invariants), with the main difference being computational efficiency. 

It is important to note that explicitly constructing a minimal operator basis—i.e., listing a complete set of independent operators class by class—is significantly more challenging than simply counting them. Although the number of independent terms remains the same, the operators can be expressed in different forms depending on the chosen conventions. One may, for instance, prefer to express the basis using the smallest possible number of terms, resulting in a compact representation. Alternatively, it may be advantageous to expand the terms to make flavour symmetries manifest; this is particularly useful in contexts where flavour structure plays a critical role (as in the dimension-seven basis of Refs.~\cite{Liao:2016hru,Lehman:2014jma}).

Another common choice involves whether to express operators explicitly as real or complex terms. In general, a subclass of operators can be written in real form if its field content is closed under complex conjugation. For example, operators of the type $ \ell^2 H^2 D$ involve the field set $ \{\ell, \ell^\dagger, H, H^\dagger\} $, which allows for real combinations. By contrast, operators like $ \ell e H^3 $ typically yield complex structures, and our convention treats terms with $\{\ell, e^\dagger, H, H^\dagger, H^\dagger\}$ as distinct from those with $ \{\ell^\dagger, e, H, H, H^\dagger\} $. One could, in principle, form real linear combinations across such types, but this approach is rarely adopted in the literature.

Finally, one may choose between using standard field strength tensors $F $ and $ \widetilde{F} $ or adopting a chiral convention in which the left- and right-handed combinations are defined as $ F_{L,R} = F \mp \ii \widetilde{F} $. While the number of operators remains unchanged under this choice, the chiral basis yields operators with well-defined helicity, which can be advantageous in specific phenomenological analyses.

\section{Operator counting algorithm}\label{sec:CountingOperators}

Once all possible relations between operators are understood, one can determine the minimal number of independent interactions required to define an \gls{eft}. This process has been automated in various tools such as \texttt{Sim2Int}~\cite{Fonseca:2017lem} and \texttt{basisgen}~\cite{Criado:2019ugp}, which we employ to reproduce the number of independent \gls{smeft} operators and the minimal number of terms in each class. In addition to the physical operators, we also consider redundant ones, as they play an important role in the renormalisation framework discussed in the next chapter.

If an operator class contains no repeated fields or derivatives, the number of invariant terms under global and gauge symmetries can be obtained using standard group-theoretical techniques~\cite{Slansky:1981yr,Cahn2014,Cheng1985}.

For classes involving repeated fields, Ref.~\cite{Fonseca:2019yya} describes a systematic method to count independent operators, which can be summarised as follows:

\begin{itemize}
\item For $m $ repeated fields, symmetries are represented by permutation elements $ \pi_{\mathcal{G}_i} \in S_m $, where $ \mathcal{G}_i $ includes global, gauge, and flavour symmetries.
\item The full symmetry of the operator is encoded in a tensor product representation of the symmetric group: $ \pi_{\text{Lorentz}} \otimes \pi_{SU(3)_c} \otimes \pi_{SU(2)_L} \otimes \pi_\text{fields} $, with $ \pi_\text{fields} $ representing field (anti)commutation properties.
\item This combined representation is decomposed into irreducible representations of $ S_m $. The greatest multiplicity among all the irreps is equal to the minimum number of terms. The total number of irreps (including multiplicity) gives the number of independent operators.
\item For every set of repeated fields, this decomposition is carried out separately.
\item The final operator count for a given subclass is obtained by multiplying the counts from each repeated subset.
\end{itemize}

If operators include derivatives, each derivative acting on a field may be considered a distinct field with its own Lorentz transformation properties. When multiple derivatives act on the same field, their antisymmetric part can be rewritten as a field strength tensor via the commutator of covariant derivatives:
\begin{equation}
[D_\mu,D_\nu]=-\ii g_1 \frac{Y}{2} B_{\mu\nu} -\ii g_2 \frac{\sigma^I}{2} W^I_{\mu\nu} -\ii g_3 \frac{\lambda^A}{2} G^A_{\mu\nu}\,.
\end{equation}
In such cases, we typically discard the antisymmetric component and substitute it with the corresponding field strength tensor, which belongs to a different operator class. To avoid overcounting, derivatives are decomposed into irreducible representations of the Lorentz group. It has been shown~\cite{Fonseca:2019yya} that only the highest-spin irreps of the derivative expansion contribute to genuinely independent structures, while the lower-spin components correspond to redundancies from \glspl{eom}. These can be retained in a Green’s basis but are otherwise removed for minimality.

\gls{ibp} redundancies are handled by treating derivatives as a dummy field $ \mathfrak{D} $. In this formalism, an operator type with $ n $ derivatives is decomposed into subtypes with $ k $ instances of $ \mathfrak{D} $ and $ n-k $ standard derivatives. If each subtype contains $ N(n,k) $ operators, then the total number of independent structures is given by the alternating sum~\cite{Fonseca:2019yya}:
\begin{equation}
N_\mathcal{O}=\sum_{k=0}^{k=n}(-1)^kN(n,k)
\end{equation}

These methods are implemented in several modern tools. \texttt{Sim2Int}~\cite{Fonseca:2017lem} uses the algorithm outlined above, while \texttt{basisgen}~\cite{Criado:2019ugp}  follows a comparable prescription and yields equivalent results. The Hilbert series approach, implemented in \texttt{DEFT}~\cite{Gripaios:2018zrz}, is conceptually distinct but agrees with these tools in all tested cases. Recently, on-shell methods for basis-generation have also been automated in \texttt{ABC4EFT}~\cite{Li:2022tec}. 

In Table~\ref{tab:CountingSMEFT} we show an overview of \gls{smeft} operator counting. Given the large number of operator classes, a full analysis is impractical here. Nevertheless, some general patterns can be observed. Certain classes are purely physical—especially those without derivatives. Others are purely redundant due to equations of motion. As we will see in Chapter~\ref{ch:RGE}, such classes typically do not contribute to the dimension-eight \glspl{rge}, as their one-loop divergences vanish. Moreover, they usually appear only at the loop level in weakly coupled \gls{uv} completions of \gls{smeft}.

\begin{table}
\begin{center}
\begin{footnotesize}
$\begin{array}{|r|c|c|}
\hline
\text{Class} & \text{\# terms} & \text{\# operators} \Tstrut\\
\hline
\hline
\psi^2H^2 & 2 + 0 & 12 + 0 \Tstrut\\
\hline
H^2D^4 & 0 + 1 & 0 + 1 \Tstrut\\
H^4D^2 & 2 + 2 & 2 + 2 \\
H^6 & 1 + 0 & 1 + 0 \\
\psi^2D^3 & 0 + 5 & 0 + 45 \\
\psi^2HD^2 & 0 + 24 & 0 + 216 \\
\psi^2H^2D & 9 + 14 & 81 + 126 \\
\psi^2H^3 & 6 + 0 & 54 + 0 \\
\psi^4 & 38 + 0 & 2751 + 0 \\
XH^2D^2 & 0 + 2 & 0 + 2 \\
X\psi^2D & 0 + 30 & 0 + 270 \\
X\psi^2H & 16 + 0 & 144 + 0 \\
X^2D^2 & 0 + 3 & 0 + 3 \\
X^2H^2 & 8 + 0 & 8 + 0 \\
X^3 & 4 + 0 & 4 + 0 \\
\hline
\psi^2H^2D^2 & 4 + 8 & 24 + 66 \Tstrut\\
\psi^2H^3D & 2 + 0 & 18 + 0 \\
\psi^2H^4 & 2 + 0 & 12 + 0 \\
\psi^4D & 6 + 8 & 276 + 858 \\
\psi^4H & 18 + 0 & 1188 + 0 \\
X\psi^2H^2 & 4 + 0 & 24 + 0 \\
\hline
\hline
\text{Dimension} & \text{\# terms} & \text{\# operators} \Tstrut\\
\hline
\hline
5 & 2 + 0 & 12 + 0 \Tstrut\\
6 & 84 + 81 & 3045 + 665 \\
7 & 36 + 16 & 1542 + 924 \\
8 & 1019 + 1642 & 44807 + 66197\\
\hline
\end{array}
$
$
\begin{array}{|r|c|c|}
\hline
\text{Class} & \text{\# terms} & \text{\# operators} \Tstrut\\
\hline
\hline
H^2D^6 & 0 + 1 & 0 + 1 \Tstrut\\
H^4D^4 & 3 + 10 & 3 + 10 \\
H^6D^2 & 2 + 2 & 2 + 2 \\
H^8 & 1 + 0 & 1 + 0 \\
\psi^2D^5 & 0 + 5 & 0 + 45 \\
\psi^2HD^4 & 0 + 54 & 0 + 486 \\
\psi^2H^2D^3 & 16 + 135 & 144 + 1215 \\
\psi^2H^3D^2 & 36 + 60 & 324 + 540 \\
\psi^2H^4D & 13 + 14 & 117 + 126 \\
\psi^2H^5 & 6 + 0 & 54 + 0 \\
\psi^4D^2 & 67 + 461 & 4923 + 36549 \\
\psi^4HD& 168 + 270 & 13338 + 22092 \\
\psi^4H^2 & 93 + 0 & 6603 + 0 \\
XH^2D^4  & 0 + 6 & 0 + 6 \\
XH^4D^2  & 6 + 4 & 6 + 4 \\
X\psi^2D^3& 0 + 100 & 0 + 900 \\
X\psi^2HD^2 & 48 + 208 & 432 + 1872 \\
X\psi^2H^2D  & 92 + 66 & 828 + 594 \\
X\psi^2H^3 & 22 + 0 & 198 + 0 \\
X\psi^4 & 216 + 0 & 16380 + 0 \\
X^2D^4  & 0 + 3 & 0 + 3 \\
X^2H^2D^2  & 18 + 44 & 18 + 44 \\
X^2H^4 & 10 + 0 & 10 + 0 \\
X^2\psi^2D & 57 + 188 & 513 + 1692 \\
X^2\psi^2H & 96 + 0 & 864 + 0 \\
X^3D^2 & 0 + 16 & 0 + 16 \\
X^3H^2  & 6 + 0 & 6 + 0 \\
X^4 & 43 + 0 & 43 + 0 \\
\hline
\end{array}$
\end{footnotesize}
\end{center}
\caption{\label{tab:CountingSMEFT} Number of \gls{smeft} operators up to dimension eight, for three fermion generations ($ n_f = 3$). Each class is listed with the number of independent \emph{terms} and \emph{operators}, split as physical + redundant (due to \gls{eom}) real operators and terms. The counting was obtained using the \texttt{Sym2Int}~\cite{Fonseca:2017lem} package. Upper-left: Classes of dimension-five, -six and -seven operators. Right: Classes of dimension-eight operators. Lower-left: Summary of all classes.
}
\end{table}

\section{Offshell independence in momentum space\footnote{This section contains original work from the thesis.}}

One of the central challenges in working with \glspl{eft} is the construction of a minimal and independent operator basis. While many modern matching and running procedures do not require the basis to be explicitly fixed, the presence of redundant operators—arising from \glspl{eom}, \gls{ibp}, or algebraic identities—necessitates careful treatment in practical computations. This is especially true when one computes counterterms or matches a \gls{uv} theory to an \gls{eft}.

The key question we address in this section is: given a set of operators, how can one determine whether they are linearly independent? If dependencies exist, how can one obtain a minimal independent set? While automated tools (such as those implementing the algorithm described in Section~\ref{sec:CountingOperators}) can provide the number of independent operators, they do not yield explicit operator expressions. To this end, we propose a constructive method that tests operator independence using off-shell Green’s functions at tree level—what we refer to as a \textbf{Green’s Basis}.

Several motivations support the development of this approach:
\begin{enumerate}
\item In standard quantum field theory calculations using Feynman diagrams, matrix elements are derived from connected and amputated diagrams. The number of such diagrams increases rapidly with the number of external legs, whereas \gls{1pi} diagrams are fewer, making the computation more manageable.
\item In the path integral formulation of matching~\cite{Gaillard:1985uh,Cheyette:1987qz,Henning:2014wua}, the resulting \gls{eft} generally includes operators that are redundant due to \gls{ibp}, field redefinitions, or other identities. A straightforward way to simplify the \gls{eft} is to match off-shell amplitudes at tree level onto a known basis of independent Green’s functions.
\item Helicity amplitude methods have proven effective in computing certain anomalous dimensions strictly on-shell~\cite{Caron-Huot:2016cwu,Bern:2019wie,Yang:2019vag,Baratella:2020lzz,Bern:2020ikv,EliasMiro:2020tdv,Baratella:2020dvw,Jiang:2020mhe,AccettulliHuber:2021uoa,Baratella:2021guc}. However, their applicability is limited in scenarios involving operator mixing across different mass dimensions or in amplitudes generated by operators with fewer external legs than the processes they contribute to.
\end{enumerate}

The method developed in~\cite{Chala:2021cgt} is based on the momentum-space representation of operators and tree-level amplitudes. To determine operator dependencies, we evaluate Green’s functions for processes involving insertions of the candidate operators. By restricting our analysis to \gls{1pi} diagrams, we exclude contributions from one-particle-reducible diagrams and, implicitly, keep \gls{eom}-induced redundancies. This off-shell framework naturally leads to a basis where operators are independent up to field redefinitions, that can later be removed by onshell relations (See Section~\ref{sec:OnshellRelations}).

The amplitudes derived in this way are expressed as linear combinations of \textit{kinemtic invariants}--Lorentz-invariant products of external momenta and polarisation vectors (or spinor structures in fermionic cases). Therefore, if we have a set of operators $\{\mathcal{O}^{(p)}\}$ contributing to the process $a\rightarrow b$, the resulting amplitudes can be parametrised at tree level as:
\begin{equation}
\mathcal{A}(a\rightarrow b)=c^{(p)} \sum_{p,m} f_p^m \kappa_m\,,
\end{equation}
where $c^{(p)}$ are the Wilson Coefficient of the irrelevant operators, $\kappa_m$ are the kinematic invariants and $f_p^m$, the \textit{amplitude matrix}, is a set of numerical coefficients and SM couplings derived from the Feynman Rules of the Lagrangian. 

The linear independence of the operators is then equivalent to the linear independence of the vectors. Thus, testing operator independence reduces to computing the rank of the amplitude matrix. If the rank equals the number of operators (or the dimension of the set of kinematic invariants, whichever is smaller), the operators are independent.

Importantly, a non-maximal rank in a specific process does not necessarily imply operator dependence—it could result from accidental symmetries. Therefore, to conclusively establish operator dependence, one must test multiple processes.

The various redundancies discussed in Section~\ref{sec:dependancies} can be understood as linear independence of the amplitude matrix in momentum space. There are only three differences: 
\begin{itemize} \label{list:IndependencyConditions}
\item \gls{ibp} manifests as momentum conservation. The effect of \gls{ibp} can be implemented by removing one of the external momenta from the set of kinematic invariants. 
\item \gls{eom} relations are not considered in this off-shell approach. 
\item Redundancies by Schouten identities, which relate to four-vector independence in four dimensions, are avoided by restricting the contractions with only four independent momenta and/or polarisation vectors when there is a larger number. 
\end{itemize}

\paragraph{Integration by Parts}

The interplay between \gls{ibp} identities and momentum conservation can be elucidated by examining the derivative expansion of a total derivative operator constructed from two fields $\phi$ and $\chi$, implicitly contracted with some tensor structure $\widetilde{g}_\mu$ (internal indices are suppressed for clarity).

Expanding the total derivative yields the identity referenced in Eq.~\eqref{eq:IBPref}. Upon performing a Fourier transform of the Lagrangian, and considering that the derivative acts only on the adjacent field, the resulting expression naturally simplifies to a statement of momentum conservation:
\begin{align}
D^\mu(\phi\widetilde{g}_\mu\chi)&=D^\mu\phi\widetilde{g}_\mu\chi+\phi\widetilde{g}_\mu D^\mu\chi=0,\\
p^\mu_\text{total}&=p^\mu_\phi+p^\mu_\chi=0.
\end{align}
This reasoning can be readily generalized to systems involving arbitrary field content.

As an example, let's review the case shown in \ref{sec:ibp} from the momentum space perspective. Let us consider the same dimension-eight six-Higgs operators:
\begin{align}
\mathcal{O}_1 &=\left( H^\dagger H\right)D_\mu\left( H^\dagger H\right)D^\mu\left( H^\dagger H\right),\\
\mathcal{O}_2 &=\left( H^\dagger H\right)^2\left(D^2 H^\dagger H+ H^\dagger D^2 H\right),\\
\mathcal{O}_3 &=\left( H^\dagger H\right)^2\left(D_\mu  H^\dagger  D^\mu H\right).
\end{align}
The \gls{1pi} amplitude for $H^0(p_1)\mapsto H^0 (p_2)H^+(p_3) H^-(p_4) H^+(p_5) H^- (p_6)$ reads:
\begin{align}\nonumber
 \mathcal{A} &= 2 i\, c_1 (2\kappa_{13}+2\kappa_{14}+2\kappa_{15}+2\kappa_{16}-2\kappa_{23}-2\kappa_{24}-2\kappa_{25}-2\kappa_{26}\\\nonumber&\,\,\,\,\,\,\,\,\,\,\,\,\,\,\,\,-\kappa_{34}-2\kappa_{35}-\kappa_{36}-\kappa_{45}-2\kappa_{46}-\kappa_{56})\\[0.1cm]\nonumber
 &- 4 i \,c_2 (\kappa_{11}+\kappa_{22}+\kappa_{33}+\kappa_{44}+\kappa_{55}+\kappa_{66})\\[0.1cm]
 &+ 2i\, c_3 (2\kappa_{12}-\kappa_{34}-\kappa_{36}-\kappa_{45}-\kappa_{56})\,,
\end{align}

where $\kappa_{ij} = p_i\cdot p_j$ and $c_n,\,n=1,2,3$ are the \gls{wc} of the operators.

Apparently, the amplitude matrix $f$ associated with this process is of rank 3. To illustrate this, consider the submatrix $\hat{f}$, corresponding to the invariants $\kappa_{11}$, $\kappa_{12}$, $\kappa_{13}$ which takes the form:
\begin{equation}
 \hat{f} = \begin{bmatrix}0 & 0 & 4i\\ -4i & 0 & 0\\0 & 4i &0 \end{bmatrix}\Longrightarrow 3\geq\rm{Rank}(f)\geq \rm{Rank}(\hat{f}) = 3\,.
\end{equation}

However, the set of kinematic invariants chosen is not linearly independent due to momentum conservation, specifically $p_1 = p_2+p_3+p_4+p_5+p_6$. As a consequence, the invariants $\kappa_{i1}$ can always be expressed in terms of the others and thus eliminated. Taking this constraint into account, we obtain instead:\pagebreak
\begin{align}\nonumber
 \mathcal{A} &= 2i\,c_1 (2 \kappa_{33}+3 \kappa_{43}+2 \kappa_{44}+2 \kappa_{53} +3\kappa_{54}+2\kappa_{55}+3\kappa_{63}+2\kappa_{64}+3\kappa_{65}+2\kappa_{66} ) \\\nonumber
 &-8i\,c_2 (\kappa_{22}+\kappa_{32}+\kappa_{33}+\kappa_{42}+\kappa_{43}+\kappa_{44}+\kappa_{52}+\kappa_{53}+\kappa_{54}+\kappa_{55}+\kappa_{62}+\kappa_{63}\\\nonumber&\,\,\,\,\,\,\,\,\,\,\,\,\,+\kappa_{64}+\kappa_{65}+\kappa_{66})\\
 &+2i \, c_3 (2\kappa_{22}+2\kappa_{32}+2\kappa_{42}-\kappa_{43}+2\kappa_{52}-\kappa_{54}+2\kappa_{62}-\kappa_{63}-\kappa_{65})\,.
\end{align}
The corresponding matrix is found to have rank 2. This can be readily seen from the fact that the first and third rows in the expression above sum to minus one half of the second row, i.e., $\mathcal{O}_2 = -2(\mathcal{O}_1+\mathcal{O}_3)$ holds. At the level of the Lagrangian, this linear dependence originates from the identity given in~\eqref{eq:IBPExample}, derived in Section~\ref{sec:ibp}. 
\paragraph{A more involved example}
We now turn to a slightly more sophisticated example. When gauge bosons are included, the structure of the process remains conceptually analogous. Consider, for instance, the following set of operators:
\begin{align}
 \mathcal{O}_1 &= D_\mu ( H^\dagger H) D^\nu B^{\mu\rho} B_{\nu\rho}\,,\\
 \mathcal{O}_2 &= (D^2 H^\dagger H + H^\dagger D^2 H) B^{\nu\rho} B_{\nu\rho}\,,\\
 \mathcal{O}_3 &= D_\mu H^\dagger D^\mu H B^{\nu\rho} B_{\nu\rho}\,.
\end{align}

The amplitude for the process $H^0(p_1)\mapsto H^0(p_2) B(p_3) B(p_4)$ takes the following form:
\begin{align}\nonumber
 \mathcal{A} &= -i c_1 (\kappa_{3334} + 2\kappa_{3434} + \kappa_{3444} -\kappa'_{4333}-2 \kappa'_{4334}-\kappa'_{4344})\\[0.1cm]\nonumber\,
 &+ 4i c_2 (2 \kappa_{2234} + 2\kappa_{2334} + 2\kappa_{2434} + \kappa_{3334} + 2\kappa_{3434}+\kappa_{3444}-2\kappa'_{4322}-2\kappa'_{4323}\\\nonumber
 &\,\,\,\,\,\,\,\,\,\,\,\,\,\,-2\kappa'_{4324}-\kappa'_{4333}-2\kappa-2\kappa'_{4334}-\kappa_{4344})\\[0.1cm]
 &-4i c_3 (\kappa_{2234}+\kappa_{2334}+\kappa_{2434}-\kappa'_{4322}-\kappa_{4323}-\kappa_{4324})\,.
\end{align}
Here, $p_1$ has been eliminated using momentum conservation. The relevant kinematic invariants are defined as\pagebreak
\begin{align}
\kappa_{ijkl} = (\varepsilon_3\cdot\varepsilon_4)(p_i\cdot p_j)(p_k\cdot p_l)\,,\nonumber\\\kappa'_{ijkl} = (\varepsilon_3\cdot p_i)(\varepsilon_4\cdot p_j)(p_k\cdot p_l)\,,
\end{align} 
where $\varepsilon$ denotes a polarization vector.

The matrix constructed from these invariants has rank 2, implying that one of the operators can be expressed as a linear combination of the other two. Indeed, it is straightforward to verify from the expression above that 
$$\mathcal{O}_1 = -\frac{1}{4}\mathcal{O}_2 -\frac{1}{2}\mathcal{O}_3.$$
This linear dependence corresponds to the following relation at the level of the Lagrangian:
\begin{align}\nonumber
 \mathcal{O}_1 &= -D_\mu( H^\dagger H) D^\mu B^{\rho\nu} B_{\nu\rho} - D_\mu ( H^\dagger H) D^\rho B^{\nu\mu} B_{\nu\rho}\\\nonumber
 &= -D_\mu( H^\dagger H) D^\mu B^{\rho\nu} B_{\nu\rho} - D_\mu ( H^\dagger H) \mc{D^\nu} B^{\mu\rho} B_{\nu\rho}\\\nonumber
 &= -D_\mu( H^\dagger H) D^\mu B^{\rho\nu} B_{\nu\rho} - \mathcal{O}_1\\\nonumber
 \Rightarrow \mathcal{O}_1 &= -\frac{1}{2} D_\mu( H^\dagger H) D^\mu B^{\rho\nu} B_{\nu\rho}\\\nonumber
 &= \frac{1}{2}D^2( H^\dagger H) B^{\rho\nu} B^{\nu\rho}-\frac{1}{2}D_\mu( H^\dagger H) B^{\rho\nu} D^\mu B_{\nu\rho}\,\\\nonumber
 &= \frac{1}{2}D^2( H^\dagger H) B^{\rho\nu} B_{\nu\rho} - \mathcal{O}_1\\\nonumber
 \Rightarrow \mathcal{O}_1 &=  \frac{1}{4} D^2( H^\dagger H) B^{\rho\nu} B_{\nu\rho}\,\\\nonumber
 &= -\frac{1}{4} (D^2 H^\dagger H+  H^\dagger D^2 H) B^{\nu\rho} B_{\nu\rho} - \frac{1}{4} (2 D_\mu H^\dagger D^\mu H) B^{\nu\rho} B_{\nu\rho}\\
 &= -\frac{1}{4} \mathcal{O}_2 - \frac{1}{2}\mathcal{O}_3\,.
\end{align}
In the first equality, we have made use of the Bianchi identity,
$$D_\mu B^{\nu\rho} + D_\nu B^{\rho\mu}+D_\rho B^{\mu\nu}=0.$$
In the second step, we have relabeled indices as $\nu\leftrightarrow\rho$ in the final operator. In the fifth equality, the derivative acting on $B^{\rho\nu}$has been integrated by parts. In the penultimate step, the derivative was explicitly expanded. Throughout the derivation, we have also used the antisymmetry of the field strength tensor, $B^{\nu\rho} = -B^{\rho\nu}$.

This example illustrates that off-shell redundancies can be efficiently identified in momentum space, even when their discovery through purely algebraic manipulations becomes nontrivial—such as in scenarios involving a large number of fields or a large operator basis.

\paragraph{Schouten identities}

As an example illustrating the implementation of Schouten identities in momentum space, consider the following set of operators:
\begin{align}
 \mathcal{O}_1 &= \ii (D_\mu H^\dagger\sigma^I D_\nu H-D_\nu H^\dagger \sigma^I D_\mu H) B^{\mu}_{\,\,\rho} \widetilde{W}^{I\nu\rho} \,,\\
 \mathcal{O}_2 &= \ii (D_\mu H^\dagger\sigma^I D_\nu H-D_\nu H^\dagger \sigma^I D_\mu H) \widetilde{B}^{\mu}_{\,\,\rho} W^{I\nu\rho} \,.
\end{align}

These operators are related by the identity given in Eq.~\eqref{eq:4dim1}.

We now proceed to compute and impose momentum conservation in the amplitude for the following process:
\begin{equation}
H^0(p_1)\mapsto~H^0(p_2)W^3(p_3) B(p_4).
\end{equation}
The resulting expression is:
\begin{align}\nonumber
 \mathcal{A} &= c_1 (-\kappa_{323443}-\kappa_{323444}+\kappa_{343424}+\kappa_{342334}+\kappa_{342344})\\
 &+ c_2 (-\kappa_{423433}-\kappa_{423434}-\kappa_{343423}+\kappa_{342433}+\kappa_{342434})\,.
\end{align}
In this case, the kinematic invariants are defined as $\kappa_{ijklmn} = \epsilon(\varepsilon_i,p_j,p_k,p_l)~(\varepsilon_m\cdot~p_n)$ and $\kappa'_{ijklmn} = \epsilon(\varepsilon_i,\varepsilon_j,p_k,p_l) (p_m\cdot p_n)$ where $\epsilon$ denotes the Levi-Civita symbol, and $\varepsilon$ represents polarization vectors. At first glance, one might conclude that the corresponding operators are linearly independent, as the matrix constructed from these invariants has rank 2.

However, as anticipated in the discussion of conditions for operator independence (see~\ref{list:IndependencyConditions}), we must account for Schouten identities. These identities constrain contractions involving more than four independent vectors in four-dimensional spacetime. To implement this, we construct kinematic invariants using contractions involving at most four linearly independent momenta and/or polarization vectors. In this example, after eliminating $p_1$ via momentum conservation, we are left with three independent momenta and two polarization vectors, which satisfy the following relation:
\begin{equation}
p_4 = a_1 \varepsilon_3+a_2\varepsilon_4 +a_3 p_2+a_4 p_3,
\end{equation}
for some real coefficients $a_i, i=1,...,4$. Incorporating this constraint, the amplitude takes the form:
\begin{align}\nonumber
 \mathcal{A} &= (c_1+c_2) \bigg[a_3 \kappa_{342323} + a_3 (1+a_4) \kappa_{342323}+a_4(1+a_4) \kappa_{342333} \\
 &\,\,\,\,\,\,\,\,\,\,\,\,\,\,\,\,\,\,\,\,+a_1 a_3 \kappa_{342332} + a_1(1+2 a_4) \kappa_{342333} +a_2 a_4 \kappa_{342343} + a_1^2 \kappa_{342333} \bigg]\,.
\end{align}
From this expression, it is evident that the two operators are related—in fact, they are identical up to an \textit{evanescent} term, which vanishes in $d=4$ dimensions.

As a final remark, the kinematic invariants depend solely on the process under consideration—that is, the external states involved—and the overall power counting of the operator coefficients. 
As previously noted, the invariants may change when the diagrams are evaluated on-shell; however, apart from this, once the field content of the theory is specified and the relevant kinematic invariants are determined, the method can be applied consistently~\cite{Chala:2024llp,LopezMiras:2025gar}.

\section{A Green's basis of bosonic operators\footnote{This section contains original work from the thesis.}}\label{sec:GB}

There is a fundamental distinction between determining the number of independent operators in an effective field theory and constructing their explicit forms. Due to the freedom to perform linear transformations—i.e., basis rotations—among operators, multiple representations of the same physical content can exist. In the context of a Green's basis, there is often an additional degree of freedom in choosing which operators are labelled as redundant and which are retained as physical.

While the final form of an operator basis can vary depending on conventions, the crucial objective is to identify one valid, complete, and independent basis. Once this is achieved, other bases will be related via appropriate transformations (rotations).

In the construction of such bases, the Hilbert Series is highly effective for enumerating invariants, but it does not provide the explicit form of the operators. For this task, the so-called `traditional' method—i.e., the direct construction of Lorentz- and gauge-invariant operators followed by elimination of redundancies—can be adapted to produce independent operator sets systematically, organised by field content and mass dimension.

Notably, systematic implementations of this method have only emerged in recent years, and its extension to the construction of Green's bases happened \textbf{after} the beginning of this thesis\footnote{See, for instance, the developments of the dimension-eight~\cite{Ren:2022tvi} and dimension-seven~\cite{Zhang:2023kvw} Green’s bases.}. Before these advancements, operator bases were assembled `manually', often through iterative procedures that identified and removed redundancies one by one. In many cases, the full operator count was unknown before this construction, leading to incremental updates as new dependencies were discovered\footnote{The evolution can be traced in the arXiv revisions of the Warsaw basis~\cite{Grzadkowski:2010es} and the dimension-seven basis of \cite{Lehman:2014jma}, which was later refined by other authors \cite{Liao:2016hru}.}.

A systematic approach to constructing physical bases was introduced in \cite{Li:2020gnx} and later extended to Green's bases in \cite{Ren:2022tvi}. This method leverages the spin-helicity formalism to recast operator structures into spinor variables, facilitating the use of momentum-space independence criteria, such as those listed in (\ref{list:IndependencyConditions}). Furthermore, it incorporates the full set of symmetries described in Section~\ref{sec:CountingOperators}, applied directly to the spinor representation. While this framework is rigorous and comprehensive, its implementation is technically involved and computationally demanding.

Before the publication of those results, an alternative and considerably simpler method—particularly well-suited for bosonic operators and lower-dimensional cases—was presented in \cite{Chala:2021cgt}. This method, developed as part of the original work in this thesis, provides an accessible route to constructing a Green's basis without sacrificing rigour. Relying on the momentum-space algorithm for identifying independent operator structures (as summarized in~\ref{list:IndependencyConditions}) and on established operator counting results in \gls{smeft} (see Table~\ref{tab:CountingSMEFT}), the method is conceptually straightforward: one generates all candidate operator structures consistent with the symmetries and then tests their linear independence by evaluating amplitudes for relevant processes.

Crucially, we showed in Section~\ref{sec:CountingOperators} that to establish the independence of a set of operators belonging to a given class, it suffices to consider physical processes that involve only the fields present in that class. Table~\ref{tab:Processes} lists the specific processes used to construct a Green's basis for bosonic operators. The physical operators were selected to match those appearing in the basis proposed by \cite{Murphy:2020rsh}, while the redundant operators were chosen to be real-valued for consistency and convenience.

\begin{table}[h]
\begin{center}
$
\begin{array}{|c|c|c|}
\hline
\text{Type} & \text{\# operators} & \text{Process} \Tstrut\\
\hline
\hline
BH^2D^4 & 3 &H^0(p_1)\rightarrow H^0(p_2)B(p_3) \Tstrut\\
WH^2D^4 & 3 &H^0(p_1)\rightarrow H^-(p_2)W^+(p_3) \\
B^2H^2D^2 & 12 &H^0(p_1)\rightarrow H^0(p_2)B(p_3)B(p_4) \\
W^2H^2D^2 & 19 &H^0(p_1)\rightarrow H^0(p_2)W^+(p_3)W^-(p_4) \\
WBH^2D^2 & 19 &H^0(p_1)\rightarrow H^-(p_2)W^+(p_3)B(p_4) \\
G^2H^2D^2 & 12 & H^0(p_1)\rightarrow H^0(p_2)G^A(p_3)G^B(p_4) \\
W^2BD^2 & 4 & B(p_1)\rightarrow W^+(p_2)W^-(p_3) \\
G^2BD^2 & 4 & B(p_1)\rightarrow G^A(p_2)G^B(p_3) \\
W^3D^2 & 4 & W_3(p_1)\rightarrow W^+(p_2)W^-(p_3) \\
G^3 & 4 &  G^A(p_1)\rightarrow G^B(p_2)G^C(p_3) \\
B^2D^4 & 1 & \text{Not needed} \\
W^2D^4 & 1 & \text{Not needed} \\
G^2D^4 & 1 & \text{Not needed} \\
\hline
\end{array}
$
\end{center}
\caption{\label{tab:Processes}Processes used in \cite{Chala:2021cgt} to prove off-shell independence of bosonic dimension eight operators. Columns show the subclasses (types) of operators, the number of elements in the Green's Basis and the process used to check the maximal rank of the amplitude matrix. Class $X^2D^4$ does not require computations since each subclass contains only one independent operator. Other bosonic dimension eight classes with redundant operators ($H^6D^2$, $H^4D^4$, $XH^4D^2$) had already been computed in Ref.~\cite{Chala:2021pll}. 
}
\end{table}

The operator bases computed to date are summarized in Table~\ref{tab:ListOfBases}. The explicit list of operators used in our computations is provided in Appendix~\ref{app:TableOfOperators}. In particular, the Green's basis for dimension-eight bosonic operators was constructed using the methodology outlined in the previous section. This basis includes 89 physical operators, coinciding with those in Ref.\cite{Murphy:2020rsh}, along with 86 redundant operators. While some of the redundant operators had been previously derived in Ref.\cite{Chala:2021pll}, many were presented in Ref.~\cite{Chala:2021cgt} for the first time and are original contributions of this work.

\subsection{List of original operators in the Green's basis.}

The off-shell independent operators were obtained by evaluating the amplitudes of the processes listed in Table~\ref{tab:Processes} and verifying that the resulting matrix has a rank equal to the expected number of independent operators. The latter was computed using the tools \texttt{Sym2Int}~\cite{Fonseca:2017lem} and \texttt{basisgen}~\cite{Criado:2019ugp}, and is reported for all operator classes up to dimension eight in Table~\ref{tab:CountingSMEFT}. Notably, the number of operators increases significantly in the presence of fermionic fields; for this reason, our analysis is restricted to the purely bosonic sector\footnote{Also, this work was partially motivated by the intention to compute the \glspl{rge} of bosonic dimension-eight opearators in \cite{DasBakshi:2022mwk}.}.

For completeness, the interaction terms listed below also include the physical operators (in the relevant classes) as defined in Ref.~\cite{Murphy:2020rsh}, using the same naming conventions for consistency.

\subsection{Operators in the class $X  H^2 D^4$}
There are 3 real terms for $X=B$ and 3 more for $X=W$.
In the first case, it suffices to compute the amplitude for the process $ H^0(p_1)\to H^0(p_2) B(p_3)$, while in the second case only $ H^0(p_1)\to H^-(p_2) W^+(p_2)$ is needed.

\subsubsection{$X=B$}
%
%
\begin{align}
 \mathcal{O}_{8;B H^2 D^4}^{(1)} &= i (D_\nu H^\dagger D^2 H-D^2 H^\dagger D_\nu H) D_\mu B^{\mu\nu}\,,\\
\mathcal{O}_{8;B H^2 D^4}^{(2)} &= (D_\nu H^\dagger D^2 H+D^2 H^\dagger D_\nu H) D_\mu B^{\mu\nu}\,,\\
\mathcal{O}_{8;B H^2 D^4}^{(3)} &= i (D_\rho D_\nu H^\dagger D^\rho H-D^\rho H^\dagger D_\rho D_\nu H) D_\mu B^{\mu\nu}\,.
\end{align}

\subsubsection{$X=W$}
%
\begin{align}
 \mathcal{O}_{8;W H^2 D^4}^{(1)} &= i (D_\nu H^\dagger \sigma^I D^2 H-D^2 H^\dagger\sigma^I D_\nu H) D_\mu W^{I\mu\nu}\,,\\
  \mathcal{O}_{8;W H^2 D^4}^{(2)} &= (D_\nu H^\dagger \sigma^I D^2 H+D^2 H^\dagger\sigma^I D_\nu H) D_\mu W^{I\mu\nu}\,,\\
 \mathcal{O}_{8;B H^2 D^4}^{(3)} &= i (D_\rho D_\nu H^\dagger\sigma^I D^\rho H-D^\rho H^\dagger \sigma^I D_\rho D_\nu H) D_\mu W^{I\mu\nu}\,.
\end{align}

\subsection{Operators in the class $X^2  H^2 D^2$}
There are 12 independent operators for $X^2=B^2$ and $G^2=B^2$, $19$ for $X^2=W^2$ and also $19$ for $X^2=WB$.
One can check the independence of the operators below by evaluating the amplitudes $ H^0(p_1)\to H^0(p_2)B(p_3)B(p_4)$,  $ H^0(p_1)\to H^0(p_2)W^+(p_3)W^-(p_4)$ and  $ H^0(p_1)\to H^-(p_2)W^+(p_3)B(p_4)$, respectively.

\subsubsection{$X^2 = B^2$}
\begin{align}
 \mathcal{O}_{8;B^2 H^2 D^2}^{(1)} &= (D^\mu H^\dagger D_\nu H) B_{\mu\rho} B^{\nu\rho}\,,\\
 \mathcal{O}_{8;B^2 H^2 D^2}^{(2)} &= (D^\mu H^\dagger D_\mu H) B_{\nu\rho} B^{\nu\rho}\,,\\
 \mathcal{O}_{8;B^2 H^2 D^2}^{(3)} &= (D^\mu H^\dagger D_\mu H) B_{\nu\rho} \widetilde{B}^{\nu\rho}\,,\\
 \mathcal{O}_{8;B^2 H^2 D^2}^{(4)} &= (D_\mu H^\dagger  H+ H^\dagger D_\mu H) D_\nu B^{\mu\rho} B^\nu_{\,\,\rho}\,,\\
 \mathcal{O}_{8;B^2 H^2 D^2}^{(5)} &= i ( H^\dagger D_\mu D_\nu H-D_\mu D_\nu  H^\dagger  H) B^{\mu\rho} B^{\nu}_{\,\,\rho}\,,\\
\mathcal{O}_{8;B^2 H^2 D^2}^{(6)} &=  H^\dagger H D_\mu D_\nu B^{\mu\rho} B^{\nu}_{\,\,\rho}\,,\\
\mathcal{O}_{8;B^2 H^2 D^2}^{(7)} &= i ( H^\dagger D_\nu H-D_\nu H^\dagger H) D_\mu B^{\mu\rho} B^{\nu}_{\,\,\rho}\,,\\
\mathcal{O}_{8;B^2 H^2 D^2}^{(8)} &= ( H^\dagger D_\nu H+D_\nu H^\dagger H) D_\mu B^{\mu\rho} B^\nu_{\,\,\rho}\,,\\
\mathcal{O}_{8;B^2 H^2 D^2}^{(9)} &= ( H^\dagger D^2 H+D^2 H^\dagger H) B^{\nu\rho} \widetilde{B}_{\nu\rho}\,,\\
\mathcal{O}_{8;B^2 H^2 D^2}^{(10)} &= i ( H^\dagger D^2 H-D^2 H^\dagger H) B^{\nu\rho} \widetilde{B}_{\nu\rho}\,\\
\mathcal{O}_{8;B^2 H^2 D^2}^{(11)} &= ( H^\dagger D_\nu H+D_\nu H^\dagger H) D_\mu B^{\mu\rho} \widetilde{B}^{\nu}_{\,\,\rho}\,\\
\mathcal{O}_{8;B^2 H^2 D^2}^{(12)} &= i( H^\dagger D_\nu H - D_\nu H^\dagger H) D_\mu B^{\mu\rho} \widetilde{B}^{\nu}_{\,\,\rho}\,.
\end{align}

\subsubsection{$X^2 = W^2$}
\begin{align}
 \mathcal{O}_{8;W^2 H^2 D^2}^{(1)} &= (D^\mu H^\dagger D^\nu H) W_{\mu\rho}^I W_\nu^{I\rho}\,,\\
 \mathcal{O}_{8;W^2 H^2 D^2}^{(2)} &= (D^\mu H^\dagger D_\mu H) W_{\nu\rho}^I W^{I\nu\rho}\,,\\
 \mathcal{O}_{8;W^2 H^2 D^2}^{(3)} &= (D^\mu H^\dagger D_\mu H) W_{\nu\rho}^I \widetilde{W}^{I\nu\rho}\,,\\
 \mathcal{O}_{8;W^2 H^2 D^2}^{(4)} &= i\epsilon^{IJK}(D^\mu H^\dagger\sigma^I D^\nu H) W_{\mu\rho}^J W_\nu^{K\rho}\,,\\
 \mathcal{O}_{8;W^2 H^2 D^2}^{(5)} &= \epsilon^{IJK}(D^\mu H^\dagger\sigma^I D^\nu H) (W_{\mu\rho}^J \widetilde{W}_\nu^{K\rho}-\widetilde{W}^J_{\mu\rho} W_\nu^{K\rho})\,,\\
 \mathcal{O}_{8;W^2 H^2 D^2}^{(6)} &= i\epsilon^{IJK}(D^\mu H^\dagger\sigma^I D^\nu H) (W_{\mu\rho}^J \widetilde{W}_\nu^{K\rho}+\widetilde{W}^J_{\mu\rho} W_\nu^{K\rho})\,,\\
 \mathcal{O}_{8;W^2 H^2 D^2}^{(7)} &= i\epsilon^{IJK}( H^\dagger \sigma^I D^\nu H-D^\nu H^\dagger \sigma^I  H) D_\mu W^{J\mu\rho}\widetilde{W}^{K}_{\nu\rho}\,,\\
 \mathcal{O}_{8;W^2 H^2 D^2}^{(8)} &= \epsilon^{IJK}  H^\dagger\sigma^I H D_\nu D_\mu W^{J\mu\rho}\widetilde{W}^{K\nu}_{\,\,\,\,\rho}\,,\\
 \mathcal{O}_{8;W^2 H^2 D^2}^{(9)} &= i ( H^\dagger D_\nu H-D_\nu H^\dagger  H) D_\mu W^{I\mu\rho} \widetilde{W}^{I\nu}_{\,\,\,\,\rho}\,,\\
 \mathcal{O}_{8;W^2 H^2 D^2}^{(10)} &= ( H^\dagger D_\nu H+D_\nu H^\dagger  H) D_\mu W^{I\mu\rho} \widetilde{W}^{I\nu}_{\,\,\,\,\rho}\,,\\
 \mathcal{O}_{8;W^2 H^2 D^2}^{(11)} &= ( H^\dagger D_\nu H+D_\nu H^\dagger  H) D_\mu W^{I\mu\rho} W^{I\nu}_{\,\,\,\,\rho}\,,\\
 \mathcal{O}_{8;W^2 H^2 D^2}^{(12)} &= i( H^\dagger D_\nu H-D_\nu H^\dagger  H) D_\mu W^{I\mu\rho} W^{I\nu}_{\,\,\,\,\rho}\,,\\
 \mathcal{O}_{8;W^2 H^2 D^2}^{(13)} &=  H^\dagger H D_\mu W^{I\mu\rho} D_\nu W^{I\nu}_{\,\,\,\,\rho}\,,\\
 \mathcal{O}_{8;W^2 H^2 D^2}^{(14)} &= (D_\mu H^\dagger H+ H^\dagger D_\mu H) W^{I\nu\rho} D^\mu W^{I}_{\nu\rho}\,,\\
 \mathcal{O}_{8;W^2 H^2 D^2}^{(15)} &= i(D_\mu H^\dagger H- H^\dagger D_\mu H) W^{I\nu\rho} D^\mu W^{I}_{\nu\rho}\,,\\
 \mathcal{O}_{8;W^2 H^2 D^2}^{(16)} &= (D_\mu H^\dagger H+ H^\dagger D_\mu H) D^\mu W^{I\nu\rho}\widetilde{W}^{I}_{\nu\rho}\,,\\
 \mathcal{O}_{8;W^2 H^2 D^2}^{(17)} &= i(D_\mu H^\dagger H- H^\dagger D_\mu H) D^\mu W^{I\nu\rho}\widetilde{W}^{I}_{\nu\rho}\,,
 \\
 \mathcal{O}_{8;W^2 H^2 D^2}^{(18)} &= \epsilon^{IJK}( H^\dagger \sigma^I D^\nu H+D^\nu H^\dagger \sigma^I  H) D_\mu W^{J\mu\rho}W^{K}_{\nu\rho}\,,\\
 \mathcal{O}_{8;W^2 H^2 D^2}^{(19)} &= i\epsilon^{IJK}( H^\dagger \sigma^I D^\nu H-D^\nu H^\dagger \sigma^I  H) D_\mu W^{J\mu\rho}W^{K}_{\nu\rho}\,.
\end{align}

\subsubsection{$X^2 = WB$}
\begin{align}
 \mathcal{O}_{8;WB H^2D^2}^{(1)} &= (D^\mu H^\dagger\sigma^I D_\mu H) B_{\nu\rho}W^{I\nu\rho}\,,\\
 \mathcal{O}_{8;WB H^2D^2}^{(2)} &= (D^\mu H^\dagger\sigma^I D_\mu H) B_{\nu\rho}\widetilde{W}^{I\nu\rho}\,,\\
 \mathcal{O}_{8;WB H^2D^2}^{(3)} &= i(D^\mu H^\dagger\sigma^I D^\nu H) (B_{\mu\rho}W^{I\,\rho}_{\nu}-B_{\nu\rho} W^{I\,\rho}_\mu)\,,\\
 \mathcal{O}_{8;WB H^2D^2}^{(4)} &= (D^\mu H^\dagger\sigma^I D^\nu H) (B_{\mu\rho}W^{I\,\rho}_{\nu}+B_{\nu\rho} W^{I\,\rho}_\mu)\,,\\
 \mathcal{O}_{8;WB H^2D^2}^{(5)} &= i(D^\mu H^\dagger\sigma^I D^\nu H) (B_{\mu\rho}\widetilde{W}^{I\,\rho}_{\nu}-B_{\nu\rho} \widetilde{W}^{I\,\rho}_\mu)\,,\\
 \mathcal{O}_{8;WB H^2D^2}^{(6)} &=  (D^\mu H^\dagger\sigma^I D^\nu H) (B_{\mu\rho}\widetilde{W}^{I\,\rho}_{\nu}+B_{\nu\rho} \widetilde{W}^{I\,\rho}_\mu)\,,\\
 \mathcal{O}_{8;WB H^2D^2}^{(7)} &= i( H^\dagger\sigma^I D^\mu H-D^\mu H^\dagger\sigma^I H) D_\mu B^{\nu\rho} W^{I}_{\nu\rho}\,,\\
 \mathcal{O}_{8;WB H^2D^2}^{(8)} &= ( H^\dagger\sigma^I D^\nu H+D^\nu H^\dagger\sigma^I H) D_\mu B^{\mu\rho} W^{I}_{\nu\rho}\,,\\
 \mathcal{O}_{8;WB H^2D^2}^{(9)} &= i( H^\dagger\sigma^I D^\nu H-D^\nu H^\dagger\sigma^I H) D_\mu B^{\mu\rho} W^{I}_{\nu\rho}\,,\\
 \mathcal{O}_{8;WB H^2D^2}^{(10)} &= ( H^\dagger\sigma^I H) D^\mu B_{\mu\rho} D_\nu W^{I\nu\rho} \,,\\
 \mathcal{O}_{8;WB H^2D^2}^{(11)} &= (D_\nu H^\dagger\sigma^I H+ H^\dagger\sigma^I D_\nu H) B_{\mu\rho} D^\mu W^{I\nu\rho}\,,\\
 \mathcal{O}_{8;WB H^2D^2}^{(12)} &= i(D_\nu H^\dagger\sigma^I H- H^\dagger\sigma^I D_\nu H) B_{\mu\rho} D^\mu W^{I\nu\rho}\,,\\
 \mathcal{O}_{8;WB H^2D^2}^{(13)} &= ( H^\dagger\sigma^I H) B_{\mu\rho}  D_\nu D^\mu W^{I\nu\rho} \,,\\
 \mathcal{O}_{8;WB H^2D^2}^{(14)} &= i(D_\nu H^\dagger\sigma^I H- H^\dagger\sigma^I D_\nu H) D^\mu B_{\mu\rho} \widetilde{W}^{I\nu\rho} \,,\\
 \mathcal{O}_{8;WB H^2D^2}^{(15)} &= i ( H^\dagger\sigma^I D_\mu H-D_\mu H^\dagger\sigma^I H) D^\mu B_{\nu\rho} \widetilde{W}^{I\nu\rho}\,,\\
 \mathcal{O}_{8;WB H^2D^2}^{(16)} &= ( H^\dagger\sigma^I H) (D^2 B^{\nu\rho}) \widetilde{W}^{I}_{\nu\rho}\,,\\
 \mathcal{O}_{8;WB H^2D^2}^{(17)} &= ( H^\dagger\sigma^I H) (D^\rho D_\mu W^{I\mu\nu}) \widetilde{B}_{\nu\rho}\,,\\
 \mathcal{O}_{8;WB H^2D^2}^{(18)} &= i(D^\nu H^\dagger\sigma^I H- H^\dagger\sigma^I D^\nu H) \widetilde{B}^{\mu\rho} D_\mu W^I_{\nu\rho}\,,\\
 \mathcal{O}_{8;WB H^2D^2}^{(19)} &= (D^\nu H^\dagger\sigma^I H+ H^\dagger\sigma^I D^\nu H) \widetilde{B}^{\mu\rho} D_\mu W^I_{\nu\rho}\,.
\end{align}

\subsubsection{$X^2 = G^2$}
\begin{align}
 \mathcal{O}_{8;G^2 H^2 D^2}^{(1)} &= (D^\mu H^\dagger D_\nu H) G^A_{\mu\rho} G^{A\nu\rho}\,,\\
 \mathcal{O}_{8;G^2 H^2 D^2}^{(2)} &= (D^\mu H^\dagger D_\mu H) G^A_{\nu\rho} G^{A\nu\rho}\,,\\
 \mathcal{O}_{8;G^2 H^2 D^2}^{(3)} &= (D^\mu H^\dagger D_\mu H) G^A_{\nu\rho} \widetilde{G}^{A\nu\rho}\,,\\
 \mathcal{O}_{8;G^2 H^2 D^2}^{(4)} &= (D_\mu H^\dagger  H+ H^\dagger D_\mu H) D_\nu G^{A\mu\rho} G^{A\nu}_{\,\,\rho}\,,\\
 \mathcal{O}_{8;G^2 H^2 D^2}^{(5)} &= i ( H^\dagger D_\mu D_\nu H-D_\mu D_\nu  H^\dagger  H) G^{A\mu\rho} G^{A\nu}_{\,\,\rho}\,,\\
\mathcal{O}_{8;G^2 H^2 D^2}^{(6)} &=  H^\dagger H D_\mu D_\nu G^{A\mu\rho} G^{A\nu}_{\,\,\rho}\,,\\
\mathcal{O}_{8;G^2 H^2 D^2}^{(7)} &= i ( H^\dagger D_\nu H-D_\nu H^\dagger H) D_\mu G^{A\mu\rho} G^{A\nu}_{\,\,\rho}\,,\\
\mathcal{O}_{8;G^2 H^2 D^2}^{(8)} &= ( H^\dagger D_\nu H+D_\nu H^\dagger H) D_\mu G^{A\mu\rho} G^{A\nu}_{\,\,\rho}\,,\\
\mathcal{O}_{8;G^2 H^2 D^2}^{(9)} &= ( H^\dagger D^2 H+D^2 H^\dagger H) G^{A\nu\rho} \widetilde{G}^{A\nu}_{\,\,\rho}\,,\\
\mathcal{O}_{8;G^2 H^2 D^2}^{(10)} &= i ( H^\dagger D^2 H-D^2 H^\dagger H) G^{A\nu\rho} \widetilde{G}^A_{\nu\rho}\,\\
\mathcal{O}_{8;G^2 H^2 D^2}^{(11)} &= ( H^\dagger D_\nu H+D_\nu H^\dagger H) D_\mu G^{A\mu\rho} \widetilde{G}^{A\nu}_{\,\,\rho}\,\\
\mathcal{O}_{8;G^2 H^2 D^2}^{(12)} &= i( H^\dagger D_\nu H - D_\nu H^\dagger H) D_\mu G^{A\mu\rho} \widetilde{G}^{A\nu}_{\,\,\rho}\,.
\end{align}

\subsection{Operators in the class $X^3 D^2$}
In this case, there are 4 operators for each of the combinations $X^3=W^2B$, $X^3=G^2B$, $X^3=W^3$ and $X^3=G^3$. The (CP-conserving) $W^3$ and $G^3$ operators were previously presented in Ref.~\cite{Quevillon:2018mfl}. For the test, again, only one amplitude is needed for each combination to manifest their independence. For example: $B(p_1)\to W^+(p_2) W^-(p_3)$ and $B(p_1)\to G(p_2) G(p_3)$.
\subsubsection{$X^3=W^2 B$}
\begin{align}
 \mathcal{O}_{8;W^2B D^2}^{(1)} &= B_{\mu\nu} D_\rho W^{I\mu\nu} D_\sigma W^{I\rho\sigma}\,,\\
 \mathcal{O}_{8;W^2B D^2}^{(2)} &= B_{\mu\nu} (D^2 W^{I\mu\rho}) W^{I\nu}_{\,\,\,\,\,\,\rho}\,,\\
 \mathcal{O}_{8;W^2B D^2}^{(3)} &= \widetilde{B}_{\mu\nu} D_\rho W^{I\mu\nu} D_\sigma W^{I\rho\sigma}\,,\\
 \mathcal{O}_{8;W^2B D^2}^{(4)} &= \widetilde{B}_{\mu\nu} (D^2 W^{I\mu\rho}) W^{I\nu}_{\,\,\,\,\,\,\rho}\,.
\end{align}

\subsubsection{$X^3=G^2 B$}
\begin{align}
 \mathcal{O}_{8;G^2B D^2}^{(1)} &= B_{\mu\nu} D_\rho G^{A\mu\nu} D_\sigma G^{A\rho\sigma}\,,\\
 \mathcal{O}_{8;G^2B D^2}^{(2)} &= B_{\mu\nu} (D^2 G^{A\mu\rho}) G^{A\nu}_{\,\,\,\,\,\,\rho}\,,\\
 \mathcal{O}_{8;G^2B D^2}^{(3)} &= \widetilde{B}_{\mu\nu} D_\rho G^{A\mu\nu} D_\sigma G^{A\rho\sigma}\,,\\
 \mathcal{O}_{8;G^2B D^2}^{(4)} &= \widetilde{B}_{\mu\nu} (D^2 G^{A\mu\rho}) G^{A\nu}_{\,\,\,\,\,\,\rho}\,.
\end{align}

\subsubsection{$X^3 = W^3$}
\begin{align}
 \mathcal{O}_{8;W^3 D^2}^{(1)} &= \epsilon^{IJK} W^I_{\mu\nu} D_\rho W^{J\mu\nu} D_\sigma W^{K\rho\sigma}\,,\\
\mathcal{O}_{8;W^3 D^2}^{(2)} &= \epsilon^{IJK} W^I_{\mu\nu} D_\rho W^{J\rho\mu} D_\sigma W^{K\sigma\nu}\,,\\
 \mathcal{O}_{8;W^3 D^2}^{(3)} &= \epsilon^{IJK}\widetilde{W}^I_{\mu\nu} D_\rho W^{J\mu\nu} D_\sigma W^{K\rho\sigma}\,,\\
\mathcal{O}_{8;W^3 D^2}^{(4)} &= \epsilon^{IJK} \widetilde{W}^I_{\mu\nu} D_\rho W^{J\rho\mu} D_\sigma W^{K\sigma\nu}\,,\\
\end{align}

\subsubsection{$X^3 = G^3$}
\begin{align}
 \mathcal{O}_{8;G^3 D^2}^{(1)} &= f^{ABC} G^A_{\mu\nu} D_\rho G^{B\mu\nu} D_\sigma G^{C\rho\sigma}\,,\\
\mathcal{O}_{8;G^3 D^2}^{(2)} &= f^{ABC} G^A_{\mu\nu} D_\rho G^{B\rho\mu} D_\sigma G^{C\sigma\nu} \;, \\
 \mathcal{O}_{8;G^3 D^2}^{(3)} &= f^{ABC}\widetilde{G}^A_{\mu\nu} D_\rho G^{B\mu\nu} D_\sigma G^{C\rho\sigma}\,,\\
\mathcal{O}_{8;G^3 D^2}^{(4)} &= f^{ABC} \widetilde{G}^A_{\mu\nu} D_\rho G^{B\rho\mu} D_\sigma G^{C\sigma\nu} \;, 
\end{align}

\subsection{Operators in the class $X^2 D^4$}
In this class, there is only 1 operator per category, $X=B,W,G$. So the independence of operators is obvious.

\subsubsection{$X=B$}
\begin{align}
 \mathcal{O}_{8;B^2 D^4} &= (D_\sigma D_\mu B^{\mu\nu}) (D^\sigma D^\rho B_{\rho\nu})\,.
\end{align}

\subsubsection{$X=W$}
\begin{align}
 \mathcal{O}_{8;W^2 D^4} &= (D_\sigma D_\mu W^{I\mu\nu}) (D^\sigma D^\rho W^I_{\rho\nu})\,.
\end{align}

\subsubsection{$X=G$}
\begin{align}
 \mathcal{O}_{8;G^2 D^4} &= (D_\sigma D_\mu G^{A\mu\nu}) (D^\sigma D^\rho G^A_{\rho\nu})\,.
\end{align}

\begin{table}[h]
\begin{center}
\resizebox{\textwidth}{!}{
\begin{tabular}{|c|c|}
\hline 
Dimension & Basis \Tstrut\\
\hline \hline
\textcolor{blue}{$d_6$} & Grzadkowski et al. \cite{Grzadkowski:2010es} \Tstrut\\ 
\textcolor{red}{$d_6$} & Gherardi, Marzocca and Venturini \cite{Gherardi:2020det} \\ 
\hline
\textcolor{blue}{$d_7$} & Lehman \cite{Lehman:2014jma} + Liao and Ma \cite{Liao:2016hru} 
\Tstrut\\ 
\textcolor{red}{$d_7$} & Zhang \cite{Zhang:2023kvw} \\ 
\hline
\textcolor{blue}{$d_8$} & Murphy \cite{Murphy:2020rsh} \footnotesize{\&} Li, Ren, Shu et al. \cite{Li:2020gnx}\Tstrut \\ 
\textcolor{red}{$d_8$} & Chala, \textbf{A.D} and Guedes \cite{Chala:2021cgt} (bosonic) \& Ren and Yu \cite{Ren:2022tvi} (GB only)\\ 
\hline 
\end{tabular}}
\end{center}
\caption{\label{tab:ListOfBases} List of tables and references up to dimension 8.  \textcolor{blue}{Blue} bases are physical, while Green's bases are shown in \textcolor{red}{red}. There are also physical bases of dimension nine \cite{Liao:2020jmn,Li:2020xlh} that follow the procedure of the dimension seven \cite{Liao:2016hru} and dimension eight \cite{Ren:2022tvi} precedents, respectively. They both build bases by applying the independence relations to sets of operators until the number is minimal, but the latter uses a systematic method generalizable to any dimension for \gls{smeft}.}
\end{table}

\section{Onshell relations}\label{sec:OnshellRelations}
A Green’s basis can be constructed to include a specific set of physical operators—those that remain after eliminating redundancies via the \gls{eom}. The relationship between a Green’s basis and the physical operators it contains is not univocal, nor is a given set of physical operators uniquely linked to a Green’s basis. Thus, when working with physical and redundant sets of operators simultaneously, the two bases have to be specified. The relation between them is typically expressed through a redefinition of the associated \glspl{wc} in the physical basis—the on-shell relations.

Consider a redundant operator of the form 
$$\mathcal{R}=\sum_p h_p(g, a_s) \mathcal{O}_p,$$
 where the \glspl{eom} relate the left-hand side to a linear combination of physical operators $\mathcal{O}_p$ on the right-hand side. The coefficients $h_p(g, a_s)$ are analytic functions of the \gls{sm} couplings $g\equiv(g_1,g_2,g_3,\lambda,m_H^2)$ and \glspl{wc} associated with physical operators, generically represented by $a_s$ here.

Now consider inserting $\mathcal{R}$  into the Lagrangian, constructed from the physical basis:
$$\mathcal{L}=\mathcal{L}_{phys}+b\mathcal{R},$$
 where $b$ is the Wilson coefficient of the redundant operator. Substituting $\mathcal{R}$ and regrouping terms by operator class yields
  $$\mathcal{L}=\sum \left(a_p+bh_p(g, a_s)\right) \mathcal{O}_p,$$ which is equivalent to performing a shift of the form
  \begin{equation}\label{eq:OnshellRelations}
  a_p\mapsto c_p= a_p+bh_p(g, a_s)\,,
  \end{equation} on the \glspl{wc} of the physical Lagrangian. This is analogous to the effect of a field redefinition.

It is important to note, as discussed in  Section~\ref{sec:EoM}, that if redundant operators are inserted more than once (i.e., at higher loop order), the \glspl{eom} relations no longer hold in general, and care must be taken in such cases.

\subsection{Computation of the on-shell relations}

The on-shell relations between operators are obtained by substituting the terms proportional to the \glspl{eom}, such that the resulting operators contain a reduced number of derivatives. After suitable algebraic manipulations, these new operators can be rewritten in terms of the physical basis. However, due to the inherent freedom in choosing a basis, some redundant operators may not manifest their dependence on \glspl{eom} terms explicitly, which can render the derivation of on-shell relations a laborious task.

A practical strategy to streamline this process involves performing a matching between two Lagrangians: the one constructed from physical operators (serving as the \gls{ir} theory) and the one involving redundant operators, from which \glspl{eom} terms have been eliminated (interpreted as the \gls{uv} theory). Typically, finding all possible \gls{eom} terms is not straightforward, and it is extremely complicated for operators with two or more derivatives. There has been recent progress in the automatisation of this process~\cite{Chala:2024llp,LopezMiras:2025gar}, but most of the calculations of our onshell relations were performed by hand.

After implementing the \glspl{eom} by hand in the redundant operators, the resulting terms need to be expressed as a linear combination of the physical operators. The safest course of action is to carry out the matching at the tree level, using the same processes employed earlier to verify the independence of operators (see Table~\ref{tab:Processes}). Since the relevant kinematic invariants have already been identified in that context, setting up the corresponding system of equations becomes a straightforward procedure. This is the process we chose to obtain the on-shell relations of the new dimension-eight redundant operators presented in Ref.~\cite{Chala:2021cgt} with the previously known physical basis of Ref.~\cite{Murphy:2020rsh}.

Let us consider an explicit example. In the \gls{smeft} operator class $XH^4D^2$, there are a total of $2+1$ operators of the type $BH^4D^2$. Our goal is to determine the on-shell reduction of the operator $\mathcal{O}_{8;BH^4D^2}^{(3)}$. By examining its definition,
\begin{equation}
\mathcal{O}_{8;BH^4D^2}^{(3)}=(H^{\dag} H) D_{\nu} B^{\mu\nu} (D_\mu H^\dagger \ii  H + \text{h.c.}),
\end{equation}
we observe that the \glspl{eom} for the B field strength tensor,
$$D^\mu B_{\mu \nu} = - \frac{g_1}{2} H^\dagger \ii \overleftrightarrow{D}_\nu  H + ... \quad\text{(c.f. Eq.~\eqref{eq:EoMforSMEFT})},$$
can be applied directly. If we retain only order $\Lambda^0$ terms, the resulting expression becomes proportional to an operator in the class $H^6D^2$, that already belongs in the physical basis:
\begin{equation}
\mathcal{O}_{8;BH^4D^2}^{(3)}= - (H^{\dag} H) \frac{g_1}{2} H^\dagger \ii \overleftrightarrow{D}^\mu  H (D_\mu H^\dagger \ii  H + \text{h.c.}),
\end{equation}
which is then matched onto the structure $-g_1\mathcal{O}_{8;H^6D^2}^{(2)}$. 
Consequently, the on-shell effect of $\mathcal{O}_{8;BH^4D^2}^{(3)}$ is equivalent to a shift in the Wilson coefficient of $\mathcal{O}_{8;H^6D^2}^{(2)}$, given by:$$c_{8,H^6D^2}^{(2)}\mapsto c_{8,H^6D^2}^{(2)} - g_1 b_{8;B H ^4D^2}^{(3)}.$$
This shift can then be incorporated into the full coefficient of $\mathcal{O}_{8;H^6D^2}^{(2)}$, which are many., accounting for all other contributing operators. From the set of known bosonic operators at dimension eight, we already have the following additional contributions\footnote{See the Appendix~\ref{app:TableOfOperators} to check the definitions of all the operators.}:
\begin{align}
c_{8,H^6D^2}^{(2)}&\mapsto c_{8,H^6D^2}^{(2)}  
+\frac{1}{4} b_{8;B^2D^4}  g_1^2
    g_2^2-\frac{b_{8;B^2 H^2D^2}^{(8)}  g_1^2}{2}-2
   b_{8;B H^2D^4}^{(1)}  g_1 \lambda - \frac{1}{8} b_{8;B H ^2D^4}^{(3)}
    g_1  g_2^2 \cr
&   +2 b_{8;B H ^2D^4}^{(3)}  g_1 \lambda\,-b_{8;B H ^4D^2}^{(3)}  g_1+b_{8; H ^2D^6}  g_1^2 \lambda +2
   b_{8; H ^4D^4}^{(12)} \lambda -2 b_{8; H ^4D^4}^{(6)} \lambda
   \cr 
&   +b_{8;W^2D^4}  g_1^2
    g_2^2 +\frac{b_{8;WB H ^2D^2}^{(10)}  g_1
    g_2}{2} -\frac{ b_{8;WB H ^2D^2}^{(8)}  g_1
    g_2}{4}-\frac{3 b_{8;WB H ^2D^2}^{(11)}  g_1
    g_2}{8}\cr
    &-b_{8;WB H ^2D^2}^{(13)}  g_1  g_2-\frac{1}{2}
   b_{8;W H ^2D^4}^{(3)}  g_1^2  g_2-\frac{b_{8;W H ^4D^2}^{(7)}
    g_2}{4}\,,
\end{align}
but additional contributions arise from redundant dimension-eight fermionic operators, as well as from products of dimension-six operators.

\begin{figure}[h]
\begin{center}
\includegraphics[scale=0.13]{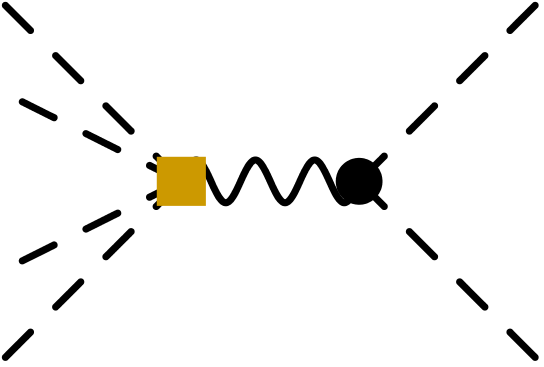}\qquad
\includegraphics[scale=0.13]{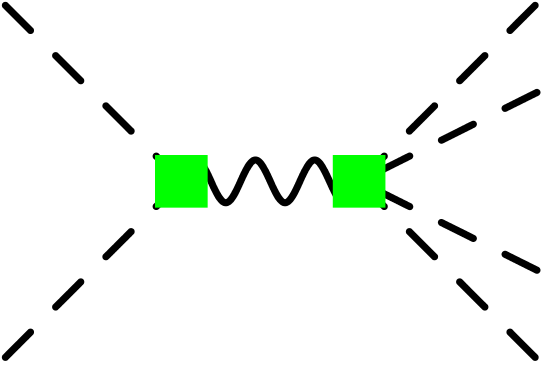}
\end{center}
\caption{\label{fig:onshelld6}Feynman diagrams illustrating the contributions of the operators $\mathcal{O}_{8;BH^4D^2}^{(3)}$ (left) and $\mathcal{O}_{6;BDH}$ (right) to the on-shell realization of $\mathcal{O}_{8;H^6D^2}^{(2)}$. On the left, the $\Lambda^0$ \gls{eom} is applied to the gauge boson leg via the gauge coupling $g_1$. On the right, the \gls{eom} at \order{2} is used, which requires the insertion of $\mathcal{O}_{6;HD}$. Black dots represent SM vertices, Green boxes represent dimension-six interactions and orange boxes represent dimension-eight interactions.}
\end{figure}

In the latter case, when pairs of dimension-six operators are setting on-shell a dimension-eight operator, it is important to note that one of the contributing operators is physical, while the other is redundant. Consider, for example, the dimension-six operator $\mathcal{O}_{6;BDH}$, defined analogously to $\mathcal{O}_{8;BH^4D^2}^{(3)}$:
\begin{equation}
\mathcal{O}_{6;BDH}= D_{\nu} B^{\mu\nu} ( H^\dagger \ii  D_\mu H +  \text{h.c.}),
\end{equation}
and the corresponding \glspl{eom} for the $B$ field strength tensor, extended to include \order{2} terms:
\begin{equation}
D^\mu B_{\mu \nu} =... + \frac{a_{6;HD}}{\Lambda^2}\frac{g_1}{2} H^\dagger \ii \overleftrightarrow{D}_\nu  H + \text{fermionic}.
\end{equation} 
Using this relation, we obtain an additional contribution to the \gls{wc} shift, as computed in Ref.~\cite{Chala:2021pll}:
\begin{equation}
c_{8,H^6D^2}^{(2)}\mapsto c_{8,H^6D^2}^{(2)} + a_{6,HD}(g_1 b_{6;BDH} - g_2b_{6;WDH} -b_{6;HD}^\prime).
\end{equation}
As previously discussed, \gls{eom} substitutions are not valid for multiple insertions of redundant operators. However, in the example under consideration (see Figure~\ref{fig:onshelld6}), each contribution involves only a single insertion of a redundant operator.

The full list of on-shell relations is too long to write, but the ones used for calculations have already been published in different articles \cite{Gherardi:2020det, Chala:2021pll, Chala:2021cgt,Bakshi:2024wzz}. There are also on-shell relations for dimension-seven Green's Basis that we did not need here \cite{Zhang:2023kvw}.
\chapter{Renormalisation Group Equations}\label{ch:RGE}

We now arrive at the central topic of this thesis: the \glspl{rge}. In this section, we will synthesize the information presented in previous sections to explain this powerful tool of \gls{qft}. Conceptually, the key question we seek to address is: What is the relationship between two measurements of the same observable at different energy scales? In particle physics, this typically refers to scattering processes occurring at accelerators with varying center-of-mass energies. Renormalisation provides a framework to relate these measurements to parameters in the Lagrangian. If the energy of the process always remains within the energy range of a single \gls{eft}, only one theory should be required to describe the experiment. In this case, assuming the experiment and theory remain the same, the measured Lagrangian parameter would (in principle) be identical, but at different scales. So, what is the mathematical relationship between these two values of the parameter?

\section{Definition}
\subsection{Callan-Symanzyk Equation}\label{sec:CallanSymanzyk}

Recall the \gls{msbar} scheme defined in Section~\ref{sec:MSbar}. As discussed earlier, the bare and renormalised parameters are related by an expression that explicitly includes a dependence on the renormalisation scale $\mu$ for the renormalised parameter (see Eq.~\eqref{eq:ScaleDependence}). However, we did not focus on this aspect at that point. This scale dependence is a key feature of minimal subtraction schemes, which we will soon explore after formalizing the \glspl{rge}.

For example, we take the Green's function computed for the \gls{wfr} of the right-handed leptons, as introduced in Section~\ref{sec:Divergences}. In this case, the relation between the bare Green's function $G^0(p)$ and the renormalised Green's function 
$G(p)$\footnote{From now on, we will drop the subscript $R$ from renormalised quantities. Bare quantities will be distinguished by the superscript $0$.} is:
\begin{equation}
\ii G^0(p) = \left\langle \overline{e}^0 e^0 \right\rangle = \mu^{2\epsilon} Z_{e}(\mu)\left\langle \overline{e} e \right\rangle,
\end{equation}
where the factor $\mu^{2\epsilon}$ compensates for the shift in the mass dimension of the electron fields in $d=4-2\epsilon$ dimensions compared to $d=4$. The renormalised Green's function is scale-dependent due to the running of the renormalised couplings. This implies that the counterterms are also scale-dependent, as they are functions of the renormalised couplings and masses.

Since the bare Green's function and the bare fields are not scale-dependent, the renormalised Green's function must acquire scale dependence to compensate for both the renormalisation factor $Z_{e}=1+\delta_{e}(\mu)$ and the  $\mu^{2\epsilon}$ factor. This compensating scale dependence is mathematically expressed by the Callan-Symanzik equation~\cite{Callan:1970yg,Symanzik:1970rt}:
\begin{equation}\label{eq:CallanSymanzyk}
%
\mu\frac{{\rm{d}}G^0}{{\rm{d}}\mu}\overset{!}{=} 0 \Rightarrow 0=\mu^{2\epsilon}Z_{e}G(\mu)\left[2\epsilon + \frac{1}{G(\mu)}\mu\frac{{\rm{d}}G(\mu)}{{\rm{d}}\mu} + \frac{1}{Z_e(\mu)}\mu\frac{{\rm{d}}Z_{e}(\mu)}{{\rm{d}}\mu}\right],
\end{equation}
where the scale dependence of the Green's function is unknown.

The next step is to express this relationship as a \gls{rge} for the Green's function $G(p)=\left\langle \overline{e} e \right\rangle$: 
\begin{equation}
\mu\frac{{\rm{d}}G(\mu)}{{\rm{d}}\mu}=\beta_G(y_e(\mu),g_1(\mu),G(\mu))
\end{equation}
where $\beta_G$ is the \textit{beta function}, which governs the scale dependence of $G(\mu)$.

Thus, the \gls{rge} is a first-order linear differential equation. Beta functions are defined by isolating the first-derivative term in the differential equations for the Green's function—computed up to the desired loop level and involving only renormalised quantities. In the case of \gls{wfr} for the electron, the Green's function depends on the Yukawa couplings $y^e$ and the gauge coupling $g_1$, as deduced from the Feynman diagrams in Figure~\ref{fig:WFRexample}. To find an exact solution for the \gls{rge} (if it exists), we would need the \glspl{rge} of $y_e$ and $g_1$, as well.

The \glspl{rge} can be defined at any loop order. Higher-loop corrections are obtained by including additional terms in the counterterms and inserting them into the Callan-Symanzik equation, solving perturbatively.

Before delving into a simplified derivation of the beta functions, there are some important caveats to discuss.

\subsection{Running couplings\label{sec:RunningCouplings}}

When the beta function is different from zero, we say the Green's Function (or whatever object is being considered) is \textit{running}. There is a difference between renormalising and running. 
Renormalisation involves absorbing divergences into redefined (renormalised) parameters and fields, while running refers to the scale dependence of these renormalised parameters, as governed by the \glspl{rge}.

We can compute the running of any renormalisable quantity in the Lagrangian, including couplings and operators, as well as Green's functions. In particular, when studying the running of operators in both the \gls{sm} and beyond, we observe a relationship between the running of the coefficients of these operators. The starting point for this analysis is the scale independence of the bare Lagrangian. Specifically, the bare Lagrangian $\mathcal{L}^0$ satisfies:
\begin{equation}
\mu\frac{{\rm{d}}\mathcal{L}^0}{{\rm{d}}\mu}=0.
\end{equation}
This condition implies:
\begin{equation}
\sum_p (\mu\frac{{\rm{d}}c^0_p}{{\rm{d}}\mu}\mathcal{O}^0_p + c^0_p\mu\frac{{\rm{d}}\mathcal{O}^0_p}{{\rm{d}}\mu})=0,
\end{equation}
which leads to the relation:
\begin{equation}
\mu\frac{{\rm{d}}c^0_p}{{\rm{d}}\mu}=-\mu\frac{{\rm{d}}\mathcal{O}^0_p}{{\rm{d}}\mu},
\end{equation}
where $c^0_p$ are the bare \glspl{wc} and $\mathcal{O}^0_p$ are the corresponding operators. Here, we are assuming that all the operators are independent.

From this, we can deduce that the running of \gls{wc} is intimately connected with the running of the operators themselves. In the \gls{eft} formalism, it is customary to write the \glspl{rge} for the \glspl{wc}, as they encapsulate the energy scale dependence of the operators.

\subsection{Anomalous dimension}
Given the beta function of a coupling or \gls{wc} $c$, we define the anomalous dimension $\gamma$ as:
\begin{equation}
\gamma = \beta/c,
\end{equation}
where $c$ is evaluated at tree-level. Thus, $\gamma$ is determined by the loop corrections to the quantity being considered, just like the beta functions. The term `anomalous dimension' refers to the modification in the power counting of fields due to the effects of \gls{dimreg}, which alters the spacetime dimension to regulate divergences.

To see this, recall the definition of power counting and its relation with mass dimension (see Section~\ref{sec:PowerCounting}). By definition of the power counting, there is a `dilatation' symmetry in the Lagrangian~\cite{Schwartz2013}: The invariance under the rescaling of all dimensionful quantities (including fields, couplings, and derivatives) by a dimensionless factor $\lambda$, with the rescaling given by $g\mapsto \lambda^{[g]} g$, where $[g]$, represents the mass dimension of the quantity $g$. Since this is true for all dimensions, we can compare the shift in the mass dimension when going from $d=4-2\epsilon$ dimensions to $d=4$ dimensions. 

Dimensional continuation (analytical continuation of the spacetime dimension) modifies the power counting of dimensionful quantities. For example, the change in the mass dimension for a quantity $g$ introduces a factor of $\mu^{\gamma\epsilon}$ in its definition, as seen in Eq.~\eqref{eq:DimensionShift}:
\begin{equation}
\left\lbrace g\right\rbrace_{d=4-2\epsilon}=\mu^{\gamma\epsilon}\left\lbrace g\right\rbrace_{d=4}.
\end{equation} 
Here,$\gamma\epsilon$ represents the difference in mass dimensions between $d=4-2\epsilon$ and $d=4$.

Taking the scale derivative with respect to $\mu$, we find the tree-level \glspl{rge} for the quantity $g$:
\begin{equation}
\mu\frac{{\rm{d}}g}{{\rm{d}}\mu} = \gamma \epsilon g \equiv ([g]_{d=4-2\epsilon}-[g]_{d=4}) g.
\end{equation}

This equation shows that the tree-level anomalous dimension represents the distortion in the power counting of a quantity when the spacetime dimension is changed. At higher loop levels, the anomalous dimension also accounts for the scale introduced by loop integrals, which contribute additional terms to the \glspl{rge}.
\subsection{Applications of running}

In \gls{qft}, the \glspl{rge} have two main applications: resumming logarithmic terms and expressing the values of couplings at different energy scales. The procedure involves calculating the value of an observable at a given scale based on its known value at a lower energy scale.

\begin{figure}
\begin{center}
\includegraphics[scale=0.13]{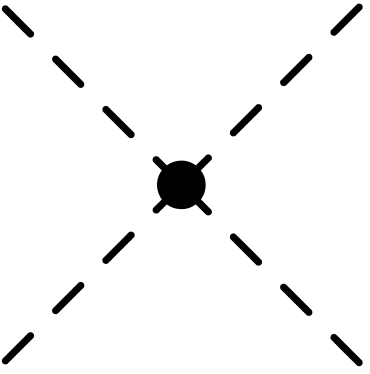}\qquad
\includegraphics[scale=0.13]{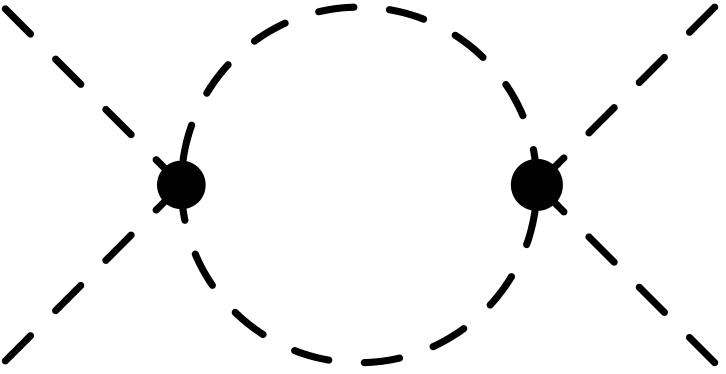}\qquad
\includegraphics[scale=0.13]{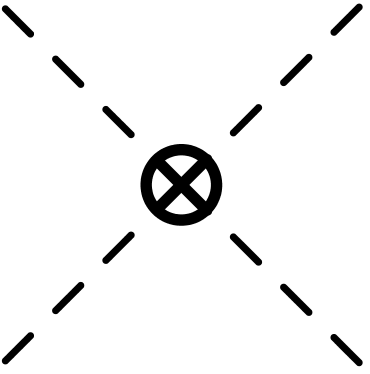}
\end{center}
\caption{\label{fig:RunningApps}Diagrams contributing to the amplitude of $H^+H^0\rightarrow H^+H^0$ at tree-level and one-loop. Only diagrams proportional to  $\lambda$ are included here. The black dots represent SM vertices, and the black cross represents the counterterm.}
\end{figure}

For example\footnote{Adapted from Ref.~\cite{Cohen:2019wxr}}, let us consider the scattering  process $H^+H^0\rightarrow H^+H^0$ in the \gls{sm}. At tree level, the amplitude is straightforward. At one-loop, we can use tools like \texttt{FeynRules}~\cite{Alloul:2013bka}, \texttt{FeynArts}~\cite{Hahn:2000kx} and \texttt{FormCalc}~\cite{Hahn:1998yk} to perform the computation. For simplicity, we focus on the self-renormalisation of the Higgs quartic coupling $\lambda$, considering only the terms proportional to $\lambda$ (see Figure~\ref{fig:RunningApps}). 

To compute physical observables at one loop, we need the finite part of the amplitude, as the counterterm cancels the divergences. At tree level, the amplitude is given by:
\begin{equation}
\mathcal{A}^{TL}[\mu]=-2\lambda.
\end{equation}
At one-loop, the amplitude takes the form:
\begin{equation}\label{ec:ProblematicLog}
\mathcal{A}^{1L}[\mu]=\frac{3\lambda^2}{2\pi^2} \left(\ln \left(\frac{\mu^2}{m_H^2}\right)+\frac{2}{3}\right),
\end{equation}
where the ratio $\mu/m_H$ arises naturally from the loop integral solved using \gls{dimreg} and the \gls{msbar} scheme.

Assuming the values of the couplings are known at a smaller, well-explored scale $\Lambda_{s}$, we aim to evaluate them at a higher scale $\Lambda_{c}$, which corresponds to future collider energies. This allows us to make predictions for experiments at $\Lambda_{c}$ and compare them with the data at the low-energy scale $\Lambda_{s}$. Directly evaluating both amplitudes at $\mu=\Lambda_c$ could lead to a non-perturbative result if the separation between the high scale and the theory scale is too large (i.e., $\Lambda_c \gg m_H$).
To address this, we compute the \glspl{rge} for the Higgs quartic coupling $\lambda$. Using the \gls{msbar} counterterm:
\begin{equation}
\delta_\lambda=-\frac{3\lambda}{2\pi^2 \epsilon},
\end{equation}
and the Callan-Symanzik equation (Eq.~\eqref{eq:CallanSymanzyk}), we obtain:
\begin{equation}
\mu\frac{{\rm{d}}\lambda}{{\rm{d}}\mu} = 2\epsilon\lambda[TL] + \frac{3\lambda^2}{\pi^2}[1L].
\end{equation}

Setting $\epsilon\rightarrow 0$ and solving these equations for constant coefficients leads to a logarithmic dependence on the energy scale:
\begin{align}
\mu\frac{{\rm{d}}\lambda}{{\rm{d}}\mu}&=\frac{3\lambda^2}{\pi^2}\Rightarrow\\
\frac{1}{\lambda(\mu)}-\frac{1}{\lambda(\Lambda)}&=\frac{3}{\pi^2}\ln \left(\Lambda/\mu\right)\Rightarrow\\
\lambda(\mu)&=\lambda(\Lambda)\left(1+\lambda(\Lambda)\frac{3}{\pi^2}\ln \left(\Lambda/\mu\right)\right)^{-1}.
\end{align}
Here, $\Lambda$ represents the integration constant, determined by imposing a boundary condition on the coupling at a known scale.

In this simplified case, we can express the coupling $\lambda$ at the low scale $\Lambda_s$ in terms of its value at the high energy scale $\Lambda_c$. The amplitude at low scale can then be `run' up to the high-energy value. If we run the tree-level amplitude, we obtain the `\gls{rg}-improved' result at \gls{lo}:

\begin{equation}
\mathcal{A}^{LL+LO}=-2\lambda(\mu=\Lambda_s)=-2\lambda(\Lambda_c)\left(1+\lambda(\Lambda_c)\frac{3}{\pi^2}\ln \left(\Lambda_c/\Lambda_s\right)\right)^{-1}\,.
\end{equation}
This expression includes a logarithmic term that remains well-behaved, even if the collider scale is significantly separated from the theory scale $m_H$, provided the experiment is performed at a scale $\Lambda_{s}$ that is not too far from $\Lambda_{c}$. 

We can also apply the \gls{ll} expansion in the \gls{nlo} expression by inserting the solution of the RGE into the one-loop amplitude. The result remains well-behaved:
\begin{align}\label{eq:WellBehavedLog}
\mathcal{A}^{LL+NLO}&=\frac{3(\lambda[\mu=\Lambda_s])^2}{2\pi^2} \left(\ln \left(\frac{\Lambda_s^2}{m_H^2}\right)+\frac{2}{3}\right)\cr
&=\frac{3\lambda^2[\Lambda_c]}{2\pi^2} \left(1+\lambda[\Lambda_c]\frac{3}{\pi^2}\ln \left(\Lambda_c/\Lambda_s\right)\right)^{-2}\left(\ln \left(\frac{\Lambda_s^2}{m_H^2}\right)+\frac{2}{3}\right).
\end{align}

The key point is that there are two perturbative expansions: one in the coupling $\lambda$ and another in the product of $\lambda$ and the logarithm of the energy scales, $\lambda \ln( \Lambda_{c}/\Lambda_{s})$, which is known as the \gls{ll} expansion. If both expansions are of the same order, perturbation theory breaks down. However, when they are well-separated, this formulation absorbs the divergences from the logarithms of distant scales. The \gls{ll}+\gls{nlo} amplitude is related to the \gls{nlo} expression expanded at $\mu=\Lambda_c\approx\Lambda_{s}$, which implies that the problematic logarithm in Eq.~\eqref{ec:ProblematicLog} has been resummed into the well-behaved expression in Eq.~\eqref{eq:WellBehavedLog}.

\subsection{Mixing and power counting}

The structure of \gls{smeft} \glspl{rge} requires detailed consideration of operator renormalisation beyond the case of single operators. In \glspl{qft}, loop diagrams with operator insertions can generate divergent amplitudes contributing to distinct operator structures. As a result, the corresponding counterterms imply a matrix structure in the \glspl{rge}, reflecting operator mixing, which significantly influences the evolution of \glspl{wc}.

When the insertion of an operator $\mathcal{O}_t$ in a loop induces divergences in a different operator $\mathcal{O}_s$, the renormalisation procedure necessitates including counterterms for $\mathcal{O}_s$, even if it was not present in the bare Lagrangian. This leads to \glspl{rge} of the form:

\begin{equation}
\mu\frac{{\rm{d}}c_s}{{\rm{d}}\mu}=\sum_t \gamma_{st}c_t + \mathcal{O}(c^2),
\end{equation}
where $c_s$ and $c_t$ are \glspl{wc}, and $\gamma_{st}$ is the \gls{adm} encoding the mixing. The $\mathcal{O}(c^2)$ term reflects contributions from higher-loop or multiple insertions.

As discussed in Section~\ref{sec:DefiningRenormalizability}, regulating divergences in higher-dimensional operators sometimes requires operators of even higher dimension. In the \gls{eft} framework, truncating the Lagrangian at a given operator dimension fixes the calculational precision. However, loop-level diagrams involving multiple insertions of lower-dimensional operators can contribute to higher-dimensional structures if consistent with the power counting. For example, one-loop diagrams with two insertions of dimension-six operators can generate contributions of the form:

\begin{equation}
\frac{\beta_p}{\Lambda^4} \sim \left(\frac{c_s c_t}{\Lambda^4}\right),
\end{equation}
which is, in principle, comparable in magnitude to the one-loop corrections for dimension-eight operators.

While some all-dimensional predictions of the \gls{adm} are possible in special cases (e.g., the \glspl{rge} $H^n$ and $\ell^2 H^n$ operators \cite{Liao:2017amb} are known for all dimensions in \gls{smeft}), in general, explicit computation order by order remains the most reliable method for determining these effects.

\section{Algorithm and computation}

Our goal is to derive a general expression for the \glspl{rge} that can be applied systematically to \gls{smeft} and other \glspl{eft}. Consider an operator and its corresponding \gls{wc} in the bare Lagrangian:
\begin{equation}
\mathcal{L}^{0} \supset \frac{c^{0}}{\Lambda^{r-4}}\mathcal{O}^{0}.
\end{equation}
After renormalisation, divergences are absorbed into counterterms, encoded in the renormalisation constant $ Z_\mathcal{O} $ (defined as  $c^{0}=Z_\mathcal{O}^{-1} c$)  , and a factor of $\mu^\gamma $ appears to compensate for the dimensional mismatch introduced by \gls{dimreg}. The renormalised operator can then be written in terms of the bare one as:
\begin{equation}\label{eq:bareWC}
Z_\mathcal{O}^{-1}\mu^{\gamma}\frac{c}{\Lambda^{r-4}}\mathcal{O}^{0} = Z_\mathcal{O}^{-1}\mu^{\gamma}\frac{c}{\Lambda^{r-4}} Z_\phi^{-1}\mathcal{O}\,,
\end{equation}
where $Z_\phi $ accounts for the \gls{wfr} of the fields appearing in the operator $ \mathcal{O}$\footnote{As mentioned in Section~\ref{sec:RunningCouplings}, we assume the scale dependence in $\mathcal{O}$ is due only to the renormalisation of the fields, while the intrinsic scale dependence (due to the divergences of the operator itself) are encoded in the \gls{wc} $c$ and its counterterm $Z_\mathcal{O}$. Thus, in our notation, $\mu\frac{{\rm{d}}\mathcal{O}}{{\rm{d}}\mu}=0$ while $\mu\frac{{\rm{d}}c}{{\rm{d}}\mu}\neq 0$. }. Using the standard definitions $ \varphi^{0} = Z_\varphi^{-1} \varphi $ for scalars and $ \chi^{0} = Z_\chi^{-\frac{1}{2}} \chi $ for fermions or gauge bosons, we express:
\begin{equation}
Z_\phi=\prod_{\phi=\varphi,\chi} Z_\varphi^{n_\varphi}Z_\chi^{n_\chi/2}\,,
\end{equation}
where $ n_\phi $ denotes the number of each field $ \phi $ in $ \mathcal{O} $. For instance, for an operator like $ \ell^2 H^2 $, we would have $ Z_\phi = Z_H^{2} Z_\ell^{1} $.

From Eq.~\eqref{eq:bareWC}, we can extract the bare coefficient $ c^{0} $ in terms of renormalised quantities. Applying the Callan-Symanzik equation~\eqref{eq:CallanSymanzyk} and expanding the renormalisation constants as $ Z = 1 + \delta $, we obtain:
\begin{equation}
\mu\frac{{\rm{d}}c^0}{{\rm{d}}\mu}=0\Rightarrow \mu\frac{{\rm{d}}c}{{\rm{d}}\mu}= -\gamma^{(\text{TL})} \epsilon c + c \mu\frac{{\rm{d}}}{{\rm{d}}\mu} (\delta_\mathcal{O}+\delta_\phi) \,,
\end{equation}
where $\gamma^{(\text{TL})}\epsilon$ represents the tree-level anomalous dimension of $c$, and the second term originates from the scale dependence of the counterterms. In the limit $\epsilon \to 0 $, the first term vanishes, and only the loop-induced running remains.

Since the explicit $\mu$-dependence has been factored out, the scale dependence in the counterterms arises solely from their dependence on running couplings. This allows us to re-express the derivative as:

\begin{equation}
\beta_c^{(1\text{L})}=c \sum_m \beta^{(\text{TL})}_m \frac{{\rm{d}}}{{\rm{d}}x_m}(\delta_\mathcal{O}+\delta_\phi)\,,
\end{equation}
where the $ x_m $ denote the running parameters, and $\beta_m^{(\text{TL})} $ are their \gls{lo} beta functions. The contributions from \gls{wfr}, encoded in $\delta_\phi $, can be computed separately and added to the operator beta function.

We now express the \gls{rge} for the Wilson coefficient as: 
\begin{equation}\label{eq:RGEdefinition}
\beta =-c \sum_m \gamma^{(\text{TL})}_m x_m \frac{d}{dx_m} \left(\frac{\tilde{c}}{c}\right) + \text{WFR},
\end{equation}

where $ \tilde{c} = - c \delta_\mathcal{O}$ denotes the coefficient of the divergence at one loop (i.e., the one-loop contribution to the operator's counterterm) in the \gls{msbar} scheme. This expression captures the dependence of the \gls{wc} on the running parameters via the structure of the divergent terms.

\subsection{Preliminary considerations}

Once the general expression for the beta function in Eq.~\eqref{eq:RGEdefinition} is established, several important considerations must be addressed before calculating the divergences. Operator insertions lead to a proliferation of diagrams, some of which are highly nontrivial to evaluate. A systematic approach is therefore essential to identify which contributions are necessary and which can be safely ignored. Key points include:

\begin{itemize}
\item Power counting determines which operator insertions are relevant. For a complete result at a given order, all contributions with the same power-counting suppression must be included.
\item Lower-dimensional operators may receive higher-order corrections from multiple insertions of marginal or irrelevant operators.
\item Not all \gls{adm} elements need to be computed. Some contributions vanish due to symmetry or structural arguments.
\item Operator selection can be optimized when the \gls{uv} theory naturally suppresses certain interactions.
\end{itemize}

The first point concerns operator mixing. In principle, mixing occurs among all operators within a given dimensional class and between different classes via multiple insertions. While listing participating operators is straightforward at low dimensions, the computation of their associated diagrams becomes increasingly demanding. Fortunately, the various contributions to the beta function are additive, allowing the calculation to be modularized.

Although renormalisation typically focuses on higher-dimensional operators, lower-dimensional operators can also receive loop-level corrections. Such corrections must carry dimensionful suppression factors due to power counting. In \gls{smeft}, where the Higgs vacuum expectation value and mass $m_H$ provide the only low-energy scales, this implies that such contributions often arise through Higgs insertions, typically in the form of loops involving Higgs fields. Nonetheless, each case requires explicit analysis to determine relevance.

Non-renormalisation theorems~\cite{Cheung:2015aba} provide predictive power by identifying zeros in the \gls{adm} without explicit calculation. While currently limited—mostly applying to linear renormalisation of dimension-six and dimension-eight operators at one loop—they are still useful for simplifying computations. Structural arguments based on operator content, such as mismatched field content or quantum numbers, can also be used to anticipate vanishing contributions.

In weakly coupled \gls{uv} completions of \gls{smeft}, not all operator classes are generated at tree level. As shown in~\cite{Craig:2019wmo}, some operators only arise at loop level. Their insertions into \glspl{rge} introduce additional loop suppression, which can justify their exclusion from beta function calculations—though some authors choose to retain them for completeness. Additionally, many \gls{uv} completions do not include \gls{lnv} operators at low energies due to the large scale (typically $\Lambda\sim 10^{10}\,\text{GeV}$) expected for these models~\cite{Herrero-Garcia:2019czj}, offering further grounds for their omission in practical computations.

\subsection{Offshell diagrammatical approach}

Historically, the \glspl{rge} of \gls{smeft} have been computed diagrammatically and off-shell. Results up to $\mathcal{O}(\Lambda^{-3})$ have been obtained using this approach. We adopt an off-shell formalism, as \gls{1pi} diagrams are typically easier to organize and compute, despite introducing additional redundancies. Rather than removing the redundancies case by case, our approach instead employs a Green's basis to systematically absorb redundancies, followed by the application of \glspl{eom}.

We applied this method to dimension-eight operators, leading to original results~\cite{Bakshi:2024wzz,DasBakshi:2022mwk,DasBakshi:2023htx} (see also~\cite{Chala:2021pll}, which was the first computation using this approach). Other groups have similarly used it to renormalise dimension-seven operators in 2023~\cite{Zhang:2023kvw}. In recent years, alternative techniques have emerged. Functional renormalisation provides a robust non-diagrammatic method, and promising results have been obtained via unitarity cuts~\cite{AccettulliHuber:2021uoa} and geometrical approaches based on the space of operators~\cite{Helset:2022pde,Assi:2023zid,Assi:2025fsm}.

The core idea of the off-shell method is to work within the framework of renormalised perturbation theory, computing only the divergences of \gls{1pi} diagrams. As discussed in Section~\ref{sec:GB}, this necessitates including additional operators to absorb divergences, which are later removed via on-shell relations. Amplitudes are expressed as linear combinations of kinematic invariants, determined by the operator’s field content and power counting. Constructing the independent kinematic structures using metric tensors, spinors, and the Levi-Civita symbol becomes straightforward under this framework.

We now present the method in general terms, with specific examples and refinements order by order in power counting, in subsequent sections. Consider the \gls{smeft} Lagrangian in renormalised perturbation theory, including physical operators from Table~\ref{app:TableOfOperators}:

\begin{equation}
\mathcal{L}_{UV}=\mathcal{L}_{SM} +\sum_{r=5}^{n}\sum_{q}\sum_{p=1}^{n_{q}}\frac{c_{r;\,q}^{(p)}}{\Lambda^{r-4}}\mathcal{O}_{r;\,q}^{(p)}.
\end{equation}

Here, $c_{r,q}^{(p)}$ are the \glspl{wc} of the $p$-th operator in class $q$ with mass dimension $r$, chosen to be dimensionless by explicit power counting. In the low-energy theory, we take the same physical operators, with coefficients $a_{r,q}^{(p)}$. Then, redundant operators $\mathcal{R}_{r,q}^{(p)}$ and their coefficients $b_{r,q}^{(p)}$ are added to absorb off-shell divergences. Each operator has an associated counterterm $Z_{r,q}^{(p)}$, expanded perturbatively as $Z_{r,q}^{(p)} = 1 + \delta_{r,q}^{(p)}$.

In the case that the operator term has flavour indices, the counterterm should respect the symmetries of the coefficient, while admitting flavour-dependent contributions. In that case, the counterterm also has flavour indices, so the following notation is preferable:
\begin{equation}
[Z_{r;\,q}^{(p)}]_{\alpha\beta\dots}[a_{r;\,q}^{(p)}]_{\alpha\beta\dots}=[a_{r;\,q}^{(p)}]_{\alpha\beta\dots}+[\delta a_{r;\,q}^{(p)}]_{\alpha\beta\dots}.
\end{equation} 

In any case, counterterms are fixed by the divergent parts of one-loop \gls{1pi} diagrams using \gls{dimreg} and the \gls{msbar} scheme:
\begin{align}\tilde{a}_{r;\,q}^{(p)}&=-\delta_{r;\,q}^{(p)}a_{r;\,q}^{(p)}\,,\\
\tilde{b}_{r;\,q}^{(p)}&=-\delta_{r;\,q}^{(p)}b_{r;\,q}^{(p)}\,,
\end{align}
where contributions from \gls{wfr} are omitted at this stage. Thus, it is remarkably simpler to work directly with the divergences. Once computed, redundant coefficients are removed using on-shell relations. The physical divergences are shifted according to the on-shell relations, as defined in Eq.~\eqref{eq:OnshellRelations}:
\begin{equation}\tilde{a}_{r;\,q}^{(p)}\mapsto \tilde{c}_{r;\,q}^{(p)}=\tilde{a}_{r;\,q}^{(p)}-\sum_s h(m_H^2,\lambda,g_i, \left\lbrace a_{r;\,q}^{(p)}\right\rbrace)\tilde{b}_{r;\,q}^{(s)}\,,
\end{equation}
and these are inserted into the \gls{rge} as defined in Eq.~\eqref{eq:RGEdefinition}.

The tree-level anomalous dimension is derived from the field content and spacetime dimensionality. For an operator $\mathcal{O} = X^{n_X} \psi^{n_\psi} H^{n_H} D^{n_D}$ in $d = 4 - 2\epsilon$ dimensions, we obtain:
\begin{equation}
\gamma^{(\text{TL})}=n_X+n_\psi+n_H-2\equiv n_\phi-2,
\end{equation}
where $n_\phi$ is the number of fields in the operator and the subtraction of 2 accounts for the overall mass dimension of the Lagrangian term in $d$ dimensions.
\gls{wfr} is typically included after computing all operator mixings. It contributes to self-renormalisation via \gls{sm} counterterms.

All diagrammatic computations are carried out using the tools \texttt{FeynRules}~\cite{Alloul:2013bka}, \texttt{FeynArts}~\cite{Hahn:2000kx}, and \texttt{FormCalc}~\cite{Hahn:1998yk}, with \texttt{MatchMakerEFT}~\cite{Carmona:2021xtq} used for cross-checks. These tools automate Feynman rule generation and diagram evaluation. The Background Field Method~\cite{Abbott:1980hw} is applied manually when necessary. \texttt{MatchMakerEFT} also performs one-loop matching and \gls{rge} extraction, though its default implementation includes only dimension-six \glspl{rge} up to $\mathcal{O}(\Lambda^{-2})$.

Counterterms are computed using \gls{dimreg} and the \gls{msbar} scheme. Since we restrict ourselves to one-loop computations, evanescent operators are not included; they contribute only finite parts and thus do not affect the \glspl{rge}.

We have outlined the general renormalisation strategy. In subsequent sections, we explore explicit examples across different operator classes, summarising known \glspl{rge} at lower orders and comparing them with our results at $\mathcal{O}(\Lambda^{-4})$. We highlight universal features as well as complications that arise only at higher orders in the power counting expansion.

\section{Renormalisation up to first order in the cutoff}

The first irrelevant operator encountered in the \gls{smeft} is the Weinberg operator, along with its Hermitian conjugate:
\begin{align}
[\mathcal{O}_{5;\,\ell H}]_{\alpha\beta}&=\epsilon_{ij}\epsilon_{kl}(\ell^i_\alpha)^\top C\ell^k_\beta H^j H^l\;,\\
[\mathcal{O}^\dagger_{5;\,\ell H}]_{\alpha\beta}&=\epsilon_{ij}\epsilon_{kl}\overline{\ell}^i_\alpha C (\overline{\ell}^k_\beta)^\top (H^*)^j (H^*)^l\;,
\end{align}
which correspond to the operator $\mathcal{Q}_{\nu\nu}$ in the basis of Ref.~\cite{Grzadkowski:2010es}.

This operator plays a central role in seesaw mechanisms, where it emerges upon integrating out heavy fields—for example right-handed neutrinos—that couple to light \gls{sm} leptons via Yukawa interactions. After \gls{ewsb}, the operator defined above generates a Majorana mass term for the left-handed neutrinos:
\begin{equation}
[c_{5;\,\ell H}]_{\alpha\beta}[\mathcal{O}_{5;\,\ell H}]_{\alpha\beta}\rightarrow \frac{v^2}{2} [c_{5;\,\ell H}]_{\alpha\beta} (\overline{\nu_L})_\alpha^c (\nu_L)_\beta\,,
\end{equation}
and analogously for the Hermitian conjugate. Here, $(\nu_L)_\alpha^c = \ii \sigma^2 (\nu_L)_\alpha^*$ denotes the charge-conjugated field. The \gls{rge} running leads to radiative corrections to the light neutrino masses generated via this mechanism.

From the standpoint of power counting, dimension-five operators can only mix among themselves at order $\mathcal{O}(\Lambda^{-1})$. There is no operator mixing into dimension-four (renormalisable) terms in the \gls{sm} at this order, either. The corresponding \gls{rge} can be schematically expressed as:
\begin{equation}
16\pi^2\mu\frac{{\rm{d}}}{{\rm{d}}\mu}c_5=\gamma^{(d5)}c_5\,,
\end{equation}
where $\gamma^{(d5)}$ is the anomalous dimension matrix governing the evolution of the operator coefficients. 

The one-loop renormalisation of this operator was first computed in Refs.~\cite{Chankowski:1993tx,Babu:1993qv}, with both analyses yielding consistent results. A later study in Ref.~\cite{Antusch:2001ck} revisited the calculation and corrected a missing numerical factor.\footnote{Possibly related to a vertex diagram miscalculation, although this is not explicitly discussed in the original papers.} These results can now be easily reproduced using modern tools such as FeynRules, FeynArts, and FormCalc, following our methodology.
 
\begin{figure}[h]
\begin{center}
\includegraphics[scale=0.13]{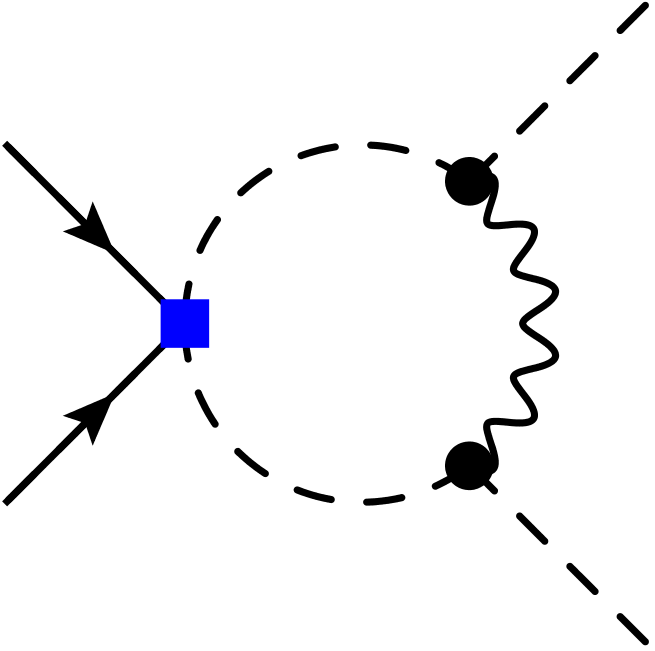}\qquad
\includegraphics[scale=0.13]{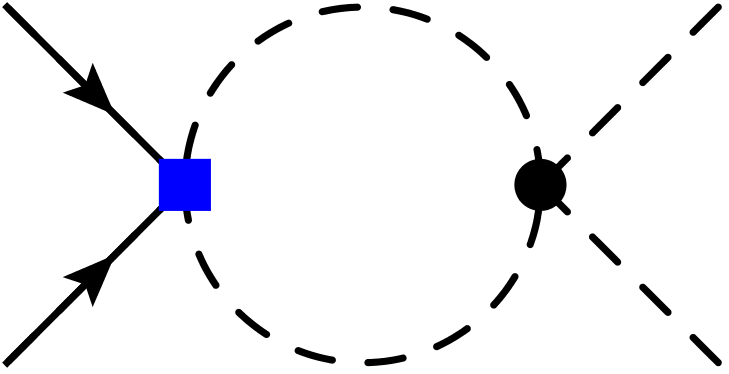}\qquad
\includegraphics[scale=0.13]{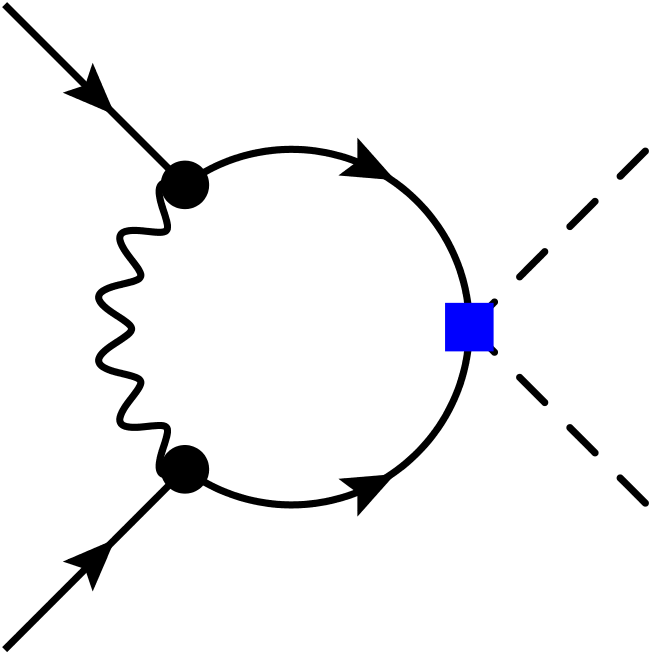}\\ \vspace{1cm}
\includegraphics[scale=0.13]{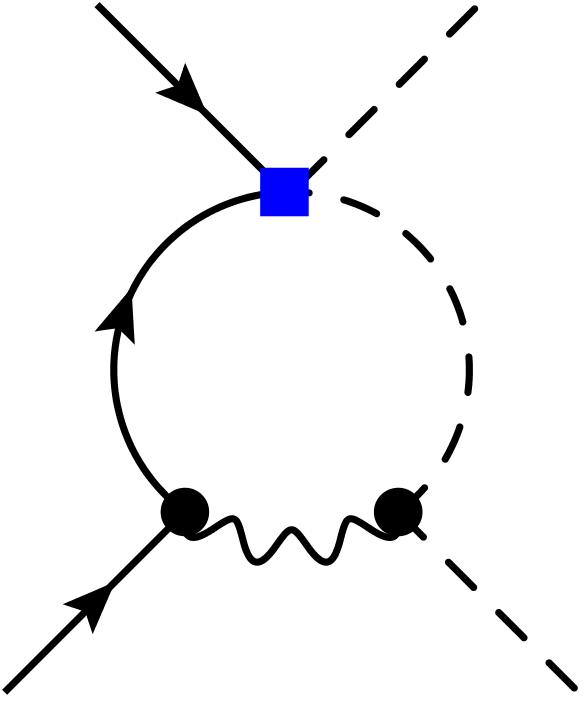}\qquad
\includegraphics[scale=0.13]{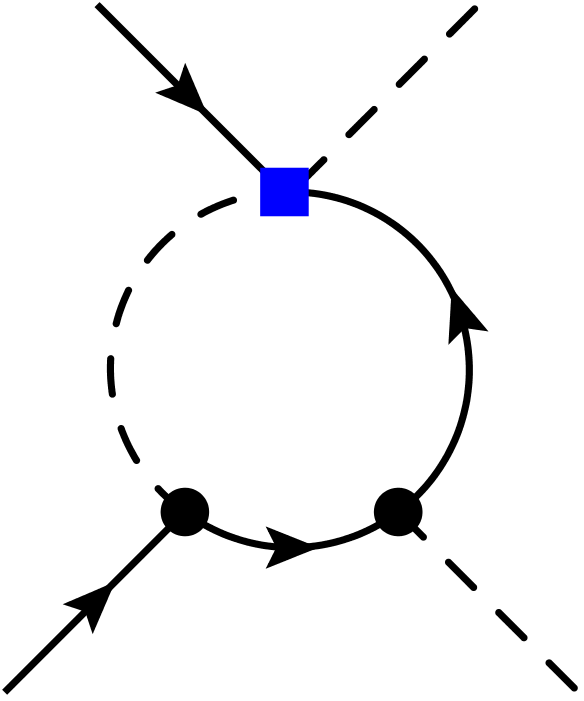}
\end{center}
\caption{\label{fig:d5div}One-loop Feynman diagrams contributing to the renormalisation of the Weinberg operator. There are 21 diagrams in total, accounting for all relevant internal field insertions and channels. Black dots indicate \gls{sm} vertices, while blue boxes denote insertions of dimension-five operators.}
\end{figure}

To extract the counterterm associated with $[c_{5;,\ell H}]{\alpha\beta}$, we consider the process $(e_L)_\alpha (\nu_L)_\beta\rightarrow~H^0H^- $. At one loop, 21 Feynman diagrams contribute to the amplitude, as shown in Fig.~\ref{fig:d5div}. The divergent part of the amplitude reads:
\begin{align}\label{eq:d5divergence}
16\pi^2\epsilon[\tilde{c}_{;\,\ell H}]_{\alpha\beta}&=-\frac{g_1^2}{4} [c_{5;\,\ell H}]_{\alpha\beta} + \frac{3 g_2^2}{4} [c_{5;\,\ell H}]_{\alpha\beta} -2\lambda [c_{5;\,\ell H}]_{\alpha\beta}\nonumber\\ 
&+ \left((\left[c_{5;\ell H}\right]\left[y^e\right]\left[y^e\right]^\dagger)+(\left[c_{5;\ell H}\right]\left[y^e\right]\left[y^e\right]^\dagger)^\top\right)_{\alpha\beta}\,.
\end{align}

At this order, there are no one-loop connected reducible diagrams beyond those contributing to \gls{wfr}. Consequently, the full \gls{rge} for the operator coefficient can be derived using the divergence in Eq.~\eqref{eq:d5divergence} and the counterterms for the Higgs and lepton fields from Ref.~\cite{Antusch:2001ck}: 
\begin{align}
(\delta_\ell)_{i} &=-\frac{\left([y^e]^* [y^e]^\top \right)}{16\pi^2 \epsilon}-\frac{g_1^2}{32\pi^2 \epsilon}\mathbbm{1}-\frac{g_2^2}{32\pi^2 \epsilon}\mathbbm{1}\,,\\
\delta_H&=-\frac{\text{Tr}\left([y^e] [y^e]^\dagger \right)}{8\pi^2 \epsilon}-3\frac{\text{Tr}\left([y^u] [y^u]^\dagger \right)}{8\pi^2 \epsilon}-3\frac{\text{Tr}\left([y^d] [y^d]^\dagger \right)}{8\pi^2 \epsilon}-\frac{g_1^2}{16\pi^2 \epsilon}-\frac{3g_2^2}{16\pi^2 \epsilon},
\end{align}
where the matrices of $(\delta_\ell)_{i}$ refer to flavour space. In our notation\footnote{Ref.~\cite{Antusch:2001ck} uses $\kappa_{\alpha\beta}=[c_{5;\ell H}]_{\alpha\beta}+[c^\dagger_{5;\ell H}]_{\alpha\beta}$}, we obtain:
\begin{align}
16\pi^2\frac{{\mathrm{d}}}{{\mathrm{d\mu}}}\left[c_{5;\ell H}\right]_{\alpha\beta}&=\left(-3g_2^2 + 4\lambda +\Tr{2(y^e)^\dagger(y^e)+6(y^u)^\dagger(y^u)+6(y^d)^\dagger(y^d)}\right)\left[c_{5;\ell H}\right]_{\alpha\beta}\nonumber\\
& -\frac{3}{2}\left((\left[c_{5;\ell H}\right]\left[y^e\right]\left[y^e\right]^\dagger)+(\left[c_{5;\ell H}\right]\left[y^e\right]\left[y^e\right]^\dagger)^\top\right)_{\alpha\beta}.
\end{align}

It is worth noting that all the numerical coefficients in this \gls{rge} are of order one, in agreement with expectations from naturalness and effective field theory dimensional analysis.

\section{Renormalisation up to second order in the cutoff}
At the next order in the \gls{smeft} expansion, we encounter operators of dimension six. In this regime, the anomalous dimension governing the scale dependence of the \glspl{wc} $c_{6;p}$ can be written as:
\begin{equation}
16\pi^2\mu\frac{{\rm{d}}}{{\rm{d}}\mu}c_{6;p}=\gamma_{ps}^{(d6)}c_{6;s}+\gamma_{p}^{(d5^2)}c_5^2\,.
\end{equation}
Here, $\gamma^{(d6)}$ encodes the mixing between dimension-six operators, while $\gamma_{p}^{(d5^2)}$ represents the contribution from insertions of two dimension-five operators.

A few years after the introduction of the Warsaw basis, the authors of Refs.~\cite{Jenkins:2013zja,Jenkins:2013wua,Alonso:2013hga,Alonso:2014zka} carried out a detailed analysis of the \glspl{rge} of these operators, including their mutual mixing. In total, there are 63 independent operators (for one generation) at dimension six, including four that violate baryon and lepton numbers. This gives rise to $63^2=3969$ possible entries in the \gls{adm} (excluding flavour indices), making the full computation of all loop diagrams a substantial task. As a result, it is advisable to first analyse the operator mixing class by class before focusing on individual operators.

Using \gls{nda}, the authors of Ref.~\cite{Jenkins:2013sda} were able to identify which entries in the \gls{adm} could be non-vanishing. However, explicit one-loop computations revealed more vanishing entries than \gls{nda} had predicted. These additional zeros arose either due to the absence of contributing Feynman diagrams or the finiteness of those diagrams. In several cases, cancellations occurred between diagrammatic divergences and counterterms associated with the \gls{eom}. Such cancellations are more naturally understood using on-shell methods, we will also encounter them when discussing dimension-eight operators.

The complete set of \glspl{rge} for the dimension-six operators is distributed across four major references~\cite{Jenkins:2013zja,Jenkins:2013wua,Alonso:2013hga,Alonso:2014zka}, organized by the type of contribution: (1) those involving the Higgs quartic coupling $\lambda$, (2) Yukawa couplings, (3) gauge couplings, and (4) operators that violate baryon and lepton number.

We present below a summary of the non-vanishing anomalous dimensions based on the explicit one-loop results, supplemented by the \gls{nda}-based expectations (see Table~3 in Ref.~\cite{Alonso:2013hga}). In particular, some entries deviate from the \gls{nda}-expected magnitude $\gamma\sim~\mathcal{O}(1)$. This behavior is especially prominent in the \gls{rge} for the six-Higgs operator $\mathcal{O}_H=(H^\dagger H)^3$, which includes terms such as~\cite{Alonso:2013hga}:
\begin{equation}
16\pi^2\frac{{\mathrm{d}}}{{\mathrm{d\mu}}}c_{6;H}=108\lambda c_{6;H} -(48g_1^4 +12g_1^2g_2^2)c_{6;HB} + \dots
\end{equation}
While this effect does not appear at dimension five, it becomes more significant as we move to higher-dimensional operators.

	\begin{table}[h]
		\begin{center}
			\resizebox{\textwidth}{!}{
				\begin{tabular}{|c||cccccccc|}
				\hline
					$\gamma_{mn}\left(1/\Lambda^2\right)$ & $X^3$ & $H^6$ & $H^4D^2$ & $X^2H^2$ & $\psi^2H^3$ & $X\psi^2H$ & $\psi^2H^2D$ &$\psi^4$ \Tstrut\cr
					\hline
					\hline
					$X^3$ & $g^2$ & 0 & 0 & $\slashed{0}$ & 0 & 0 & 0 & 0  \Tstrut\cr
					$H^6$ & $\slashed{0}$ & $y^2$ & $\lambda g^2$ & $g^4$ & $y^3$ & 0 & $ \lambda y^2$ & 0\cr
					$H^4D^2$ & $\slashed{0}$ & 0 & $y^2$ & $\slashed{0}$ & $\slashed{0}$ & 0 & $y^2$ & 0  \cr 
					$X^2H^2$ & $\slashed{0}$ & 0 & 0 & $y^2$ & 0 & $yg$ & 0 & 0 \cr
					$\psi^2H^3$ & $\slashed{0}$ & 0 & $y^3$ & $\slashed{0}$ & $y^2$ & $y^2 g$ & $y^3$ & $y^3$ \cr
					$X\psi^2H$ & $\slashed{0}$ & 0 & 0 & $yg$ & $\slashed{0}$ & $y^2$ & $\slashed{0}$ & $yg$ \cr
					$\psi^2H^2D$ & $\slashed{0}$ & 0 & $y^2$ & $\slashed{0}$ & $\slashed{0}$ & $\slashed{0}$ & $y^2$ & $y^2$ \cr
					$\psi^4$ & $\slashed{0}$ & 0 & 0 & 0 & 0 & $\slashed{0}$ & $y^2$ & $y^2$ \cr
					\hline
			\end{tabular}}
		\end{center}
		\caption{\label{tab:ADMd6}Anomalous dimension matrix for dimension-six operators. Rows correspond to operator classes receiving corrections; columns indicate source operator classes. $\slashed{0}$ indicates a vanishing entry where \gls{nda} suggests a non-zero contribution. See~\cite{Jenkins:2013zja,Jenkins:2013wua,Alonso:2013hga,Alonso:2014zka} for complete results.}
	\end{table}


In addition to the mixing of dimension-six operators among themselves, there exists another contribution at \order{2}: the insertion of two dimension-five operators into one-loop diagrams. These contributions can also renormalise dimension-six operators. This effect was computed in Ref.~\cite{Davidson:2018zuo} for the \gls{sm} and a two-Higgs-doublet model\footnote{Note that the definitions of the Weinberg operator and its Hermitian conjugate in Ref.~\cite{Davidson:2018zuo} differ from the conventions used here.}. 
The resulting additional terms in the \glspl{rge} of dimension-six operators are:\pagebreak
\begin{align}\label{eq:d5d5tod6}
16\pi^2\frac{{\mathrm{d}}}{{\mathrm{d\mu}}}[c_{6;H\ell}^{(1)}]_{\alpha\beta}&= -\frac{3}{2}\left([c_{5;\ell H}^{\dagger}][c_{5;\ell H}]\right)_{\alpha\beta},\\
16\pi^2\frac{{\mathrm{d}}}{{\mathrm{d\mu}}}[c_{6;H\ell}^{(3)}]_{\alpha\beta}&= \left([c_{5;\ell H}^{\dagger}][c_{5;\ell H}]\right)_{\alpha\beta},\\
16\pi^2\frac{{\mathrm{d}}}{{\mathrm{d\mu}}}[c_{6;eH}]_{\alpha\beta}&= \frac{3}{2}\left([c_{5;\ell H}^{\dagger}][c_{5;\ell H}][y^e]\right)_{\alpha\beta},\\
16\pi^2\frac{{\mathrm{d}}}{{\mathrm{d\mu}}}[c_{6;\ell\ell}]_{\alpha\beta\gamma\rho}&= -\frac{1}{2}[c_{5;\ell H}^{\dagger}]_{\alpha\gamma}[c_{5;\ell H}]_{\beta\sigma}.
\end{align}
In the above, we have made use of the symmetry of the Weinberg operator, 
\begin{equation}
[c_{5;\ell H}^{(\dagger)}]_{\alpha\beta}=[c_{5;\ell H}^{(\dagger)}]_{\beta\alpha},
\end{equation} 
to simplify expressions.

\section{Renormalisation up to third order in the cutoff}

In this case, the anomalous dimension governing the scale dependence of dimension-seven operators takes the form:
\begin{equation}
16\pi^2\mu\frac{{\rm{d}}}{{\rm{d}}\mu}c_{7;p}=\gamma_{ps}^{(d7)}c_{7;s}+\gamma_{p}^{(d5^3)}c_5^3+ \gamma_{ps}^{(d5d6)}c_5c_{6;s}\,.
\end{equation}

The primary interest in dimension-seven operators arises from their contributions to the neutrino mass matrix, similar to the dimension-five Weinberg operator. However, at dimension seven, the number of operator classes increases, and not all contribute to neutrino masses at tree level. Consequently, the \glspl{rge} for these operators were completed more recently.

The first results concerning their self-mixing were published in Ref.~\cite{Liao:2019tep}, focusing on \gls{lnv} and \gls{bnv} sectors. Before that, Ref.~\cite{Liao:2017amb} had analysed potentially vanishing entries in the \gls{adm} using techniques inspired by the non-renormalisation theorems of Ref.~\cite{Cheung:2015aba} and \gls{nda}~\cite{Jenkins:2013sda}. This approach allowed them to identify and discard many diagrams expected to yield zero contributions, thus streamlining the full computation. However, these methods only apply to single insertions of irrelevant operators and sometimes predict non-zero mixing where no one-loop diagram exists (e.g., mixing of $\psi^2H^4$ into other classes).

We summarize the structure of the \gls{adm}, grouped by operator class, in Table~\ref{tab:ADMd7}.

	\begin{table}[h]
		\begin{center}
			\resizebox{\textwidth}{!}{
				\begin{tabular}{|c||cccccccc|}
				\hline
				\multirow{2}*{$\gamma_{mn}\left(1/\Lambda^3\right)$}&&&&&&& $(L=1)$ & $(L=1)$\Tstrut\cr
					 & $\psi^2H^4$ & $\psi^2H^3D$ & $\psi^2H^2D^2$ & $X\psi^2H^2$ & $\psi^4H$ & $\psi^4D$ & $\psi^4H$ & $\psi^4D$\cr
					\hline\hline
					$\psi^2H^4$ & $y^2$ & $y^3$ & $y^4$ & $y^2g^2$ & $y^3$ & 0 &0&0\Tstrut\cr
					$\psi^2H^3D$ & 0 & $y^2$ & $y^3$ & $\slashed{0}$ & $y^2$ & 0 &0&0\cr
					$\psi^2H^2D^2$ & 0 & 0 & $y^2$ & 0 & 0 & $y^2$ &0&0\cr 
					$X\psi^2H^2$ & 0 & $y$ & $y^2$ & $y^2$ & $y$ & 0 &0&0\cr
					$\psi^4H$ & 0 & 0 & $y^3$ & $yg^3$ & $y^2$ & $y^3$ &0&0\cr
					$\psi^4D$ & 0 & 0 & $y^2$ & 0 & 0 & $y^2$ &0&0\cr
					$(L=1)\;\psi^4H\hphantom{(L=1)}$ & 0 & 0 & 0 & 0 & 0 & 0 & $y^2$ & $y^3$ \cr
					$(L=1)\;\psi^4D\hphantom{(L=1)}$ & 0 & 0 & 0 & 0 & 0 & 0 & 0 & $y^2$ \cr
					\hline
					$\psi^2H^2$ & $m_H^2$ & $y m_H^2$ & $y^2 m_H^2$ & $0$ & $y m_H^2$ & 0 &0&0\Tstrut\cr
					\hline
			\end{tabular}}
		\end{center}
		\caption{\label{tab:ADMd7} Anomalous dimension matrix of dimension seven operators. All operators have lepton number $L=2$ except the last two columns and rows. The columns represent the greatest terms from each contribution mixing into the operators in each row. $\slashed{0}$ represents a vanishing entry where the Equations of Motion cancel the off-shell divergence. See  \cite{Liao:2019tep} for complete RGEs.}
		
\end{table}

In addition to pure dimension-seven contributions, insertions of lower-dimension operators can also generate effects at \order{3}. Contributions relevant to neutrino mass corrections were calculated in~\cite{Chala:2021juk}, while a comprehensive treatment of double and triple insertions involving irrelevant operators was later provided in~\cite{Zhang:2023kvw}. These results were obtained using an off-shell approach with the dimension-seven Green’s basis. Table~\ref{tab:d5d6tod7} summarizes the dominant mixing effects of lower-dimensional operators into dimension-seven operators.

	\begin{table}[h]
		\begin{center}
			\resizebox{\textwidth}{!}{
				\begin{tabular}{|c||c|cccccccc|}
				\hline
					$\gamma_{55m}\left(1/\Lambda^3\right)$ & $\psi^2H^2$ & $H^6$ & $H^4D^2$ & $X^2H^2$ & $\psi^2 H^3$ & $\psi^2 H^2 D$ & $X \psi^2 H$ & $\psi^4$ & $X^3$\Tstrut\cr
					\hline
					\hline
					$\psi^2H^4$ & $1$ & $1$ & $g^2$ & $g^2$ & $g^2$ & $y^2$ & $gy$ &0&0\Tstrut\cr
					$\psi^2H^3D$ & 0 & 0 & $y$ & 0 & $1$ & $1$ & $g$ &0&0\cr
					$\psi^2H^2D^2$ & 0 & 0 & $1$ & 0 & $1$ & 0 & 0 & $1$ &0\cr 
					$X\psi^2H^2$ & 0 & 0 & 0 & $g$ & $g$ & 0 & $y$ & 0 & $g^2$\cr
					$\psi^4H$ & 0 & 0 & $y$ & 0 & $y$ & $1$ & $g$ & $y$ &0\cr
					$\psi^4D$ & 0 & 0 & 0 & 0 & 0 & 0 & 0 &0&0\cr
					$(L=1)\;\psi^4H\hphantom{(L=1)}$ & 0 & 0 & 0 & 0 & 0 & 0 & 0 & $y$ & 0 \cr
					$(L=1)\;\psi^4D\hphantom{(L=1)}$ & 0 & 0 & 0 & 0 & 0 & 0 & 0 &0&0\cr
					\hline
					$\psi^2H^2$ & 0 & $m_H^2$ & 0 & 0 & 0 & 0 & 0 &0&0\Tstrut\cr
					\hline
			\end{tabular}}
		\end{center}
		\caption{\label{tab:d5d6tod7} Anomalous dimension matrix. All dimension seven operators have lepton number $L=2$ except the last two rows. The columns represent the greatest terms from each contribution mixing into the operators in each row. See  \cite{Zhang:2023kvw} for complete RGEs.}
		
\end{table}		

Furthermore, the authors of Ref.~\cite{Zhang:2023kvw} required the second-order \glspl{rge} of dimen\-sion-six operators, since the on-shell reduction of the Green’s basis at \order{3} involved redefining physical dimension-seven operators using contributions from redundant dimension-six operators from classes $H^4D^2$, $XH^2D^2$, $\psi^2H^2D$, $\psi^2HD^2$, $H^2D^4$ and $X^2D^2$. These redundant operators become physical when inserted alongside a Weinberg operator. Interestingly, triple insertions of the Weinberg operator do not contribute directly to the dimension-seven operator \glspl{rge}. However, they induce shifts via on-shell relations involving $H^4D^2$ redundant operators, which in turn modify the \gls{wc} of the dimension-seven Weinberg-like operator. Among all lower-dimensional operators, only the Weinberg operator’s \gls{rge} is modified at \order{3}, as no valid diagrams can be drawn for other classes at this order.

This interplay illustrates how \glspl{rge} at lower dimensions are essential for the consistent renormalisation of higher-dimensional operators. We encountered a similar effect in the renormalisation of dimension-eight operators, which will be discussed in the subsequent sections.

\section{Renormalisation up to fourth order in the cutoff\footnote{This section contains original work from the thesis.}}\label{sec:RGEResults}

We adopt an off-shell diagrammatic approach, using the Green’s basis defined in the Appendices, supplemented by the necessary on-shell reduction identities. The corresponding \texttt{FeynRules}~\cite{Alloul:2013bka} model is publicly available online in \href{https://github.com/SMEFT-Dimension8-RGEs/Notebooks}{GitHub}, alongside intermediate steps such as the on-shell relations at order $\Lambda^{-4}$ and the divergences for the redundant Lagrangian. Due to the vast number of resulting equations—many of which contain long expressions—we limit the discussion to specific illustrative examples of interest. Although not explicitly discussed in this section, \gls{wfr} is systematically included in all self-renormalisation computations. Its contributions are essential for maintaining consistency across field redefinitions and operator mixing.

At present, there remain additional contributions to the \gls{adm} that have not yet been computed. In particular, while the renormalisation of all bosonic operators is complete, the renormalisation of two-fermion operators is only partially known, and the \glspl{rge} for four-fermion operators are only available assuming minimal flavour violation~\cite{Boughezal:2024zqa}.

The results presented here are original and were entirely computed by members of the Granada \gls{ftae} group. All results are available in a public GitHub repository, including the \glspl{rge} computed to date, most of the divergences and on-shell redundancies, and the implementation of the Green’s basis in a \texttt{FeynRules} model. These findings have also been published in four peer-reviewed articles~\cite{Chala:2021pll,DasBakshi:2022mwk,DasBakshi:2023htx,Bakshi:2024wzz} --note the last three are co-authored by the author of this thesis.

Considering the structure of the Green’s basis and the on-shell relations, it is noteworthy that fermionic operators do not redefine physical bosonic operators. In other words, the \glspl{wc} shifts for physical bosonic operators contain only redundant bosonic operators and no fermionic ones. This result follows directly from the \glspl{eom}~\eqref{eq:EoMforSMEFT}: as shown in Section~\ref{sec:EoM}, while the bosonic \glspl{eom} (e.g., for $H$, $B$, $W$, $G$) may include fermionic terms, the fermionic \glspl{eom} do not include purely bosonic contributions. This asymmetry stems from the conservation of fermion number in the \glspl{eom} and holds to all orders in $\Lambda$. We leverage this property to compute the mixing of bosonic and fermionic operators into bosonic operators first, simplifying the overall analysis. Furthermore, since the number of bosonic operators is significantly lower (as can be seen in Table~\ref{tab:CountingSMEFT}), they represent a smooth, introductory approach to the calculations, beginning with simpler cases involving bosonic operators to validate the method with less complex algebraic structures.

For simplicity, we exclude the insertion of operators that arise only at the loop level in weakly coupled theories~\cite{Craig:2019wmo}, as they correspond to formally two-loop contributions. However, some authors have considered these operators at lower power-counting orders (e.g., dimension-six and dimension-seven loop-generated operators). In the context of weakly coupled \gls{uv} completions of the \gls{smeft}, such operators do not contribute at tree level and are therefore omitted here. Nonetheless, dimension-eight loop-generated operators are retained in the Green’s basis, and their \glspl{rge} indicate mixing with tree-level-generated operators. The mixing of tree-level generated operators into loop-level operators is an effect already observed at dimension six\footnote{Recall Table~\ref{tab:ADMd6} shows the mixing of class $\psi^4$ into $X\psi^2H$.}, which cannot be fully addressed without explicit computation. 

Regarding the contributions at power-counting \order{4},  we identify three main categories based on the operator insertions in the loops:
\begin{itemize}
\item The insertion of one dimension-eight operator.
\item The insertion of two dimension-six operators.
\item The insertion of more than one LNV operator.
\end{itemize}

These computations can be separated for clarity, although all \gls{lnv} operator insertions are handled together to streamline the analysis. The contributions to the anomalous dimensions can be expressed as:
\begin{equation}
16\mu\frac{{\rm{d}}}{{\rm{d}}\mu}c_{8;p}=\gamma_{ps}^{(d8)}c_{8;s}+\gamma_{p}^{(d5^4)}c_5^4+ \gamma_{ps}^{(d5d7)}c_5c_{7;s}+  \gamma_{ps}^{(d5^2d6)}c_5c_5c_{6;s} +\gamma_{pst}^{(d6^2)}c_{6;s}c_{6;t}\,,
\end{equation}
where $c_{r;p}$ denotes the \glspl{wc} associated with dimension-$r$ operators, and the indices $p$ $s$, $t$, label operator structures. The various $\gamma$ coefficients represent the different contributions to the anomalous dimensions arising from operator mixing at this order.

\subsection{Insertion of dimension-eight operators}

We begin by addressing the renormalisation group mixing among dimension-eight operators. In~\cite{DasBakshi:2022mwk}, we presented for the first time a comprehensive list of \glspl{rge} for bosonic operators. The complete expressions are available in a Mathematica notebook hosted online at~\href{https://github.com/SMEFT-Dimension8-RGEs}{GitHub}. These were provided not only to facilitate the running of observables but also to enable cross-comparisons with results from other groups. Although prior results were scarce~\cite{AccettulliHuber:2021uoa}, the available cross-checks have been positive. Subsequently, another study employing a geometric approach to \gls{smeft} renormalisation~\cite{Helset:2022pde} confirmed the agreement with our findings in the overlapping results.

It is logical to start the computation with the mixing of bosonic and fermionic operators into bosonic ones, given their phenomenological relevance \cite{Maltoni:2024dpn,Grojean:2024tcw,Dawson:2022ewj,Asteriadis:2022ras,Durieux:2022hbu,ValeSilva:2022tph,Ardu:2025rqy,Adhikary:2025gdh}. Nonetheless, our decision to begin with bosonic dimension-eight operators is primarily pragmatic:
\begin{itemize}
\item Bosonic operators constitute a much smaller subset compared to fermionic ones, implying a reduced computational workload.
\item Avoiding fermionic external legs significantly limits the kinematic invariants, simplifying amplitude matching and accelerating the extraction of divergences.
\item The Green's basis for dimension-eight bosonic operators was established in \cite{Chala:2021cgt}, including all necessary on-shell relations.
\item The on-shell relations for bosonic operators do not involve fermionic operators. Therefore, the divergences of fermionic operators are not required to derive the RGEs of bosonic operators—though the reverse is not true.
\end{itemize}

Table \ref{tab:ADMd8tod8} summarises the leading contributions to the \gls{adm}. Despite their smaller number, bosonic operators already outnumber the complete dimension-six basis.

Interestingly, we observe operators induced at loop-level in weakly coupled theories being renormalised by tree-level-generated interactions. This effect, which was previously observed only in a single fermionic case at order \order{2} (specifically, the class $X\psi^2H$ renormalised by tree-level-generated $\psi^4$), now also appears among bosonic operators. Although such mixing is allowed in principle, it was scarcely studied, marking a key insight from our computation.

Many of the zero entries in the \gls{adm} are understood via non-renormalisation theorems or the absence of contributing diagrams. Some of the more subtle vanishing contributions—termed non-trivial zeros—result from cancellations involving divergences and on-shell relations. For example, the divergences of $\mathcal{O}_{8;W^2BH^2}^{(1)}$ cancel with the on-shell contribution from $\mathcal{O}_{8;WBH^2D^2}^{(13)}$\footnote{These cancellations are basis-dependent and may not appear in alternative operator bases.}. Conversely, some non-zero entries result purely from redundant operator mixing, such as the contribution of class $H^4D^4$ into $H^8$.

As in the dimension-six case, we observe large anomalous dimensions in several \glspl{rge}. Operators containing six or eight Higgs fields typically have the largest coefficients. For instance:
\begin{align}
 16\pi^2\mu \frac{\diff}{\diff \mu}c_{8;H^8} &= \frac{184}{3}\lambda^3 c_{ H^4}^{(2)} -12 g_1\lambda^2 c_{B H^4 D^2}^{(1)}-16g_2\lambda^2 c_{W H^4 D^2}^{(1)}+12 g_1^2\lambda c_{B^2 H^4}^{(1)} \nonumber \\
 &+ 36 g_2^2 \lambda c_{W^2 H^4}^{(1)} + 12 g_1 g_2\lambda c_{WB H^4}^{(1)}+48 \lambda^2 c_{ H^6 D^2}^{(2)}+192\lambda c_{ H^8} \nonumber \\
 & +24\lambda \text{Tr}\left[(c^{(1)}_{u^2  H^2 D^3} +c^{(2)}_{u^2  H^2 D^3} ) (y^u)^* (y^u) (y^u)^*(y^u) \right] \nonumber \\
 & +24\lambda \text{Tr}\left[(c^{(1)}_{q^2  H^2 D^3} +c^{(2)}_{q^2  H^2 D^3} -c^{(3)}_{q^2  H^2 D^3} -c^{(4)}_{q^2  H^2 D^3} ) (y^u) (y^u)^* (y^u) (y^u)^*\right].
\end{align}

Operators with four Higgs fields can also feature large anomalous dimensions, as seen in:
\begin{align}
 16\pi^2\mu \frac{\diff}{\diff \mu}c_{8;B H^4D^2}^{(1)} &= 12 \lambda  c_{B H^4D^2}^{(1)} + 60 g_1 \text{Tr}\left[(c^{(4)}_{q^2  H^2 D^3}) (y^u) (y^u)^*\right]\nonumber\\
 & -36 \text{Tr}\left[(c^{(1)}_{q^2 B  H^2 D}) (y^u) (y^u)^*\right] + \dots\\
 16\pi^2\mu \frac{\diff}{\diff \mu} c_{8;W H^4D^2}^{(1)} &= 44 g_2 \text{Tr}\left[(c^{(4)}_{q^2  H^2 D^3}) (y^u) (y^u)^*\right] + 48 \text{Tr}\left[(c^{(11)}_{q^2 W H^2 D}) (y^u)(y^u)^*\right] + \dots
\end{align}%
and there are more of them.

	\begin{table}[h]
		\begin{center}
			\resizebox{\textwidth}{!}{
				\begin{tabular}{|c||ccccccccccc|}
				\hline
					$\gamma_{mn}\left(1/\Lambda^4\right)$ & $H^8$ & $H^6D^2$ & $H^4D^4$ & $X^2H^4$ & $XH^4D^2$ & $X\psi^2H^3$ & $\psi^2H^2D^3$ & $\psi^2H^5$ & $\psi^2H^4D$ & $X\psi^2H^2D$ & $\psi^2H^3D^2$ \Tstrut\cr
					\hline
					\hline
					$H^8$ & $\lambda$ & $\lambda^2$ & $\lambda^3$ & $g^2\lambda$ & $g\lambda^2$ & 0 & $y^4\lambda$ & $y^3$ & $y^2\lambda$ & $y^2g\lambda$ & $y^3\lambda$ \Tstrut\cr
					$H^6D^2$ & 0 & $\lambda$ & $g^4$ & 0 & $g \lambda$ & 0 & $y^2g^2$ & 0 & $y^2$ & $y^2 g$ & $y^3$ \cr
					$H^4D^4$ & 0 & 0 & $g^2$ & 0 & 0 & 0 & $y^2$ & 0 & 0 & 0 & 0 \cr 
					$X^3H^2$ & 0 & 0 & 0 & 0 & $\slashed{0}$ & 0 & $\slashed{0}$ & 0 & 0 & $\slashed{0}$ & 0\cr
					$X^2H^4$ & 0 & 0 & $g^4$ & $g^2$ & $g^3$ & $yg$ & $y^2g^2$ & 0 & 0 & $y^2g$ & $yg^2$ \cr
					$X^2H^2D^2$ & 0 & 0 & $g^2$ & 0 & 0 & 0 & $g^2$ & 0 & 0 & 0 & 0\cr
					$XH^4D^2$ & 0 & 0 & $g^2$ & 0 & $g^2$ & 0 & $y^2 g$ & 0 & 0 & $y^2$ & $yg$\cr \hline
			\end{tabular}}
		\end{center}
		\caption{\label{tab:ADMd8tod8}Anomalous dimension matrix for dimension-eight bosonic operators. Entries indicate the leading terms in the mixing. $\slashed{0}$ represents entries vanishing due to \glspl{eom} canceling the off-shell divergences. See \cite{DasBakshi:2022mwk} for full expressions.}
	\end{table}

\subsubsection{Example: RGE of $H^4D^4$}
We now present a detailed example of how \glspl{rge} are computed at dimension eight, focusing on a representative case: the $H^4D^4$ operators. This class provides an ideal illustration due to its comparatively simple structure.

The definitions of all operators mentioned here can be found in Appendix~\ref{app:TableOfOperators}. In particular, the relevant physical operators in the $H^4D^4$ class are:
\begin{align}
c_{8;H^4D^4}^{(1)}&=(D_{\mu}  H^{\dag} D_{\nu}  H) (D^{\nu}  H^{\dag} D^{\mu}  H),\\
c_{8;H^4D^4}^{(2)}&=(D_{\mu}  H^{\dag} D_{\nu}  H) (D^{\mu}  H^{\dag} D^{\nu}  H),\\
c_{8;H^4D^4}^{(3)}&=(D^{\mu}  H^{\dag} D_{\mu}  H) (D^{\nu}  H^{\dag} D_{\nu}  H).
\end{align}

The first step is to consider the operators that contribute on-shell to this class. Using the on-shell relations presented in Ref.~\cite{Chala:2021cgt}, we find:
\begin{align}\label{eq:H4D4Onshell}
c_{8;H^4D^4}^{(1)}&=a_{8;H^4D^4}^{(1)}+g_1^2 b_{8;B^2D^4}-g_1 b_{8;BH^2D^4}^{(3)}-g_2^2 b_{8;W^2D^4}+g_2 b_{8;WH^2D^4}^{(3)},\\
c_{8;H^4D^4}^{(2)}&=a_{8;H^4D^4}^{(2)}-g_1^2 b_{8;B^2D^4}+g_1 b_{8;BH^2D^4}^{(3)}-g_2^2 b_{8;W^2D^4}+g_2 b_{8;WH^2D^4}^{(3)},\\
c_{8;H^4D^4}^{(3)}&=a_{8;H^4D^4}^{(3)}+g_2^2 b_{8;W^2D^4}-g_2 b_{8;WH^2D^4}^{(3)}.
\end{align}
These expressions show that, in addition to the $H^4D^4$ operators, we must compute the divergences of the $XH^2D^4$ and $X^2D^4$ operator classes. Notably, the $X^2D^4$ operators do not contain Higgs fields, and hence cannot be renormalised by tree-level-generated operators. This is because all such tree-level classes involve at least four Higgs fields or two Higgs fields and two fermions; on the other side, loop contractions can involve at most two field insertions from one operator, implying that the resulting diagrams will always contain at least two Higgs fields or two fermion fields.

For simplicity, we also invoke non-renormalisation theorems (see, e.g., Ref.~\cite{Cheung:2015aba}), which in this case imply that $H^4D^4$ operators can only be renormalised on-shell by tree-level-generated classes including $H^4D^4$, $\psi^2H^2D^3$ and $\psi^4$. However, at one-loop level, $\psi^4$ does not contribute to purely bosonic operators, so we only need to consider insertions of $H^4D^4$ and $\psi^2H^2D^3$ operators.

\begin{figure}
\begin{center}
\includegraphics[scale=0.13]{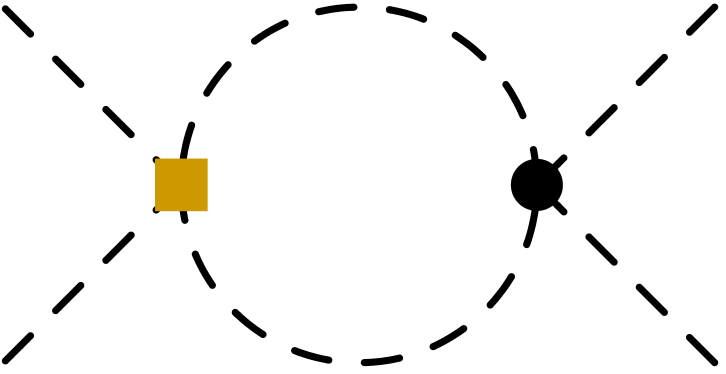}\qquad
\includegraphics[scale=0.13]{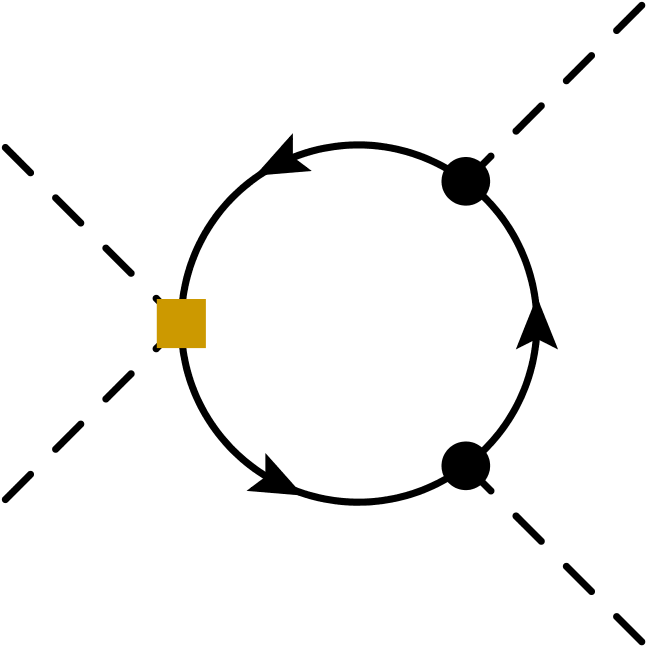}\qquad
\includegraphics[scale=0.13]{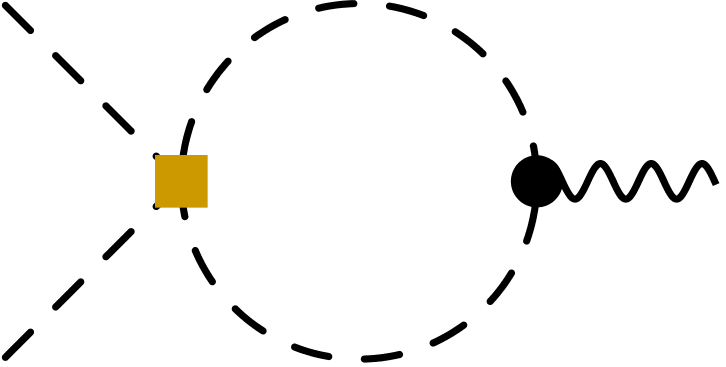}\qquad
\includegraphics[scale=0.13]{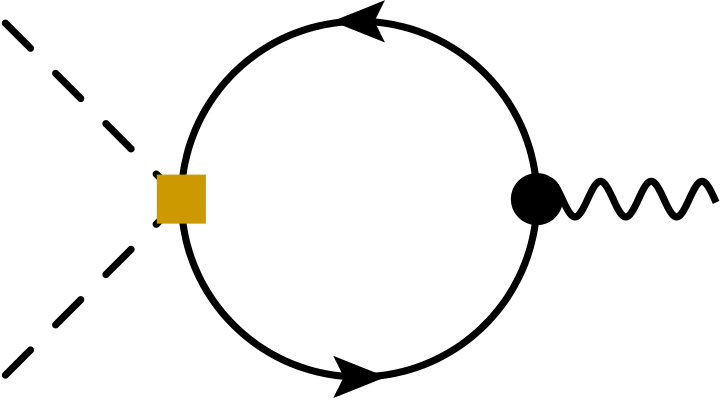}
\end{center}
\caption{\label{fig:Exampled8byd8}One-loop contributions to the off-shell amplitude for $H^4D^4$. Black dots represent SM interactions, and orange boxes denote dimension-eight operators. The left two diagrams represent direct renormalisation via $H^4D^4$ and $\psi^2H^2D^3$, while the right two contribute indirectly through on-shell relations.}
\end{figure}

In total, 19 operators need to be inserted into loop diagrams. These diagrams are similar to those shown in Figure~\ref{fig:Exampled8byd8}, and the resulting expressions are highly cumbersome, requiring symbolic computation tools for tractability. For the process $H^0H^0 \rightarrow H^+H^-$, we illustrate here only the one-loop contribution from the insertion of $\mathcal{O}_{8;H^4D^4}^{(1)}$.

The ultraviolet divergence from $\mathcal{O}_{8;H^4D^4}^{(1)}$ is:
\begin{align}
\ii\mathcal{A}^{\text{1L}}_\text{UV}&=\frac{\ii c_{8;H^4D^4}^{(1)}}{192\pi^2 \epsilon}\left(
3 (5 g_1^2-g_2^2-4 \lambda ) \kappa_{2223}
+3 (-5 g_1^2+g_2^2+4 \lambda ) \kappa_{2224}\right.\cr
&-3 (g_1^2-15 g_2^2+12 \lambda ) \kappa_{3444} 
-15 g_1^2 \kappa_{2233}
-23 g_1^2 \kappa_{2234}
-12 g_1^2 \kappa_{2323}
+26 g_1^2 \kappa_{2324}\cr
&+12 g_1^2 \kappa_{2333}
-5 g_1^2 \kappa_{2344}
+18 g_1^2 \kappa_{2424} 
-6 g_1^2 \kappa_{2433}
+14 g_1^2 \kappa_{2434}
-3 g_1^2 \kappa_{2444}\cr
&+12 g_1^2 \kappa_{3334}
-3 g_1^2 \kappa_{3344}
+12 g_1^2 \kappa_{3434}
-17 g_2^2 \kappa_{2233} 
-57 g_2^2 \kappa_{2234}
-20 g_2^2 \kappa_{2244}\cr
&-22 g_2^2 \kappa_{2323}
+138 g_2^2 \kappa_{2324}
+42 g_2^2 \kappa_{2333}
+48 g_2^2 \kappa_{2334}
-17 g_2^2 \kappa_{2344}
-28 g_2^2 \kappa_{2424}\cr
&-84 g_2^2 \kappa_{2433}
-22 g_2^2 \kappa_{2434}
+45 g_2^2 \kappa_{2444}
+42 g_2^2 \kappa_{3334}
+25 g_2^2 \kappa_{3344}
+62 g_2^2 \kappa_{3434}\cr
&+52 \lambda  \kappa_{2233}
+108 \lambda  \kappa_{2234}
+40 \lambda  \kappa_{2244}
+8 \lambda  \kappa_{2323}
+120 \lambda  \kappa_{2324}
-48 \lambda  \kappa_{2333}\cr
&-144 \lambda  \kappa_{2334}
-92 \lambda  \kappa_{2344}
-16 \lambda  \kappa_{2424}
-120 \lambda  \kappa_{2433}
-184 \lambda  \kappa_{2434}
-36 \lambda  \kappa_{2444}\cr
&-48 \lambda  \kappa_{3334}
\left. +4 \lambda  \kappa_{3344}
-88 \lambda  \kappa_{3434}\right)+\dots
\end{align}
where over 500 additional terms from insertions of other irrelevant interactions are omitted for brevity. While the following expressions are a bit shorter, it is very common to encounter such lengthy amplitudes when computing at \order{4}. It is rather impractical to analyse such bulky contributions as a whole. We focus on $\mathcal{O}_{8;H^4D^4}^{(1)}$, $\mathcal{O}_{8;H^4D^4}^{(2)}$  and $\mathcal{O}_{8;H^4D^4}^{(3)}$ in the remainder of this section to obtain the full self-renormalisation of $H^4D^4$. The rest of the contributions to these \glspl{rge} (or other contributions, for that matter) are obtained analogously.

The divergences of the $XH^2D^4$ operators by $H^4D^4$ are:
\begin{align}
\ii\mathcal{A}^{1L}_\text{UV}(H^0H^0\rightarrow B)&=\frac{\ii g_1 c_{8;H^4D^4}^{(1)}}{96\pi^2\epsilon}\left(\eta_{23333} - \eta_{32333}\right)+\frac{\ii g_1 c_{8;H^4D^4}^{(2)}}{64\pi^2\epsilon}\left(\eta_{32333}-\eta_{23333} \right)\cr
&+\frac{\ii g_1 c_{8;H^4D^4}^{(3)}}{192\pi^2\epsilon}\left(\eta_{23333} - \eta_{32333}\right),\\
\ii\mathcal{A}^{1L}_\text{UV}(H^0H^0\rightarrow W_3)&=\frac{\ii g_2 c_{8;H^4D^4}^{(2)}}{192\pi^2\epsilon}\left( \eta_{23333}-\eta_{32333}\right)\cr
&+\frac{\ii g_2 c_{8;H^4D^4}^{(3)}}{192\pi^2\epsilon}\left( \eta_{32333}-\eta_{23333}\right),
\end{align}
with the following definitions for kinematic invariants: 
\begin{align}\label{eq:KinInvariantsExample}
\kappa_{mnst}=p_m\cdot p_n p_s\cdot p_t,\\
\eta_{lmnst}=\varepsilon_3\cdot p_l p_m\cdot p_n p_s\cdot p_t.
\end{align}

The corresponding tree-level amplitudes needed to absorb the divergences (i.e., the ones entering Eq.~\eqref{eq:H4D4Onshell}) are:
\begin{align}\label{eq:IRamplitudeH4D4}
&\ii\mathcal{A}^{TL}_\text{IR}(H^0H^0 \rightarrow H^+H^-)=2\ii  \left( a_{8;H^4D^4}^{(1)}(-\kappa_{2324} + \kappa_{2433} + \kappa_{2434})\right. \cr
&+ \left.  a_{8;H^4D^4}^{(2)}(-\kappa_{2324}+\kappa_{2334} + \kappa_{2344})\right. + \left. a_{8;H^4D^4}^{(3)}(-\kappa_{2234}+\kappa_{2334}+ \kappa_{2434}) \right),\\
&\ii\mathcal{A}^{TL}_\text{IR}(H^0H^0\rightarrow B)= 2\ii \left(b_{8;BH^2D^4}^{(3)}(\eta_{32223}-\eta_{32323}-\eta_{22233}+\eta_{22333}) \right),\\
&\ii\mathcal{A}^{TL}_\text{IR}(H^0H^0\rightarrow W^3)= -2\ii \left(  b_{8;WH^2D^4}^{(3)}(\eta_{32223}-\eta_{32323}-\eta_{22233}+\eta_{22333}) \right).
\end{align}

Equating the UV and IR amplitudes leads to a system of equations, with one equation per kinematic invariant. We deliberately use a redundant basis of invariants to form an over-constrained system, enabling consistency checks of the computation. Solving the system yields the off-shell divergences:
\begin{align}
\widetilde{a}_{8;H^4D^4}^{(1)}=\frac{1}{96 \pi ^2 \epsilon }&\left( c_{8;H^4D^4}^{(1)}( 3g_1^2- 33 g_2^2 -24 \lambda) + c_{8;H^4D^4}^{(2)} (  2g_1^2-20 g_2^2- 8\lambda)\right. \cr 
+&\left.c_{8;H^4D^4}^{(3)}(  8g_1^2-8 g_2^2-8\lambda) \right),\\
\widetilde{a}_{8;H^4D^4}^{(2)}=\frac{1}{96 \pi ^2 \epsilon }&\left( c_{8;H^4D^4}^{(1)}(-8g_1^2- 14g_2^2-8 \lambda + c_{8;H^4D^4}^{(2)}(-21g_1^2 -39 g_2^2-24\lambda)\right. \cr 
+&\left.c_{8;H^4D^4}^{(3)}(-8 g_1^2-8 g_2^2-8 \lambda) \right),\\
\widetilde{a}_{8;H^4D^4}^{(3)}=\frac{1}{96 \pi ^2 \epsilon }&\left( c_{8;H^4D^4}^{(1)}(8  g_1^2+36  g_2^2-48 \lambda) + c_{8;H^4D^4}^{(2)}( 2 g_1^2+28 g_2^2-32 \lambda)\right. \cr 
+&\left.c_{8;H^4D^4}^{(3)}(3 g_1^2+25 g_2^2-80\lambda )\right),\\
\widetilde{b}_{8;BH^2D^4}^{(3)}=\frac{g_1}{192 \pi ^2 \epsilon }&(2c_{8;H^4D^4}^{(1)}-3 c_{8;H^4D^4}^{(2)}+ c_{8;H^4D^4}^{(3)} ),\\
\widetilde{b}_{8;WH^2D^4}^{(3)}=\frac{g_1}{192 \pi ^2 \epsilon }&(- c_{8;H^4D^4}^{(2)}+ c_{8;H^4D^4}^{(3)} ).
\end{align}

Substituting these into Eq.~\eqref{eq:H4D4Onshell}, we obtain the on-shell divergences, which contribute to the \glspl{rge}~\eqref{eq:RGEdefinition}. Including the tree-level anomalous dimensions $n_{H^4D^4}~=~2$, $n_{\psi^2H^2D^3} = 2$, and the \gls{wfr} term $\propto \delta_H c_{8;H^4D^4}^{(p)}$, the final \glspl{rge} are:
\begin{align}\label{sec:RGEexample1}
16\pi^2\mu \frac{\diff}{\diff \mu} c_{8;H^4D^4}^{(1)}&= c_{8;H^4D^4}^{(1)} \left(-\frac{8 }{3}g_1^2+5 g_2^2+8 \lambda\right)+c_{8;H^4D^4}^{(2)}\left(-g_1^2\frac{7 }{6}+\frac{41  }{6}g_2^2+\frac{8   }{3}\lambda\right) \cr
&+ c_{8;H^4D^4}^{(3)} \left(-\frac{5}{2}   g_1^2+\frac{5 }{2}  g_2^2+\frac{8}{3}\lambda \right)+\dots\\
16\pi^2\mu \frac{\diff}{\diff \mu} c_{8;H^4D^4}^{(2)}&=c_{8;H^4D^4}^{(1)} \left(\frac{7 }{3}g_1^2+\frac{14  }{3}g_2^2+\frac{8  }{3}\lambda \right)+c_{8;H^4D^4}^{(2)} \left(\frac{11 }{2}  g_1^2+\frac{43 }{6}g_2^2+8 \lambda\right) \cr
&+ c_{8;H^4D^4}^{(3)}  \left(\frac{5 }{2} g_1^2+\frac{5 }{2}g_2^2+\frac{8 }{3}\lambda \right)+\dots\\
16\pi^2\mu \frac{\diff}{\diff \mu} c_{8;H^4D^4}^{(3)}&=c_{8;H^4D^4}^{(1)} \left(-\frac{8}{3}g_1^2-12 g_2^2+16  \lambda\right) + c_{8;H^4D^4}^{(2)}\left(-\frac{2 }{3}g_1^2-\frac{29  }{3}  g_2^2+\frac{32 }{3}\lambda\right) \cr
&+c_{8;H^4D^4}^{(3)} \left(- 3   g_1^2-14  g_2^2+\frac{80 }{3} \lambda \right)+\dots
\end{align}

As previously mentioned, this is \textit{comparably} one of the simplest cases of renormalisation of dimension eight operators among themselves. It is remarkably lengthy, indeed, but it gets worse for classes that receive many more contributions like $H^6D^2$, which gets insertions from almost all classes both direct and indirect. 

Apart from the pedagogical approach to the computation of \glspl{rge}, this example serves as a sample of the delicate work behind Ref.~\cite{DasBakshi:2022mwk} as well as a reminder of the considerable difficulty posed by the ambitious project of computing the whole \gls{adm} at \order{4}\hspace{-0.5em}.

\subsection{Insertion of two dimension-six operators}

Ref.~\cite{Chala:2021pll} was the first to systematically renormalise dimension-eight interactions, taking into account the mixing of two dimension-six operators into a bosonic dimension-eight operator. After completing the remaining contributions to the \glspl{rge} of bosonic operators, we turned our attention once again to the renormalisation of fermionic operators, beginning with the insertion of two dimension-six operators~\cite{Bakshi:2024wzz}. This required the extension of the Green's Basis with fermionic operators. We used a modified version of \cite{Ren:2022tvi}'s basis, modified to include the physical operators of \cite{Murphy:2020rsh}. We also computed the corresponding onshell relations, needed for the \glspl{rge}. The combined results of the bosonic and fermionic \glspl{rge} are presented in Tables~\ref{tab:ADMd6d6tod8BOS} and \ref{tab:ADMd6d6tod8FER}, respectively. We remind all of the results are available in \href{https://github.com/SMEFT-Dimension8-RGEs}{GitHub}.

Unlike single-operator insertions, there are no general non-renormalisation theorems for multiple insertions\footnote{These were only derived recently~\cite{Liao:2025npz}.}. On the other hand, diagrammatic leg-counting becomes a powerful tool to anticipate possible contributions. As such, many combinations are trivially zero due to the absence of valid Feynman diagrams. The method involves pairing two dimension-six operators (as well as \gls{sm} interactions) and determining whether they can form a divergent Feynman diagram that matches a dimension-eight operator. This analysis must be performed case by case for each pair of insertions.

Most of the zeros in Tables~\ref{tab:ADMd6d6tod8BOS} and \ref{tab:ADMd6d6tod8FER} arise from the absence of valid diagrams. However, we remark on the absence of non-trivial zeros that result from accidental cancellations among off-shell divergences contributing to the same physical operator, when considered on-shell. Note the onshell-relations generate contributions from a broad number of operators, Table~\ref{tab:reductions} shows the onshell contributions of fermionic operators due to redundancies. These contributions amount to nonvanishing terms in the \glspl{rge} of physical operators. In total, 11 classes are renormalised by insertions of pairs of dimension-six operators, excluding the \glspl{rge} of lower-dimensional coefficients.

\begin{table}[h]
\begin{center}
\resizebox{\textwidth}{!}{
 \begin{tabular}{|c||c@{\hspace{0.2em}}c@{\hspace{0.2em}}c@{\hspace{0.2em}}c@{\hspace{0.2em}}|c@{\hspace{0.2em}}c@{\hspace{0.2em}}c@{\hspace{0.2em}}c@{\hspace{0.2em}}c@{\hspace{0.2em}}c@{\hspace{0.2em}}c|}
\cline{2-12}
\multicolumn{1}{c||}{} &  $H^4 D^2$ & $\psi^2  H^2 D$ & $\psi^2 H D^2 $ & $X  H^2 D^2$ &  $ H^4 D^4 $&  $H^6 D^2 $&$ \psi^2  H^2 D^3 $&$ \psi^2  H^3 D^2$ & $\psi^2  H^4 D$ &$ X \psi^2  H^2 D$ & $X  H^4 D^2 $\Tstrut\cr \hline\hline
$\psi^2  H^2 D $&  &  &  &  & & & \checkmark&&&&\Tstrut\cr
$\psi^2 H^3$ & \checkmark & \checkmark & \checkmark &  & \checkmark &  & \checkmark & \checkmark&&&\cr\hline
$\psi^2  H^2 D^3 $ &  &  &  &  & & & \checkmark&&&&\Tstrut\cr
$\psi^2 H^3 D^2 $&  &  & \checkmark &  & \checkmark &  & \checkmark & \checkmark & & &\cr
$\psi^2 H^4 D $& \checkmark & \checkmark & \checkmark & \checkmark &  &  & \checkmark & \checkmark & & \checkmark & \checkmark \cr
$\psi^2 H^5$ & \checkmark & \checkmark & \checkmark & \checkmark & \checkmark & \checkmark & \checkmark & \checkmark & \checkmark & & \checkmark\cr
$X \psi^2  H^2 D $&  &  &  &  & & & \checkmark & & & &\checkmark\cr
$X \psi^2 H^3 $&  &  & &  &  &  & \checkmark & \checkmark & & \checkmark &\cr\hline
\end{tabular}}
\end{center}
 \caption{\label{tab:reductions}Green's functions (columns) that, on-shell, contribute to the renormalisation of the different physical operators (rows) are indicated with $\checkmark$. Dimension-six and -eight interactions are separated by vertical and horizontal lines.}

\end{table}

It is noteworthy that large anomalous dimensions are again observed, despite the loop suppression. As in the previous section, classes of operators involving more than six Higgs fields yield the largest anomalous dimensions, and sizable contributions are also found in the four-Higgs sector. Moreover, the RGEs of lower-dimensional operators at \order{4} also exhibit large anomalous dimensions, including contributions from the fermionic sector. For example:
\begin{align}
16\pi^2\mu \frac{\diff}{\diff \mu} \lambda&=m_H^4(5c_{HD}^2-24c_{HD}c_{H\square}+24c_{H\square}^2)+\dots\\
16\pi^2\mu \frac{\diff}{\diff \mu} [c_{eH}]_{\alpha\beta}&=-48m_H^2[c_{eH}]_{\alpha\beta}c_{H\square}+ 24m_H^2 c_{H\square}[c_{\ell edq}]_{\alpha\beta\gamma\delta}[y^d]_{\delta\gamma})+\dots\\
%
16\pi^2\mu \frac{\diff}{\diff \mu} c_{6;H}&=\frac{m_H^2}{24}(37g_1^2-15g_2^2+840\lambda)c_{HD}^2-48m_H^2\text{Tr}\left[c_{H\ell}(y^e)(y^e)^\dagger\right]c_{H\square}+\dots
\end{align}

We also verify that none of the one-loop generated dimension-eight operators are renormalised by pairs of tree-level generated dimension-six interactions. This is because tree-level-generated vertices involve at least two Higgs fields. Consequently, when inserted in pairs, they necessarily produce operators with four or more Higgs legs, whereas loop-level-generated dimension-eight operators typically contain at most two Higgs fields.

\subsubsection{Example: RGE of $H^4D^4$}
We now complete the example from the previous section by analyzing the contribution of dimension-six operator pairs to the RGE of the $H^4D^4$ operator. In this case, since non-renormalisation theorems are not applicable, we first carry out a preliminary inspection to reduce the number of candidate insertions. In particular, any pair of tree-level dimension-six operators necessarily contributes to interactions with four or more external fields, implying that operators like $X^2D^4$ and $XH^2D^4$ cannot be renormalised via these insertions.

\begin{figure}[h]
\begin{center}
\includegraphics[scale=0.13]{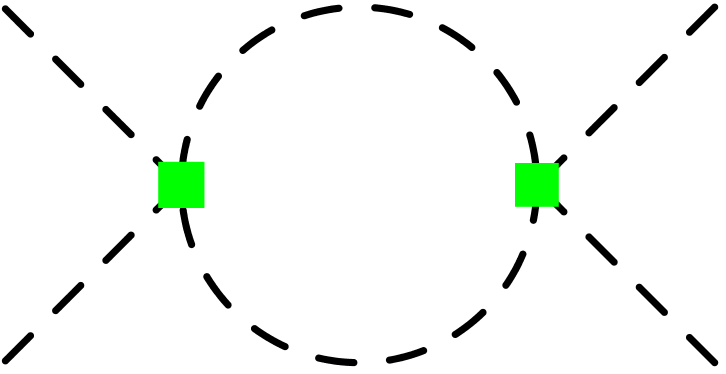}\qquad
\includegraphics[scale=0.13]{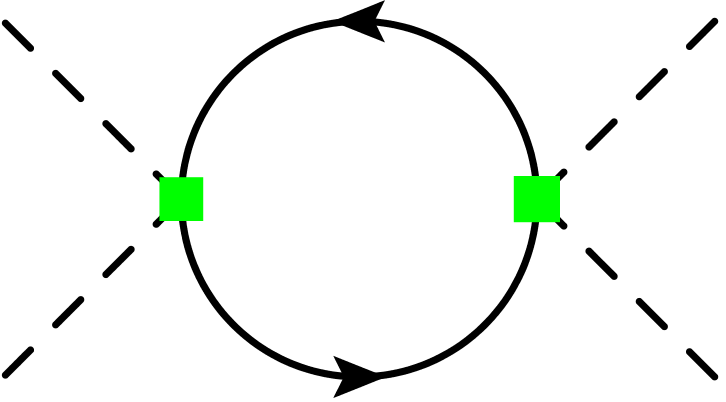}
\end{center}
\caption{\label{fig:Exampled8byd6}One-loop contributions to the $H^4D^4$ amplitude from pairs of dimension-six insertions. Green boxes represent dimension-six operator insertions.}
\end{figure}

Matching external Higgs legs to form valid diagrams with four external Higgs fields reveals that only combinations of two $H^4D^2$ operators or two $\psi^2H^2D$ operators can contribute, as illustrated in Fig.~\ref{fig:Exampled8byd6}.

We compute the one-loop amplitude for the process $H^0 H^0 \rightarrow H^+ H^-$, restricting to insertions of $H^4D^2$ operators for simplicity. The resulting amplitude is:
\begin{align}
\ii\mathcal{A}^\text{1L}&=\frac{\ii}{192 \pi ^2 \epsilon } \left(6  c_{HD}^2 \kappa _{2222}-6  c_{HD} \kappa _{2223} (4 c_{H\square}+ c_{HD})+24  c_{H\square}  c_{HD} \kappa _{2224}\right.
-18  c_{HD}^2 \kappa _{2224}\cr
&+48  c_{H\square}^2 \kappa _{2233}-24  c_{H\square}  c_{HD} \kappa _{2233}+6  c_{HD}^2 \kappa _{2233}
+128  c_{H\square}^2 \kappa _{2234}
 -88  c_{H\square}  c_{HD} \kappa _{2234}\cr
 &+10  c_{HD}^2 \kappa _{2234}+48  c_{H\square}^2 \kappa _{2244}-48  c_{H\square}  c_{HD} \kappa _{2244}
+12  c_{HD}^2 \kappa _{2244}+24  c_{H\square}  c_{HD} \kappa _{2323}\cr
&+12  c_{HD}^2 \kappa _{2323}-128  c_{H\square}^2 \kappa _{2324}+64  c_{H\square}  c_{HD} \kappa _{2324}
-28  c_{HD}^2 \kappa _{2324}-48  c_{H\square}^2 \kappa _{2333}\cr
&+24  c_{H\square}  c_{HD} \kappa _{2333}-18  c_{HD}^2 \kappa _{2333}
-64  c_{H\square}^2 \kappa _{2334}+128  c_{H\square}  c_{HD} \kappa _{2334}-8  c_{HD}^2 \kappa _{2334}\cr
&+16  c_{H\square}^2 \kappa _{2344}+64  c_{H\square}  c_{HD} \kappa _{2344}
+8  c_{HD}^2 \kappa _{2344}-24  c_{H\square}  c_{HD} \kappa _{2424}+24  c_{HD}^2 \kappa _{2424}\cr
&+16  c_{H\square}^2 \kappa _{2433}-56  c_{H\square}  c_{HD} \kappa _{2433}
+14  c_{HD}^2 \kappa _{2433}-64  c_{H\square}^2 \kappa _{2434}-16  c_{H\square}  c_{HD} \kappa _{2434}\cr
&+4  c_{HD}^2 \kappa _{2434}-48  c_{H\square}^2 \kappa _{2444}
+48  c_{H\square}  c_{HD} \kappa _{2444}-24  c_{HD}^2 \kappa _{2444}+96  c_{H\square}^2 \kappa _{3333}\cr
&-24  c_{H\square}  c_{HD} \kappa _{3333}
+3  c_{HD}^2 \kappa _{3333}+336  c_{H\square}^2 \kappa _{3334}-72  c_{H\square}  c_{HD} \kappa _{3334}\cr
&-6  c_{HD}^2 \kappa _{3334}
+192  c_{H\square}^2 \kappa _{3344}-48  c_{H\square}  c_{HD} \kappa _{3344}-6  c_{HD}^2 \kappa _{3344}+288  c_{H\square}^2 \kappa _{3434}\cr
&-24  c_{H\square}  c_{HD} \kappa _{3434}-12  c_{HD}^2 \kappa _{3434}+336  c_{H\square}^2 \kappa _{3444}-48  c_{H\square}  c_{HD} \kappa _{3444}\cr
&\left.-12  c_{HD}^2 \kappa _{3444}+3 \kappa _{4444} \left(32  c_{H\square}^2-8  c_{H\square}  c_{HD}+ c_{HD}^2\right)\right).
\end{align}
The kinematic invariants $\kappa_{ijkl}$ are defined as in the previous section~\eqref{eq:KinInvariantsExample}. The tree-level amplitude in the \gls{ir} limit also matches that of the previous section, allowing us to extract the off-shell divergences. In this case, they coincide with the physical divergences, as there are no \gls{eom} contributions.
\begin{align}
\widetilde{c}_{8;H^4D^4}^{(1)}&=\frac{1}{96 \pi ^2 \epsilon }(16c_{6;H\square}^2-32c_{6;H\square}  c_{6;HD}+11c_{6;HD}^2),\\
\widetilde{c}_{8;H^4D^4}^{(2)}&=\frac{1}{96 \pi ^2 \epsilon }(16c_{6;H\square}^2+16c_{6;H\square}  c_{6;HD}+5c_{6;HD}^2),\\
\widetilde{c}_{8;H^4D^4}^{(3)}&=\frac{1}{96 \pi ^2 \epsilon }(40c_{6;H\square}^2+16c_{6;H\square}  c_{6;HD}-7c_{6;HD}^2).
\end{align}

The corresponding contributions to the \glspl{rge} are then straightforwardly obtained using Eq.~\eqref{eq:RGEdefinition}:

\begin{align}
16\pi^2\mu \frac{\diff}{\diff \mu} c_{8;H^4D^4}^{(1)}&=-\frac{1}{3 }(16c_{6;H\square}^2-32c_{6;H\square}  c_{6;HD}+11c_{6;HD}^2),\\
16\pi^2\mu \frac{\diff}{\diff \mu}  c_{8;H^4D^4}^{(2)}&=-\frac{1}{3 \pi ^2 \epsilon }(16c_{6;H\square}^2+16c_{6;H\square}  c_{6;HD}+5c_{6;HD}^2),\\
16\pi^2\mu \frac{\diff}{\diff \mu}  c_{8;H^4D^4}^{(3)}&=-\frac{1}{3 \pi ^2 \epsilon }(40c_{6;H\square}^2+16c_{6;H\square}  c_{6;HD}-7c_{6;HD}^2).
\end{align}

\begin{table}[h]
\begin{center}
\begin{footnotesize}
\begin{align*}
\footnotesize
							\begin{array}{|c||c@{\hspace{0.2em}}c@{\hspace{0.2em}}c@{\hspace{0.2em}}c@{\hspace{0.2em}}c@{\hspace{0.2em}}|}
							\hline
					(\gamma_{H^8})_{nl} & H^6 & H^4D^2 & \psi^2H^3 & \psi^2H^2D & \psi^4 \Tstrut\\
					\hline
					\hline
					H^6 & 1 & \lambda & y & y^2 & 0 \Tstrut\\
					H^4D^2 & & \lambda^2 & y \lambda & y^2 \lambda & 0 \\
					\psi^2H^3 & & & y^2 & y \lambda & 0 \\
					\psi^2H^2D & & & & y^2 \lambda & 0 \\
					\psi^4 & & & & & 0\\
					\hline
			\end{array}
			&\quad
\begin{array}{|c||c@{\hspace{0.2em}}c@{\hspace{0.2em}}c@{\hspace{0.2em}}c@{\hspace{0.2em}}c@{\hspace{0.2em}}|}
\hline
					(\gamma_{H^6D^2})_{nl}& H^6 & H^4D^2 & \psi^2H^3 & \psi^2H^2D & \psi^4 \Tstrut\\
					\hline
					\hline
					H^6 & 0 & 1 & 0 & 0 & 0 \Tstrut \\
					H^4D^2 & & \lambda & y & y^2 & 0 \\
					\psi^2H^3 & & & 1 & y & 0 \\
					\psi^2H^2D & & & & y^2 & 0 \\
					\psi^4 & & & & & 0\\
					\hline
			\end{array}\\
%
\footnotesize
							\begin{array}{|c||c@{\hspace{0.2em}}c@{\hspace{0.2em}}c@{\hspace{0.2em}}c@{\hspace{0.2em}}c@{\hspace{0.2em}}|}
							\hline
					(\gamma_{H^4D^4})_{nl} & H^6 & H^4D^2 & \psi^2H^3 & \psi^2H^2D & \psi^4 \Tstrut \\
					\hline
					\hline
					H^6 & 0 & 0 & 0 & 0 & 0 \Tstrut\\
					H^4D^2 & & 1 & 0 & 0 & 0 \\
					\psi^2H^3 & & & 0 & 0 & 0 \\
					\psi^2H^2D & & & & 1 & 0 \\
					\psi^4 & & & & & 0\\
					\hline
			\end{array}
			&\quad
\begin{array}{|c||c@{\hspace{0.2em}}c@{\hspace{0.2em}}c@{\hspace{0.2em}}c@{\hspace{0.2em}}c@{\hspace{0.2em}}|}
\hline
					(\gamma_{X^2H^4})_{nl}& H^6 & H^4D^2 & \psi^2H^3 & \psi^2H^2D & \psi^4 \Tstrut\\
					\hline
					\hline
					H^6 & 0 & 0 & 0 & 0 & 0 \Tstrut\\
					H^4D^2 & & g^2 & 0 & 0 & 0 \\
					\psi^2H^3 & & & 0 & 0 & 0 \\
					\psi^2H^2D & & & & g^2 & 0 \\
					\psi^4 & & & & & 0\\
					\hline
			\end{array}\\
%
\footnotesize
							\begin{array}{|c||c@{\hspace{0.2em}}c@{\hspace{0.2em}}c@{\hspace{0.2em}}c@{\hspace{0.2em}}c@{\hspace{0.2em}}|}
							\hline
					(\gamma_{XH^4D^2})_{nl}& H^6 & H^4D^2 & \psi^2H^3 & \psi^2H^2D & \psi^4 \Tstrut\\
					\hline
					\hline
					H^6 & 0 & 0 & 0 & 0 & 0 \Tstrut\\
					H^4D^2 & & g & 0 & 0 & 0 \\
					\psi^2H^3 & & & 0 & 0 & 0 \\
					\psi^2H^2D & & & & g & 0 \\
					\psi^4 & & & & & 0\\
					\hline
			\end{array}&
			\end{align*}
						\end{footnotesize}
\end{center}
		\caption{\label{tab:ADMd6d6tod8BOS} Anomalous dimension matrix for the insertion of two dimension-six operators. The columns and rows represent the greatest terms from each contribution mixing into the bosonic operators of different classes. See  \cite{Chala:2021pll} for complete RGEs.}
				
\end{table}

\begin{table}[h]
\begin{center}
\begin{footnotesize}
			\begin{align*}
\begin{array}{|c||c@{\hspace{0.2em}}c@{\hspace{0.2em}}c@{\hspace{0.2em}}c@{\hspace{0.2em}}c@{\hspace{0.2em}}|}
\hline
					(\gamma_{X\psi^2H^3})_{nl}& H^6 & H^4D^2 & \psi^2H^3 & \psi^2H^2D & \psi^4 \Tstrut\\
					\hline
					\hline
					H^6 & 0 & 0 & 0 & 0 & 0 \Tstrut\\
					H^4D^2 & & 0 & 0 & yg & 0 \\
					\psi^2H^3 & & & 0 & g & 0 \\
					\psi^2H^2D & & & & yg & 0 \\
					\psi^4 & & & & & 0\\
					\hline
			\end{array}&\quad
%
\footnotesize
							\begin{array}{|c||c@{\hspace{0.2em}}c@{\hspace{0.2em}}c@{\hspace{0.2em}}c@{\hspace{0.2em}}c@{\hspace{0.2em}}|}
							\hline
					(\gamma_{\psi^2H^2D^3})_{nl} & H^6 & H^4D^2 & \psi^2H^3 & \psi^2H^2D & \psi^4 \Tstrut\\
					\hline
					\hline
					H^6 & 0 & 0 & 0 & 0 & 0 \Tstrut\\
					H^4D^2 & & 0 & 0 & 1 & 0 \\
					\psi^2H^3 & & & 0 & 0 & 0 \\
					\psi^2H^2D & & & & 1 & 1 \\
					\psi^4 & & & & & 0\\
					\hline
			\end{array}
			\\
\begin{array}{|c||c@{\hspace{0.2em}}c@{\hspace{0.2em}}c@{\hspace{0.2em}}c@{\hspace{0.2em}}c@{\hspace{0.2em}}|}
\hline
					(\gamma_{\psi^2H^5})_{nl} & H^6 & H^4D^2 & \psi^2H^3 & \psi^2H^2D & \psi^4 \Tstrut\\
					\hline
					\hline
					H^6 & 0 & y & 1 & y & y \Tstrut\\
					H^4D^2 & & y^3 & y^2 & y^3 & y\lambda \\
					\psi^2H^3 & & & y & y^2 & y^2 \\
					\psi^2H^2D & & & & y^3 & 0 \\
					\psi^4 & & & & & 0\\
					\hline
			\end{array}&\quad
%
\footnotesize
							\begin{array}{|c||c@{\hspace{0.2em}}c@{\hspace{0.2em}}c@{\hspace{0.2em}}c@{\hspace{0.2em}}c@{\hspace{0.2em}}|}
							\hline
					(\gamma_{\psi^2H^4D})_{nl}& H^6 & H^4D^2 & \psi^2H^3 & \psi^2H^2D & \psi^4 \Tstrut\\
					\hline
					\hline
					H^6 & 0 & 0 & 0 & 0 & 0 \Tstrut\\
					H^4D^2 & & y^2 & y & y^2 & g^2 \\
					\psi^2H^3 & & & 1 & y & y^2 \\
					\psi^2H^2D & & & & y^2 & y^2 \\
					\psi^4 & & & & & 0\\
					\hline
			\end{array}
			\\
\begin{array}{|c||c@{\hspace{0.2em}}c@{\hspace{0.2em}}c@{\hspace{0.2em}}c@{\hspace{0.2em}}c@{\hspace{0.2em}}|}
\hline
					(\gamma_{X\psi^2H^2D})_{nl}& H^6 & H^4D^2 & \psi^2H^3 & \psi^2H^2D & \psi^4 \Tstrut\\
					\hline
					\hline
					H^6 & 0 & 0 & 0 & 0 & 0 \Tstrut\\
					H^4D^2 & & 0 & 0 &g & 0 \\
					\psi^2H^3 & & & 0 & 0 & 0 \\
					\psi^2H^2D & & & & g & g \\
					\psi^4 & & & & & 0\\
					\hline
			\end{array}&\quad
%
\footnotesize
							\begin{array}{|c||c@{\hspace{0.2em}}c@{\hspace{0.2em}}c@{\hspace{0.2em}}c@{\hspace{0.2em}}c@{\hspace{0.2em}}|}
							\hline
					(\gamma_{\psi^2H^3D^2})_{nl} & H^6 & H^4D^2 & \psi^2H^3 & \psi^2H^2D & \psi^4 \Tstrut\\
					\hline
					\hline
					H^6 & 0 & 0 & 0 & 0 & 0 \Tstrut\\
					H^4D^2 & & 0 & y & yg & 0 \\
					\psi^2H^3 & & & 0 & g & 0 \\
					\psi^2H^2D & & & & yg & 0 \\
					\psi^4 & & & & & 0\\
					\hline
			\end{array}
			\end{align*}
			\end{footnotesize}
\end{center}
		\caption{\label{tab:ADMd6d6tod8FER} Anomalous dimension matrix for the insertion of two dimension-six operators. The columns and rows represent the greatest terms from each contribution mixing into the fermionic operators of different classes. See  \cite{Bakshi:2024wzz} for complete RGEs.}
		
\end{table}

\subsection{Insertion of Lepton Number Violating operators}

We now turn our attention to the insertion of \gls{lnv} operators in the renormalisation of dimension-eight operators. At \order{4}, we identify three distinct types of \gls{lnv} insertions that contribute: (i) four insertions of the Weinberg operator in a single diagram, (ii) two Weinberg operators plus one dimension-six operator, and (iii) one Weinberg operator with one dimension-seven operator. In principle, these insertions may appear either within loop diagrams or attached to external legs, effectively putting off-shell diagrams on shell.

This is a case where it is crucial to analyze the possible operator structures and diagram topologies before performing explicit calculations. In the off-shell formalism, we focus exclusively on \gls{1pi} diagrams. Any redundancies arising in the \gls{wc} are accounted for through appropriate shifts. For clarity, we restrict our analysis to the \glspl{rge} of purely bosonic operators, postponing the study of fermionic operators for future work.

Because the renormalised operators under consideration do not contain fermions, the contributing diagrams will have only bosonic external legs. However, the inserted operators all include fermionic fields and must therefore be contracted to form loops within the \gls{1pi} diagrams. Yukawa and gauge couplings appearing in these loops must also respect this constraint. As a result, there are no off-shell divergences proportional to the Higgs quartic coupling $\lambda$. Similarly, \gls{1pi} diagrams cannot be proportional to the Higgs mass parameter $m_H^2$, which prevents any direct divergent contribution to lower-dimensional operators. However, such terms could still appear indirectly through field redefinitions when applying the \gls{wc} shifts.

Regarding the fermionic loops: four-fermion operators appearing at dimension six or seven do not contribute to the \glspl{rge} of bosonic operators at one-loop order; their contributions begin at two loops, which lies beyond the scope of this analysis.

In terms of loop suppression, some dimension-six operators are generated only at one loop when integrating out \gls{uv} completions of the \gls{smeft}. Inserting such operators into another loop effectively yields a two-loop suppression, making their contributions negligible for our purposes. This particularly affects the renormalisation of dimension-eight operators involving gauge bosons.

At this stage, all candidate insertions involve operators with at least two external Higgs fields. Since two or more such insertions are required to build the relevant \gls{1pi} diagrams, every contributing diagram will have at least four Higgs fields as external legs. Hence, only dimension-eight operators with at least four Higgs fields can be renormalised via these insertions.

When analysing the mixing into dimension-eight operators with many fields, such as $H^6D^2$ or $H^8$, one might consider adding \gls{sm} vertices to increase the number of external legs. While these vertices do not affect power counting in the \gls{smeft}, their insertion introduces internal propagators that must be integrated over. This procedure is limited: increasing the number of loop momenta in the numerator eventually renders the diagram finite, as explained in Section~\ref{sec:DefiningRenormalizability}.

Let us now consider each class of insertions in more detail:
\begin{description}
\item[Four Weinberg operators:] Diagrams with four Weinberg insertions involve fer\-mion loops and yield eight Higgs external legs. These contribute to the renormalisation of $H^8$, but not to any other purely bosonic dimension-eight operators.
\item[Two Weinberg operators and one dimension-six operator:] Since the loop must contain only fermions, only fermionic dimension-six operators can be inserted. Bosonic dimension-six insertions contribute only via WC shifts upon setting the divergences on shell.
\item[One Weinberg and one dimension-seven operator:] In this case, symmetry considerations lead to significant cancellations. The Weinberg operator is symmetric in flavor indices: $$[c_{5;\ell H}]_{\alpha\beta} = [c_{5;\ell H}]_{\beta\alpha}\,.$$ In contrast, dimension-seven operators in the $X\psi^2H^2$ class are antisymmetric: $$[c_{7;\ell H X}]_{\alpha\beta} = -[c_{7;\ell H X}]_{\beta\alpha}\,.$$ As a result, contributions to bosonic operators are proportional to $$\text{Tr}[c_{5;\ell H}c_{7;\ell H X}]=0\,.$$

\end{description}

These features have also been analysed in \cite{DasBakshi:2023htx}, where the complete \gls{rge} were originally computed. Here, we summarise the large entries of the \gls{adm} schematically, which align with the conclusions drawn in the previous sections: Operators with more than six Higgs tend to gather the largest terms deviating from the naively expected value $\gamma\sim \mathcal{O}(1)$. For example:

\begin{align}
16\pi^2\mu \frac{\diff}{\diff \mu} c_{8;H^8}&= 32 \lambda \, \text{Tr}\left[ -c_{6;H \ell}^{(1)} c_{5;\ell H} c_{5;\ell H}^\dagger + c_{6;H \ell}^{(3)}c_{5;\ell H} c_{5; \ell H}^\dagger \right]+\dots\\
16\pi^2\mu \frac{\diff}{\diff \mu} c_{8;H^6D^2}^{(1)}&=- 32\lambda \text{Re}\left(\text{Tr} \left[c_{5;\ell  H}^\dagger c_{7;\ell  H D}^{(2)}\right] \right)+\dots\\
16\pi^2\mu \frac{\diff}{\diff \mu} c_{6;H}&=16 m_H^2 \, \text{Tr}\left[ (c_{6; H \ell}^{(1)}-c_{6; H \ell}^{(3)}) c_{5;\ell  H} c_{5;\ell  H}^\dagger \right] +\dots\\
16\pi^2\mu \frac{\diff}{\diff \mu} \lambda&=-8 m_H^4 \,\text{Re}\left(\text{Tr} \left[c_{5;\ell  H}^\dagger c_{7;\ell  H D}^{(2)}\right] \right).
\end{align}

\begin{figure}[h]
		\begin{center}
\includegraphics[scale=0.13]{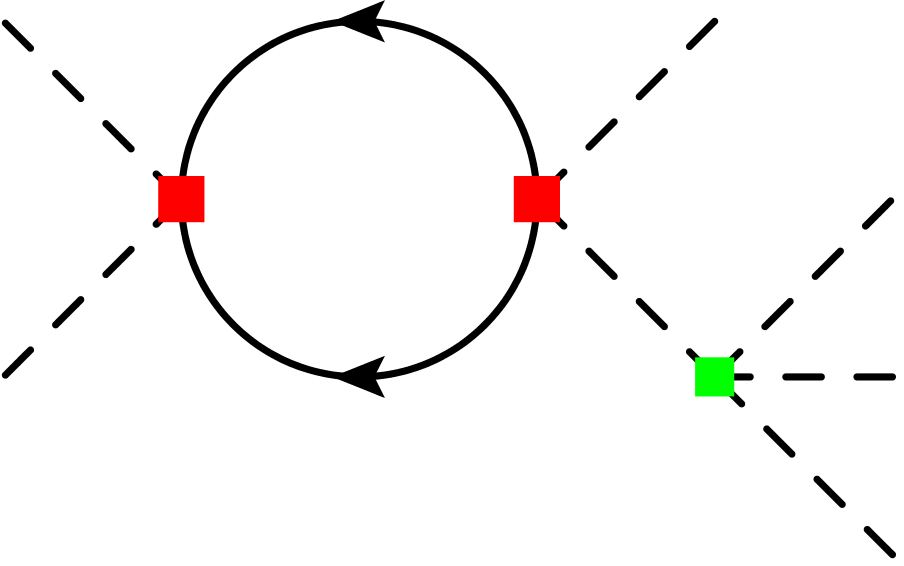} \qquad
\includegraphics[scale=0.13]{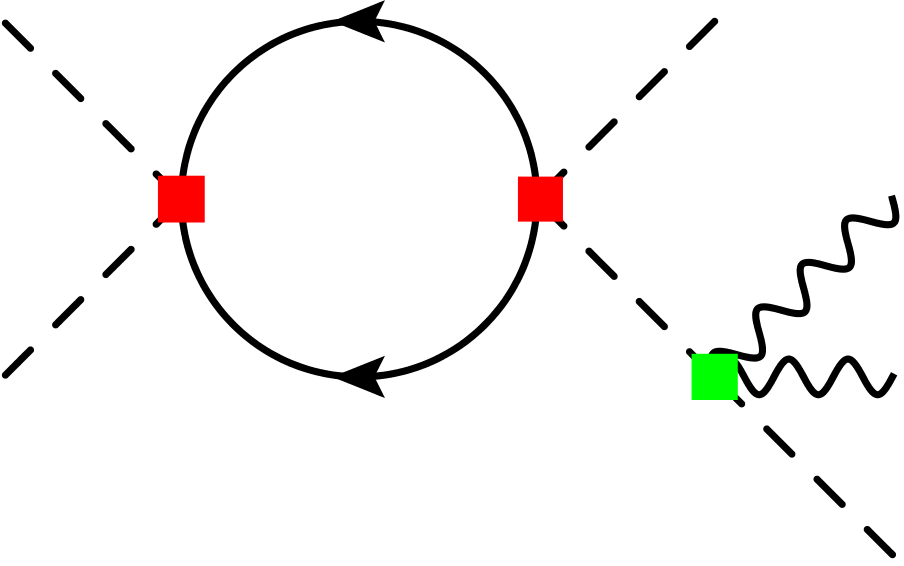} \qquad 
\includegraphics[scale=0.13]{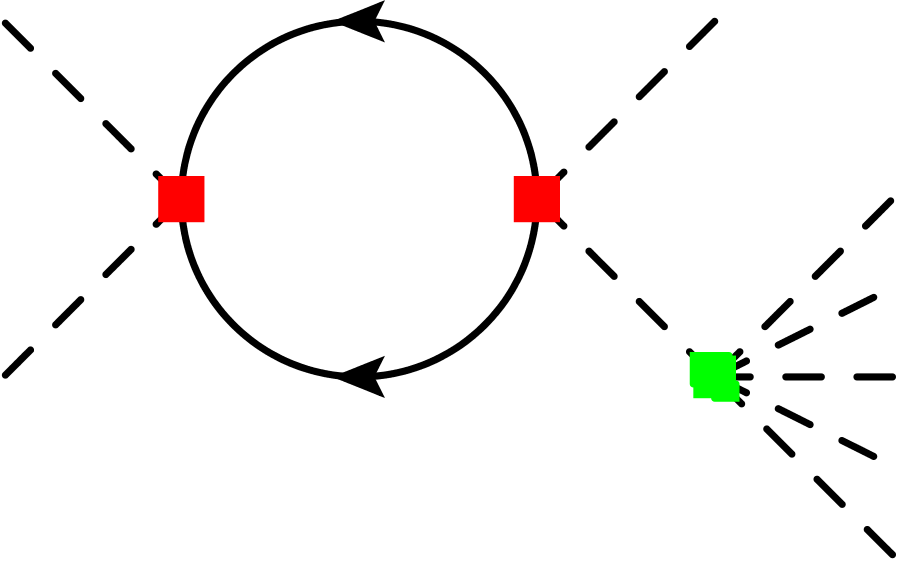} 
		\end{center}
\caption{\label{fig:zerocontrib} Diagrams with two dimension-five operators renormalising a dimension-eight operator via a dimension-six insertion. Although considered in the \glspl{rge} via on-shell relations up to \order{2} (see Appendix B from \cite{Gherardi:2020det}) and \order{2} off-shell divergences (computed directly or inferred from the on-shell divergences (G.2) in Ref.~\cite{Davidson:2018zuo}) with insertions of two Weinberg operators, these diagrams yield no net contribution. Coloured vertices represent insertions of $d_5$ (red), $d_6$ (green), and $d_7$ (blue) operators.}
	\end{figure}

Many entries of the \gls{adm} vanish, as summarised in Table~\ref{tab:ADMLNVtod8}. Some zeros arise trivially from the cancellations mentioned above. Others result from accidental (non-trivial) cancellations. For instance, one-particle-reducible diagrams involving two Weinberg operators set on-shell by a bosonic dimension-six operator (see Figure~\ref{fig:zerocontrib}) vanish upon applying unitarity cuts.
To conclude, we remark that, when restricting to \gls{lnv} insertions, loop-generated operators are not renormalised at one loop by tree-level-generated operators.

\subsubsection{Example: RGE of $H^4D^4$}
For completeness, we now compute the \gls{rge} contribution to $H^4D^4$ from \gls{lnv} insertions. This is a concise calculation, as most terms in the full equation originate from lepton number-conserving operator insertions.

According to the preceding arguments, operators like $X^2D^4$ and $XH^2D^4$ cannot be renormalised by \gls{lnv} insertions, as they involve fewer than four Higgs fields. Similarly, $H^4D^4$ cannot be renormalised by diagrams with four Weinberg operators or with two Weinberg operators plus a dimension-six insertion, as these generate more than four external Higgs legs. The only viable \gls{lnv} insertion is the combination of one Weinberg operator with a dimension-seven operator of the type $\psi^2H^2D^2$.

\begin{figure}
\begin{center}
\includegraphics[scale=0.13]{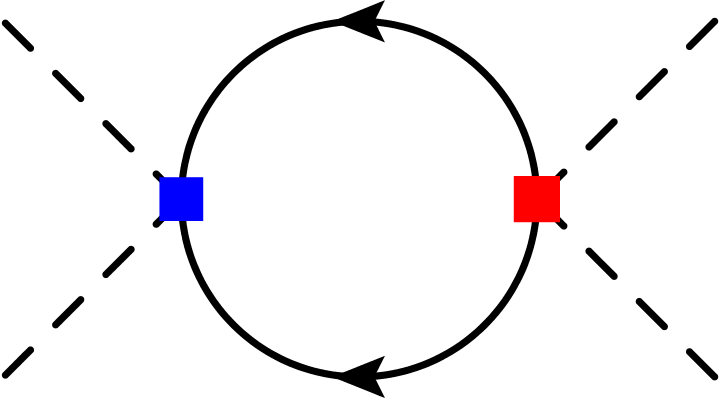}
\end{center}
\caption{\label{fig:Exampled8byLNV} One-loop diagrams contributing to $H^4D^4$ from \gls{lnv} insertions. Blue boxes denote dimension-five operators; red boxes denote dimension-seven operators.}

\end{figure}

Figure~\ref{fig:Exampled8byLNV} shows the relevant diagrams. The amplitude for the same process as the previous cases, $H^0H^0~\rightarrow~H^+H^-$, evaluates to:
\begin{equation}
\ii\mathcal{A}^\text{1L}=\frac{\ii}{8\pi^2\epsilon}\text{Re}\left(\text{Tr} \left[c_{5;\ell  H}^\dagger c_{7;\ell  H D}^{(2)}\right] \right)(\kappa_{2222}+4\kappa_{2223}+2\kappa_{2233}+4\kappa_{2323}+4\kappa_{2333}+\kappa_{3333})\,,
\end{equation}
where the $\kappa$ are the same kinematic invariants defined in earlier sections. Solving the resulting system of equations from the \gls{ir} amplitude~\eqref{eq:IRamplitudeH4D4}, we extract the divergences:
\begin{align}
 \widetilde{c}_{8;H^4D^4}^{(1)}&=0,\\
\widetilde{c}_{8;H^4D^4}^{(2)}&=\frac{1}{4\pi^2\epsilon}\text{Re}\left(\text{Tr} \left[c_{5;\ell  H}^\dagger c_{7;\ell  H D}^{(2)}\right] \right),\\
\widetilde{c}_{8;H^4D^4}^{(3)}&=0.
\end{align}
As there are no one-particle-reducible contributions, this corresponds to the full physical divergence. The \gls{rge} is thus:
\begin{align}\label{eq:RGEexample3}
16\pi^2\mu \frac{\diff}{\diff \mu} c_{8;H^4D^4}^{(1)}&=0,\\
16\pi^2\mu \frac{\diff}{\diff \mu}  c_{8;H^4D^4}^{(2)}&=-8\text{Re}\left(\text{Tr} \left[c_{5;\ell  H}^\dagger c_{7;\ell  H D}^{(2)}\right] \right),\\
16\pi^2\mu \frac{\diff}{\diff \mu}  c_{8;H^4D^4}^{(3)}&=0.
\end{align}

	\begin{table}[H]
		\begin{center}
			\resizebox{\textwidth}{!}{
				\begin{tabular}{|c||c|cc|cccc|}
				\hline
					$\gamma_{p}^{(d5^4)},\,\gamma_{ps}^{(d5^2d6)},\,\gamma_{ps}^{(d5d7)}$ & - & $H^4D^2$ & $\psi^2H^2D$ & $\psi^2H^4$ & $\psi^2H^3D$ & $\psi^2H^2D^2$ & $X\psi^2H^2$\Tstrut \cr
					\hline
					\hline
					$H^8$ & $1$ & $\lambda$ & $\lambda $ & $\lambda$ & 0 & $\lambda g^2 $ & 0\Tstrut\cr
					$H^6D^2$ & 0 & $1$ & $12$ & $1$ & $y$ & $g^2$ & 0\cr
					$H^4D^4$ & 0 & 0 & 0 & 0 & 0 & $1$ & 0 \cr 
					$X^3H^2$ & 0 & 0 & 0 & 0 & 0 & 0 & 0 \cr
					$X^2H^4$ & 0 & 0 & 0 & 0 & 0 & $g^2$ & $g$\cr
					$X^2H^2D^2$ & 0 & 0 & 0 & 0 & 0 & 0 & 0 \cr
					$XH^4D^2$ & 0 & 0 & 0 & 0 & 0 & $g$ & 0 \cr
					\hline
			\end{tabular}}
		\end{center}
\caption{\label{tab:ADMLNVtod8} \gls{adm} for dimension-eight bosonic operators. Columns correspond to insertions of the Weinberg operator and of dimension-six or dimension-seven operators. All zeros arise from the absence of diagrams or symmetry-induced cancellations. See \cite{DasBakshi:2023htx} for full details on the\gls{rge}.}

	\end{table}

\section{Applications}\label{sec:Applications}
\subsection{Positivity bounds}
Positivity bounds are mathematical inequalities among \glspl{wc}, derived from two-to-two scattering amplitudes by imposing the fundamental principles of analyticity, unitarity, and crossing symmetry of the S-matrix. These bounds constrain the allowed parameter space of \glspl{eft}, providing insight into the possible \gls{uv} completions. In the context of the \gls{smeft}, they are especially relevant due to their interplay with power counting: they typically constrain operators of dimension eight or higher, which also makes them a valuable tool for analysing the impact of subleading interactions.

In particular, positivity bounds constrain combinations of dimension-eight operators and may also affect processes involving multiple insertions of dimension-six operators, including combinations with \gls{lnv} terms. This makes them a powerful consistency check for low-energy \glspl{eft}.

While positivity bounds are derived at tree level, an important question is whether they continue to hold under \gls{rg} evolution. Using the complete set of \glspl{rge} derived in the previous Section~\ref{sec:RGEResults}, one can assess whether loop corrections preserve or violate these bounds. Schematically, one starts with a positivity inequality valid at some high scale $ \Lambda $ and evolves the \gls{wc} down to lower scales using the \glspl{rge}. If the inequality is violated at any intermediate scale, this could signal a more general instability of positivity bounds, although a deeper study would be required.

As a concrete example, consider the dimension-eight class $ H^4D^4 $, which contributes to four-Higgs scattering processes. Ref.~\cite{Chala:2021wpj} analyzed the \gls{rg} evolution of these operators and its implications for positivity bounds. The tree-level positivity constraints derived from such processes~\cite{Remmen:2019cyz} are:

\begin{align}\label{eq:PosBoundH4D4}
c_{8;H^4D^4}^{(2)}>&0\\
c_{8;H^4D^4}^{(1)}+c_{8;H^4D^4}^{(2)}>&0\\
c_{8;H^4D^4}^{(1)}+c_{8;H^4D^4}^{(2)}+c_{8;H^4D^4}^{(3)}>&0\,.
\end{align}

Assuming these bounds are satisfied at the threshold scale of \gls{smeft} $ \mu = \Lambda$, we examine their stability under \gls{rg} running. Solving the \gls{ll} \glspl{rge} gives:

\begin{equation}
16\pi^2\mu \frac{\diff}{\diff \mu}c_{8;H^4D^4}^{(p)}\equiv c_{8;H^4D^4}^{(t)}[\Lambda]\gamma_{pt}\,,
\end{equation}
where $ \gamma_{pt} $ is the \gls{adm} obtained in Eq.~\ref{sec:RGEexample1}. The running then induces the following differential inequalities at one loop:

\begin{align}
16\pi^2\mu \frac{\diff}{\diff \mu}c_{8;H^4D^4}^{(2)}\equiv c_{8;H^4D^4}^{(t)}[\Lambda]\gamma_{2t}&>0,\\
16\pi^2\mu \frac{\diff}{\diff \mu}(c_{8;H^4D^4}^{(1)}+c_{8;H^4D^4}^{(2)})\equiv \sum_{s=1,2}c_{8;H^4D^4}^{(t)}[\Lambda]\gamma_{st}&>0,\\
16\pi^2\mu \frac{\diff}{\diff \mu}(c_{8;H^4D^4}^{(1)}+c_{8;H^4D^4}^{(2)}+c_{8;H^4D^4}^{(3)})\equiv \sum_{s=1,2,3}c_{8;H^4D^4}^{(t)}[\Lambda]\gamma_{st}&>0.
\end{align}

We now test whether these \gls{rg}-evolved inequalities remain valid. For instance, consider the contribution proportional to $ g_1^2 $ in the first inequality. From the \glspl{rge} of $ c_{8;H^4D^4}^{(2)} $, one obtains:

\begin{equation}
g_1^2\left(\frac{7 c_{8;H^4D^4}^{(1)}}{3}+\frac{11 c_{8;H^4D^4}^{(2)}}{2}+\frac{5 c_{8;H^4D^4}^{(3)}}{2}\right).
\end{equation}
Rewriting this expression, we find:

\begin{equation}
\frac{19}{6}c_{8;H^4D^4}^{(2)}+\frac{7}{3}(c_{8;H^4D^4}^{(1)}+c_{8;H^4D^4}^{(2)}+c_{8;H^4D^4}^{(3)})+\frac{1}{6}c_{8;H^4D^4}^{(3)}.
\end{equation}

Although each term appears positive if Eqs.~\eqref{eq:PosBoundH4D4} are satisfied, certain values of $ c_{8;H^4D^4}^{(3)} $ can render the entire expression negative. This means that the positivity bounds can be violated under \gls{rg} running, even if they hold at the threshold scale. 

A similar analysis can be applied to other operator classes with four-field content. For example, Ref.~\cite{DasBakshi:2022mwk} considered the class $X^2H^2D^2$, using its \glspl{rge} and the associated positivity bounds~\cite{Bi:2019phv}. It was found that \gls{rg} contributions from $ H^4D^4 $ and $ \psi^2H^2D^3$ to $ X^2H^2D^2 $ preserve positivity bounds, assuming those of the contributing operator classes hold~\cite{Remmen:2019cyz,Li:2022tcz}. This supports the notion that while positivity can be violated through running in specific cases, other operator classes remain consistent under the \gls{rg} flow.

Additionally, in Ref.~\cite{DasBakshi:2023htx}, we applied the positivity bounds of the $ H^4D^4 $ class, together with the renormalisation group equations originally derived in this work, to establish a nontrivial relation among \gls{lnv} operators. Specifically, considering the \gls{rge} for $c_{8;H^4D^4}^{(2)} $ (Eq.~\eqref{eq:RGEexample3}), one obtains the following constraint:

\begin{equation}\label{eq:PosBoundLNV}
\text{Re}\left(\text{Tr} \left[c_{5;\ell  H}^\dagger c_{7;\ell  H D}^{(2)}\right] \right)>0,
\end{equation}
under the assumption that no dimension-six or dimension-eight operators are generated at tree level. This inequality represents a purely low-energy statement that imposes a positivity constraint on the \gls{uv} coefficients $ a_{5;\ell H} $ and $ a_{7;\ell H D}^{(2)} $, derived from analytic properties of the S-matrix.

This result has significant implications for \gls{uv}-complete theories such as Seesaw models. In particular, Seesaw models of type I~\cite{Yanagida:1979as, Mohapatra:1979ia} and III~\cite{Foot:1988aq} generate \gls{lnv} operators at tree level, but do not produce operators in the $ H^4D^4 $ class~\cite{Du:2022vso}. Therefore, the positivity condition in Eq.~\eqref{eq:PosBoundLNV} becomes a nontrivial test of these models. As shown in Ref.~\cite{DasBakshi:2023htx}, the bound is satisfied in both Seesaw I and III. The key observation is that these \gls{uv} completions do not generate the operator $\mathcal{O}_{\ell H D}^{(2)} $, which implies that the left-hand side of Eq.~\eqref{eq:PosBoundLNV} vanishes, thereby preserving the inequality.

This analysis illustrates how loop-level positivity bounds can bridge low-energy effective constraints with \gls{uv} dynamics, offering a novel probe of the structure of \gls{bsm} physics.

\subsection{Oblique parameters}
The oblique parameters quantify deviations from the \gls{sm} predictions for the self-energies of \gls{ew} gauge bosons, arising due to potential new physics. They play a central role in \gls{ewpt}), where they can be stringently constrained by experimental measurements. Originally introduced by Peskin and Takeuchi~\cite{Peskin:1990zt}, these parameters capture universal corrections to gauge boson propagators and are particularly effective in diagnosing the effects of heavy new physics that does not couple directly to fermions.

The vacuum polarization functions $\Pi_{X,Y}(p^2)$ describe gauge boson self-energies in the unbroken \gls{ew} phase. The traditional oblique parameters $S$, $T$, $U$ and their rescaled forms $\hat{S}$, $\hat{T}$, $\hat{U}$, as well as the more recent incorporations ($X$, $Y$, $V$, $W$) are defined as follows~\cite{Barbieri:2004qk}:
\begin{align}
\hat{S}=\frac{\alpha}{4s_\theta^2}S&=-g_2^2\Pi_{W^3B}^\prime(p=0),\\
\hat{T}=\alpha T&=\frac{g_2^2}{m_W^2}(\Pi_{W^3W^3}(p=0)-\Pi_{W^+W^-}(p=0)),\\
\hat{U}=-\frac{\alpha}{4s_\theta^2}U&=-g_2^2(\Pi_{W^3W^3}^\prime(p=0)-\Pi_{W^+W^-}^\prime(p=0)),\\
V&=\frac{g_2^2}{2m_W^2}(\Pi_{W^3W^3}^{\prime\prime}(p=0)-\Pi_{W^+W^-}^{\prime\prime}(p=0)),\\
W&=\frac{g_2^2}{2m_W^2}\Pi_{W^3W^3}^{\prime\prime}(p=0),\\
X&=\frac{g_1g_2}{2m_W^2}\Pi_{W^3B}^{\prime\prime}(p=0),\\
Y&=\frac{g_1^2}{2m_W^2}\Pi_{BB}^{\prime\prime}(p=0),\\
Z&=\frac{g_3^2}{2m_W^2}\Pi_{GG}^{\prime\prime}(p=0).
\end{align}
where $\alpha\approx 1/137$ is the fine structure constant, $s_\theta$ is the sine of the Weinberg angle, and the derivatives are with respect to $p^2$ and evaluated at $p=0$: \begin{equation}
\Pi^\prime (0) =\left.\frac{\diff \Pi(p^2)}{\diff p^2}\right|_{p^2=0}.
\end{equation}
The parameters $S$, $T$, $W$, $Y$ describe \gls{lo} deviations from \gls{sm} predictions, whereas $U$, $X$, $V$ represent subleading higher-order effects. The parameter $Z$ is defined analogously for the \gls{qcd} sector but is not an \gls{ew} observable.

In the \gls{smeft} framework, these parameters can be expressed in terms of the Wilson coefficients of universal dimension-six operators. Using the Green's basis~\cite{Gherardi:2020det}, we obtain:\pagebreak
\begin{align}
S&=16\pi^2\frac{v_T^2}{\Lambda^2}a_{6;HWB}\,,\\
\hat{T}&=-\frac{1}{2}\frac{v_T^2}{\Lambda^2}a_{6;HD}\,,\\
W&=-\frac{1}{2}\frac{m_W^2}{\Lambda^2}b_{6;2W}\,,\\
Y&=-\frac{1}{2}\frac{m_W^2}{\Lambda^2}b_{6;2B}\,,\\
Z&=-\frac{1}{2}\frac{m_W^2}{\Lambda^2}b_{6;2G}.
\end{align}
Here, $\mathcal{O}_{6;2X}$ are operators from class $X^2D^2$ that are redundant under  \gls{eom}. The parameters $m_W$ and $v_T$ denote the geometric W mass and Higgs \gls{vev}, respectively, as defined in~\cite{Hays:2020scx}. At \gls{lo}, these expressions reduce to their \gls{sm} values, with additional corrections suppressed by powers of $\Lambda$.
At \gls{nlo}, dimension-eight operators and one-loop \gls{rge} effects enter. The dimension-eight contributions to the $S$, $T$, $U$ oblique parameters are given by~\cite{Murphy:2020rsh}:
\begin{align}\label{eq:ObliqueAtD8}
S&=\frac{v_T^4}{\Lambda^4}c_{8;WBH^4}\,,\\
\hat{T}&=-\frac{1}{2}\frac{v_T^4}{\Lambda^4}c_{8;H^6D^2}^{(2)}\,,\\
U&=\frac{v_T^4}{\Lambda^4}c_{8;W^2H^4}\,.
\end{align}

The renormalisation group running at \order{4} allows us to track the quantum corrections to these parameters. It has been shown in~\cite{Chala:2021pll} that the dimension-six operators do not mix into $S$ and $U$ at this order. Instead, their renormalisation involves dimension-eight operators in the classes $H^4D^4$, $X^2H^4$, $XH^4D^2$, $X\psi^2H^3$, $\psi^2H^2D^3$, $X\psi^2H^2D$, $\psi^2H^3D^2$, as detailed in Table~\ref{tab:ADMd8tod8}.

The $T$ parameter, however, does receive corrections from insertions of dimension-six operators via the running of $\mathcal{O}_{8;H^6D^2}^{(2)}$. Notably, the operator $\mathcal{O}_{6;Hud}$ from class $\psi^2H^2D$, which does not directly contribute to $T$ at \order{2}  can induce contributions at \order{4} through self-mixing into $\mathcal{O}_{8;H^6D^2}^{(2)}$. Using the \gls{adm} from~\cite{Chala:2021pll} (presented in Table~\ref{tab:ADMd6d6tod8BOS}), along with fits of \gls{ewpd} to \gls{smeft}~\cite{deBlas:2016nqo}, bounds on coefficients such as $[\mathcal{O}_{6;Hud}]_{tb}$ can be established under the assumption that no other operators contribute.

\begin{figure}
\begin{center}
\includegraphics[scale=0.4]{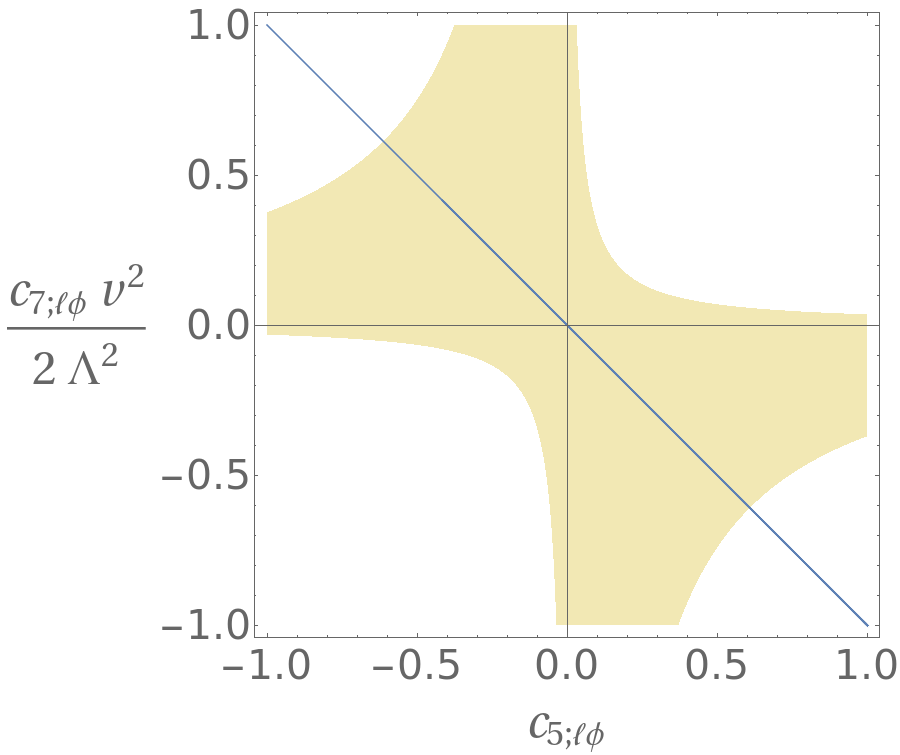}
\end{center}
\caption{\label{fig:BlindBounds}Constraints on \gls{lnv} \glspl{wc} from neutrino mass (blue line) and the T parameter (yellow region). The plot is generated with the following parameters: $\Lambda=1\,\text{TeV}$ and $v=246\,\text{GeV}$, $T=0,10\pm 0,12$~\cite{deBlas:2016nqo} and $m_\nu<0,081\,\text{eV}$~\cite{Loureiro:2018pdz}.}
\end{figure}

These constraints also impact the structure of the neutrino mass matrix. The dimension-five Weinberg operator $\mathcal{O}_{5;\ell H}$ induces a Majorana mass for neutrinos after electroweak symmetry breaking. At \order{3}, its dimension-seven analogue $\mathcal{O}_{7;\ell H}$ contributes as well. Together, they yield the relation~\cite{Loureiro:2018pdz}:
\begin{equation}\label{eq:NuMassBound}
[m_\nu] =- \frac{v^2}{\Lambda}(c_{5;\ell H}+\frac{v^2}{2\Lambda^2}c_{7;\ell H}) 
\end{equation}
This constraint implies a correlation between the coefficients $c_{5;\ell H}$ and $c_{7;\ell H}$ required to reproduce the small observed neutrino masses. However, the operator $\mathcal{O}_{8;H^6D^2}^{(2)}$ also receives loop-level contributions from these LNV operators. As shown in~\cite{DasBakshi:2023htx}, this leads to an additional contribution to the $T$ parameter~\eqref{eq:ObliqueAtD8}:

\begin{equation}
T=-\frac{1}{4\pi^2\alpha}\frac{v^4}{\Lambda^4} \ln\frac{\Lambda}{\mu} \text{Tr}\left[c_{5;\ell H}c_{7;\ell H}\right].    
\end{equation}
This relation provides a second constraint on the Wilson coefficients, effectively lifting blind directions in the mass matrix $[m_\nu]_{\alpha\beta}$.  In the one-generation limit, these conditions are illustrated in Figure~\ref{fig:BlindBounds}, showing that the coefficients
$a_{5;\ell H}$ and $a_{7;\ell H}$ cannot both take arbitrary values.

While the analyses discussed here rely on simplified assumptions about the operator content of new physics, they highlight the utility of oblique parameter constraints in \gls{smeft}. Particularly, they demonstrate how loop-level running and dimension-eight contributions can impose non-trivial constraints on otherwise unconstrained directions in parameter space. Future work involving global \gls{smeft} fits and higher-order \glspl{rge} will benefit from these insights and further develop a systematic understanding of \gls{ewpo} under running at \order{4} and in the presence of \gls{lnv}. 
\chapter{Conclusiones} \label{ch:Conclusiones}
\section{Resumen y visión general}
La investigación en física de partículas se basa fundamentalmente en observaciones experimentales y en su interpretación dentro de un marco teórico. A lo largo de las últimas décadas, el \gls{sm} ha emergido como la teoría dominante, no solo por su extraordinario poder predictivo, sino también por el respaldo constante que recibe de los datos experimentales. Aunque la evidencia favorece abrumadoramente al \gls{sm} frente a cualquier alternativa \gls{bsm}, los datos aún dejan espacio para una teoría más completa, cuyas características precisas siguen siendo desconocidas.

Los próximos experimentos, incluyendo colisionadores de partículas de próxima generación y detectores de ondas gravitacionales, buscan revelar nuevas pistas que puedan guiarnos más allá del \gls{sm}. Mientras tanto, los físicos pueden perfeccionar los análisis existentes y desarrollar nuevas técnicas para la interpretación de datos. La renormalización y las \glspl{eft} son herramientas esenciales para alcanzar estos objetivos.

En el Capítulo~\ref{ch:Renormalization}, revisamos los fundamentos de la renormalización. Aunque en sus inicios fue vista como problemática, la renormalización se ha convertido en una técnica poderosa dentro de la \gls{qft} de precisión. En particular, solo ciertos esquemas de regularización permiten eliminar las divergencias sin perder propiedades deseables de las \glspl{qft}. La combinación de la \gls{dimreg} y el esquema \gls{msbar} es ahora el enfoque estándar. Si bien la \gls{dimreg}, tal como se define en este contexto, también presenta desafíos—especialmente en amplitudes quirales—el marco es confiable a un lazo.

Para explorar modelos que difieren del \gls{sm} a altas energías pero que son consistentes con los datos a bajas energías, recurrimos a teorías no renormalizables. En el Capítulo~\ref{ch:EFT}, discutimos el uso de \glspl{eft}, que en principio requieren un número infinito de contratérminos. Sin embargo, la precisión finita de los experimentos acota efectivamente el número de contratérminos que deben considerarse. Esta característica restaura la capacidad predictiva y permite cálculos de precisión. Ilustramos esto con diversas \glspl{eft} utilizadas históricamente antes del establecimiento del \gls{sm}, muchas de las cuales siguen siendo relevantes para fenómenos a escalas sub-\gls{ew}.

Dado que nuestro interés se sitúa más allá del \glsxtrshort{vev} de Higgs, y que el \gls{sm} representa la teoría de mayor energía confirmada experimentalmente, lo tomamos como base para construir una \gls{eft}. En el Capítulo~\ref{ch:SMEFT}, describimos los beneficios de usar bases de operadores y abordamos los desafíos al identificar un conjunto completo e independiente de interacciones. Argumentamos que conservar operadores relacionados mediante ciertas redefiniciones de campo—equivalentes a aplicar las \glspl{eom} en la Lagrangiana—puede simplificar cálculos intermedios. Por ejemplo, las funciones de Green en una Lagrangiana redundante pueden calcularse usando únicamente diagramas \gls{1pi}. Aunque las redundancias deben eliminarse eventualmente, esto puede hacerse al final del cálculo aplicando relaciones en la capa de masa (on-shell). Si bien trabajar con una Lagrangiana independiente on-shell no siempre es necesario, el enfoque diagramático off-shell de la renormalización sí impone esta condición. En este marco, calculamos la primera base de Green de operadores bosónicos de dimensión ocho, publicada en Ref.~\cite{Chala:2021cgt} y detallada en la Sección~\ref{sec:GB}. Los resultados también se presentan en el Apéndice~\ref{app:TableOfOperators}, junto con ope\-radores fermiónicos de dimensión ocho. Las relaciones on-shell de esta base de Green constituyen otro resultado original que permite eliminar redundancias a favor de la base física de Ref.~\cite{Murphy:2020rsh}.

El Capítulo~\ref{ch:RGE} consolida estas ideas para sistematizar la renormalización en la \gls{smeft}. Derivamos la ecuación de Callan–Symanzik en una forma útil para la renormalización de operadores de dimensión ocho. Tras resumir el método diagramático off-shell y repasar la historia de la renormalización en la \gls{smeft}, aplicamos nuestro marco para calcular las \glspl{rge} completas de todos los operadores bosónicos, así como contribuciones sustanciales para operadores fermiónicos. Con excepción de la renormalización de operadores bosónicos mediante inserciones de operadores de dimensión seis~\cite{Chala:2021pll}, estos resultados~\cite{Bakshi:2024wzz,DasBakshi:2022mwk,DasBakshi:2023htx} son contribuciones originales de esta tesis.

Para ofrecer una visión general del estado actual de la renormalización en la \gls{smeft}—y ubicar nuestras contribuciones en contexto—remitimos al Cuadro~\ref{tab:summary}, que resume los esfuerzos realizados y en curso en este campo. Las entradas calculadas en esta tesis están destacadas.

		\begin{table}[htb!]
		\begin{center}
			\resizebox{1.\textwidth}{!}{\begin{tabular}{l|ccccccccccc}
					& $d_5$ & $d_5^2$ & $d_6$ & $d_5^3$ & $d_5\times d_6$ & $d_7$ & $d_5^4$ & $d_5^2\times d_6$ & $d_6^2$ & $d_5\times d_7$ & $d_8$\\
					\toprule
					$d_{\leq 4}$ (bosónicos) &  &  & $\gtick$~\cite{Jenkins:2013zja} &  &   &  &  &  & $\gtick$~\cite{Chala:2021pll} &  & \colorbox{blue!30}{$\gtick$~\cite{DasBakshi:2022mwk}} \\
					$d_{\leq 4}$ (fermiónicos) &  &  & $\gtick$~\cite{Jenkins:2013zja} &  &   &  &  &  & \colorbox{blue!30}{$\gtick$~\cite{Bakshi:2024wzz}} &  & $\otick$~\cite{AccettulliHuber:2021uoa,Boughezal:2024zqa,Assi:2025fsm} \\
					$d_5$ & $\gtick$~\cite{Chankowski:1993tx,Babu:1993qv,Antusch:2001ck} &  &  &  & $\gtick$~\cite{Chala:2021juk} & $\otick$~\cite{Chala:2021juk} &  &  &  & & \\
					$d_6$ (bosónicos) &  & $\gtick$~\cite{Davidson:2018zuo} & $\gtick$~\cite{Jenkins:2013zja,Jenkins:2013wua,Alonso:2013hga} & & & &  & \colorbox{blue!30}{$\gtick$~\cite{DasBakshi:2023htx}} & $\gtick$~\cite{Chala:2021pll} & \colorbox{blue!30}{$\gtick$~\cite{DasBakshi:2023htx}} & \colorbox{blue!30}{$\gtick$~\cite{DasBakshi:2022mwk}} \\
					$d_6$ (fermiónicos) &  & $\gtick$~\cite{Davidson:2018zuo} & $\gtick$~\cite{Jenkins:2013zja,Jenkins:2013wua,Alonso:2013hga,Alonso:2014zka} & & & & & $\rtick$ & \colorbox{blue!30}{$\otick$~\cite{Bakshi:2024wzz}} & $\rtick$ & $\otick$~\cite{AccettulliHuber:2021uoa,Boughezal:2024zqa,Assi:2025fsm}\\
					$d_7$ &  &  & & $\gtick$~\cite{Zhang:2023kvw} & $\gtick$~\cite{Zhang:2023kvw} & $\gtick$~\cite{Liao:2016hru,Liao:2019tep}\\
					$d_8$ (bosónicos) &  &  & & & & & \colorbox{blue!30}{$\gtick$~\cite{DasBakshi:2023htx}} & \colorbox{blue!30}{$\gtick$~\cite{DasBakshi:2023htx}} & $\gtick$~\cite{Chala:2021pll} & \colorbox{blue!30}{$\gtick$~\cite{DasBakshi:2023htx}} & \colorbox{blue!30}{$\gtick$~\cite{DasBakshi:2022mwk}}\\
					$d_8$ (fermiónicos) &  &  & & & & & $\rtick$ & $\rtick$ & \colorbox{blue!30}{$\otick$~\cite{Bakshi:2024wzz}}& $\rtick$  & $\otick$~\cite{AccettulliHuber:2021uoa,Boughezal:2024zqa,Assi:2025fsm}
					\\\bottomrule
			\end{tabular}}
			\caption{\label{tab:resumen}\it Estado del arte de la renormalización del SMEFT (adaptado de Refs.~\cite{DasBakshi:2022mwk,Chala:2021pll,DasBakshi:2023htx}). Las filas muestran los operadores renormalizados (clasificados por dimensión y estadísticas). Las columnas indican los operadores que contribuyen a la evolución del \gls{rg}. Las entradas vacías corresponden a contribuciones nulas, $\gtick$ indica que la contribución completa está disponible, $\otick$ señala que existen resultados parciales (aunque sustanciales), y $\rtick$ indica que no se conoce nada, o casi nada, al respecto.
Las contribuciones realizadas en esta tesis están marcadas con $\colorbox{blue!30}{\text{Recuadros azules}}$.}
		\end{center}
	\end{table}

La evolución (running) de los \gls{wc} en \gls{smeft} tiene una amplia gama de aplicaciones. En particular, las \glspl{rge} de operadores de dimensión seis ya se utilizan en estudios fenomenológicos. Como se discute en la Sección~\ref{sec:Applications}, se espera que las \glspl{rge} de operadores de dimensión ocho desempeñen un papel similar en física de alta precisión, e incluso podrían abrir nuevas vías conceptuales. Una de estas direcciones involucra cotas de positividad, que son restricciones derivadas de la unitariedad, causalidad y analiticidad. Dado que estas cotas son efectos de orden $\Lambda^{-4}$, el uso de \glspl{rge} para estudiar sus violaciones es particularmente pertinente. Como se señaló en la Sección~\ref{sec:Applications}, la observación de que el running del \gls{rg} puede llevar a aparentes violaciones de estas cotas podría motivar una reevaluación del marco \gls{smeft}.

\section{Líneas futuras de investigación}
De este trabajo se desprenden diversas direcciones prometedoras:

\paragraph{Completar las \glsxtrshort{rge} restantes a orden $\Lambda^{-4}$}
Una extensión natural consiste en calcular las \glspl{rge} para todos los operadores fermiónicos de dimensión ocho, incluyendo interacciones de cuatro fermiones y contribuciones con \gls{lnv} provenientes de combinaciones de operadores de menor dimensión.

\paragraph{Aclarar el papel de la \glsxtrshort{smeft} en \glsxtrshort{ewpt}}
Aunque las contribuciones de dimensión seis y ocho a los parámetros oblicuos son conocidas para operadores bosónicos, su relación con las interacciones fermiónicas sigue sin esclarecerse, dado que los parámetros oblicuos se estudian mayormente en teorías universales. Investigar esta relación profundizaría nuestra comprensión de la aplicabilidad de la \gls{smeft} en las \glspl{ewpt}.

\paragraph{Investigar la dependencia de base en las \glsxtrshort{adm}}
Se ha observado que ciertos cálculos off-shell muestran cancelaciones al evaluarse on-shell. Entender más profundamente este efecto podría esclarecer la relación entre la elección de base y las estructuras de divergencia. Esto requeriría construir nuevas bases de operadores y recalcular las \glspl{rge} a orden $\Lambda^{-4}$, lo cual es actualmente poco práctico. Resulta más factible esperar una mayor automatización de las técnicas de renormalización.

\paragraph{Comparar métodos de renormalización: off-shell, on-shell y geométrico}
Las \glspl{rge} a dimensión ocho ya se están utilizando para validar enfoques alternativos de renormalización. Una vez alcanzada la renormalización completa, una comparación sistemática de estos métodos—no solo en cuanto a resultados, sino también en eficiencia computacional—sería de gran valor. La \gls{smeft}, por su simplicidad y compatibilidad con una amplia gama de observables de baja energía, es un terreno ideal para tales estudios comparativos.

\paragraph{Reflexión final}
Sea cual sea el camino que se elija, las \glspl{rge} de dimensión ocho representan una herramienta poderosa en la búsqueda continua de \gls{np}. En un sentido más amplio, la renormalización y las \glspl{eft} han demostrado ser indispensables en la \gls{qft} y siguen moldeando la manera en que conectamos teorías de alta energía con fenómenos de baja energía. Con esta tesis, buscamos aportar a una comprensión más profunda de estas herramientas y motivar la exploración continua en este campo prometedor. 
\chapter{Conclusions} \label{ch:Conclusions}

\section{Summary and overview}

Research in particle physics relies fundamentally on experimental observations and their interpretation within a theoretical framework. Over the past decades, the \gls{sm} has emerged as the dominant theory—not only due to its remarkable predictive power but also because of the consistent support it receives from experimental data. While the evidence overwhelmingly favours the \gls{sm} over any \gls{bsm} alternatives, the data still leave room for a more complete theory, the features of which remain unknown with precision.

Upcoming experiments, including next-generation particle colliders and gravitational wave detectors, aim to uncover additional clues that may lead us beyond the \gls{sm}. In the meantime, physicists can refine existing analyses and develop new techniques for data interpretation. Renormalisation and \glspl{eft} are essential tools in achieving these goals.

In Chapter \ref{ch:Renormalization}, we reviewed the foundations of renormalisation. Although once considered problematic, renormalisation has become a powerful technique in precision \gls{qft}. In particular, only certain regularisation schemes can eliminate divergences while preserving desirable properties of \glspl{qft}. The combination of \gls{dimreg} and the 
\gls{msbar} scheme is now standard practice. While \gls{dimreg}—as defined in this context—also introduces challenges, especially with chiral amplitudes, the framework is reliable at one-loop level.

To explore models that differ from the \gls{sm} at high energies but remain consistent with low-energy evidence, we turn to non-renormalisable theories. In Chapter \ref{ch:EFT}, we discussed the use of \glspl{eft}, which may require, in principle, an infinite number of counterterms. However, the finite precision of experiments effectively bounds the number of counterterms that need to be considered. This feature restores predictivity and enables precision calculations. We illustrated this with several \glspl{eft} historically used before the \gls{sm} was established—many of which remain relevant for sub-\gls{ew} scale phenomena.

Since our interest lies beyond the Higgs \gls{vev}, and because the \gls{sm} represents the highest-energy theory currently confirmed by experiment, we take it as the foundation for building an \gls{eft}. In Chapter~\ref{ch:SMEFT}, we outlined the benefits of using operator bases and discussed the challenges involved in identifying a complete and independent set of interactions. We argued that retaining operators related by certain field redefinitions—equivalent to applying the \glspl{eom} in the Lagrangian—can simplify intermediate computations. For instance, Green’s functions in a redundant Lagrangian can be computed using only \gls{1pi} diagrams. While redundancies must eventually be removed, this can be done at the end of the calculation by applying on-shell relations. Although working with an on-shell independent Lagrangian is not always required, the off-shell diagrammatic approach to renormalisation does impose this condition. In this framework, we computed the first Green's Basis of dimension eight bosonic operators, published in Ref.~	\cite{Chala:2021cgt} and detailed in Section~\ref{sec:GB}. The results are also shown in Appendix~\ref{app:TableOfOperators} altogether with dimension-eight fermionic operators. The onshell relations of this Green's Basis are also an original result that allows to remove redundancies in favour of the physical basis of Ref.~\cite{Murphy:2020rsh}.

Chapter \ref{ch:RGE} consolidates these ideas to systematize the renormalisation of the \gls{smeft}. We derived the Callan–Symanzik equation in a useful form tailored for the renormalisation of dimension-eight operators. After summarizing the off-shell diagrammatic method and reviewing the history of \gls{smeft} renormalisation, we applied our framework to compute the complete \glspl{rge} for all bosonic operators, as well as substantial contributions to fermionic ones. Except for bosonic operator renormalisation via insertions of dimension-six operators~\cite{Chala:2021pll}, these results~\cite{Bakshi:2024wzz,DasBakshi:2022mwk,DasBakshi:2023htx} are original contributions of this thesis.

To provide an overview of the current state of \gls{smeft} renormalisation—and to place our contributions in context—we refer to Table~\ref{tab:summary}, which summarizes completed and ongoing efforts in this area. Entries computed in this thesis are highlighted.

		\begin{table}[htb!]
		\begin{center}
			\resizebox{1.\textwidth}{!}{\begin{tabular}{l|ccccccccccc}
					& $d_5$ & $d_5^2$ & $d_6$ & $d_5^3$ & $d_5\times d_6$ & $d_7$ & $d_5^4$ & $d_5^2\times d_6$ & $d_6^2$ & $d_5\times d_7$ & $d_8$\\
					\toprule
					$d_{\leq 4}$ (bosonic) &  &  & $\gtick$~\cite{Jenkins:2013zja} &  &   &  &  &  & $\gtick$~\cite{Chala:2021pll} &  & \colorbox{blue!30}{$\gtick$~\cite{DasBakshi:2022mwk}} \\
					$d_{\leq 4}$ (fermionic) &  &  & $\gtick$~\cite{Jenkins:2013zja} &  &   &  &  &  & \colorbox{blue!30}{$\gtick$~\cite{Bakshi:2024wzz}} &  & $\otick$~\cite{AccettulliHuber:2021uoa,Boughezal:2024zqa,Assi:2025fsm} \\
					$d_5$ & $\gtick$~\cite{Chankowski:1993tx,Babu:1993qv,Antusch:2001ck} &  &  &  & $\gtick$~\cite{Chala:2021juk} & $\otick$~\cite{Chala:2021juk} &  &  &  & & \\
					$d_6$ (bosonic) &  & $\gtick$~\cite{Davidson:2018zuo} & $\gtick$~\cite{Jenkins:2013zja,Jenkins:2013wua,Alonso:2013hga} & & & &  & \colorbox{blue!30}{$\gtick$~\cite{DasBakshi:2023htx}} & $\gtick$~\cite{Chala:2021pll} & \colorbox{blue!30}{$\gtick$~\cite{DasBakshi:2023htx}} & \colorbox{blue!30}{$\gtick$~\cite{DasBakshi:2022mwk}} \\
					$d_6$ (fermionic) &  & $\gtick$~\cite{Davidson:2018zuo} & $\gtick$~\cite{Jenkins:2013zja,Jenkins:2013wua,Alonso:2013hga,Alonso:2014zka} & & & & & $\rtick$ & \colorbox{blue!30}{$\otick$~\cite{Bakshi:2024wzz}} & $\rtick$ & $\otick$~\cite{AccettulliHuber:2021uoa,Boughezal:2024zqa,Assi:2025fsm}\\
					$d_7$ &  &  & & $\gtick$~\cite{Zhang:2023kvw} & $\gtick$~\cite{Zhang:2023kvw} & $\gtick$~\cite{Liao:2016hru,Liao:2019tep}\\
					$d_8$ (bosonic) &  &  & & & & & \colorbox{blue!30}{$\gtick$~\cite{DasBakshi:2023htx}} & \colorbox{blue!30}{$\gtick$~\cite{DasBakshi:2023htx}} & $\gtick$~\cite{Chala:2021pll} & \colorbox{blue!30}{$\gtick$~\cite{DasBakshi:2023htx}} & \colorbox{blue!30}{$\gtick$~\cite{DasBakshi:2022mwk}}\\
					$d_8$ (fermionic) &  &  & & & & & $\rtick$ & $\rtick$ & \colorbox{blue!30}{$\otick$~\cite{Bakshi:2024wzz}}& $\rtick$  & $\otick$~\cite{AccettulliHuber:2021uoa,Boughezal:2024zqa,Assi:2025fsm}
					\\\bottomrule
			\end{tabular}}
			\caption{\label{tab:summary}\it State of the art of the SMEFT renormalisation (adapted from Refs.\cite{DasBakshi:2022mwk,Chala:2021pll,DasBakshi:2023htx}). The rows show the renormalised operators (categorised by dimensions and statistics). The columns show the operators contributing to RG running. Blank entries vanish, $\gtick$ denotes that the complete contribution is available, $\otick$ implies that only (but substantial) partial results are present, and $\rtick$ indicates that nothing, or very little, is known. The contribution made in this thesis is marked by $\colorbox{blue!30}{\text{Blue boxes}}$.}
		\end{center}
	\end{table}

The running of \gls{wc} in \gls{smeft} has a wide range of applications. In particular, \glspl{rge} for dimension-six operators are already being used in phenomenological studies. As discussed in Section~\ref{sec:Applications}, the \glspl{rge} of dimension-eight operators are expected to play a similar role in high-precision physics and may also illuminate new conceptual avenues. One such direction involves positivity bounds—constraints derived from unitarity, causality, and analyticity. Since these bounds are inherently \order{4} effects, the use of \glspl{rge} to explore their violation is particularly relevant. As noted in Section~\ref{sec:Applications}, the observation that \gls{rg} running can lead to apparent violations of positivity bounds may prompt a reevaluation of the \gls{smeft} framework.

\section{Future directions}

Several promising research directions emerge from this work:
\paragraph{Complete the remaining  \glsxtrshort{rge} at \order{4}}
A natural extension involves computing the \glspl{rge} for all fermionic operators at dimension eight, including four-fermion interactions and \gls{lnv} contributions arising from lower-dimensional operator combinations.

\paragraph{Clarify \glsxtrshort{smeft}’s role in \glsxtrshort{ewpt}}

While the dimension-six and dimension-eight contributions to oblique parameters are known for bosonic operators, their relationship with fermionic interactions remains unclear, as oblique parameters are mostly studied in universal theories. Investigating this relationship would enhance our understanding of \gls{smeft}’s applicability to \glspl{ewpt}.

\paragraph{Investigate basis dependence in \glsxtrshort{adm}}

It has been observed that off-shell computations sometimes exhibit cancellations when expressions are taken on-shell. A deeper understanding of this effect could shed light on the interplay between basis choices and divergence structures. This would require constructing new operator bases and re-computing \glspl{rge} at \order{4}, which is currently impractical. Waiting for further automation in renormalisation techniques appears more feasible.

\paragraph{Compare renormalisation methods: off-shell, on-shell and geometrical}
The \glspl{rge} at dimension eight are already being used to validate alternative renormalisation approaches. Once full renormalisation is achieved, a systematic comparison of these methods—not only in terms of results but also in computational efficiency—would be highly valuable. \gls{smeft}, with its simplicity and compatibility with a wide range of low-energy observables, provides an ideal testing ground for such a comparative study.

\paragraph{Final Remarks} Whichever path is chosen, the \glsxtrshort{rge} at dimension eight represent a powerful tool in the ongoing search for \gls{np}. More broadly, renormalisation and \glspl{eft} have proven indispensable in \gls{qft} and continue to shape the way we bridge high-energy theories with low-energy phenomena. With this thesis, we aim to contribute to a deeper understanding of these tools and to motivate continued exploration in this promising field. 

\bibliography{PhD_bib.bib} 

\pagestyle{empty}
\appendix
\titleformat{\chapter}[display]{}{\Huge Appendix \thechapter.}{0pt}{\Huge}
\chapter{Tables of SMEFT Operators}\label{app:TableOfOperators}
	\begin{table}[h!]
		\begin{center}
			\resizebox{\textwidth}{!}{
\begin{tabular}{cclcl}
					\multicolumn{5}{c}{\rclass \large{\textbf{Dimension 8}}} \Tstrut\\[0.2cm]
					%
					\boldmath{$ H^8$} & \rlname $\mathcal{O}_{ H^8}$ & \rlop $(  H^\dagger H)^4$ &%
					\rlop & \rlop \Tstrut\\[0.2cm]
					%
					& \rhname $\mathcal{O}_{ H^6D^2}^{(1)}$ & \rhop $( H^{\dag}  H)^2 (D_{\mu}  H^{\dag} D^{\mu}  H)$ &%
					\rhname $\mathcal{O}_{ H^6D^2}^{(2)}$ & \rhop $( H^{\dag}  H) ( H^{\dag} \sigma^I  H) (D_{\mu}  H^{\dag} \sigma^I D^{\mu}  H)$ \Tstrut\\[0.2cm]
					\multirow{-2}{*}{\boldmath{$ H^6 D^2$}} & \rhname \textcolor{gray}{ $\mathcal{O}_{ H^6D^2}^{(3)}$} & \rhop \textcolor{gray}{$( H^\dagger H)^2 ( H^\dagger D^2 H + \text{h.c.})$} &%
					\rhname \textcolor{gray}{ $\mathcal{O}_{ H^6D^2}^{(4)}$} & \rhop \textcolor{gray}{$( H^\dagger H)^2 D_\mu( H^\dagger\ii \overleftrightarrow{D}^\mu H)$} \Tstrut\\[0.2cm]
					%
					& \rlname $\mathcal{O}_{ H^4D^4}^{(1)}$ & \rlop $(D_{\mu}  H^{\dag} D_{\nu}  H) (D^{\nu}  H^{\dag} D^{\mu}  H)$ &%
					\rlname $\mathcal{O}_{ H^4D^4}^{(2)}$ & \rlop $(D_{\mu}  H^{\dag} D_{\nu}  H) (D^{\mu}  H^{\dag} D^{\nu}  H)$ \Tstrut\\[0.2cm]
					& \rlname $\mathcal{O}_{ H^4D^4}^{(3)}$ & \rlop $(D^{\mu}  H^{\dag} D_{\mu}  H) (D^{\nu}  H^{\dag} D_{\nu}  H)$ &%
					\rlname \textcolor{gray} {$\mathcal{O}_{ H^4D^4}^{(4)}$} & \rlop \textcolor{gray} {$D_\mu H^\dagger D^\mu H( H^\dagger D^2 H + \text{h.c.})$} \Tstrut\\[0.2cm]
					& \rlname \textcolor{gray} {$\mathcal{O}_{ H^4D^4}^{(5)} $} & \rlop \textcolor{gray}  {$D_\mu H^\dagger D^\mu H( H^\dagger \ii D^2 H + \text{h.c.})$} &%
					\rlname \textcolor{gray} {$\mathcal{O}_{ H^4D^4}^{(6)}$} & \rlop \textcolor{gray} {$(D_\mu H^\dagger  H) (D^2 H^\dagger D_\mu H) + \text{h.c.}$} \Tstrut\\
\end{tabular}}
		\end{center}
		\caption{Green's basis of dimension-eight bosonic operators. Operators in gray are redundant. Original work, extracted from \cite{Chala:2021cgt}.
		\label{tab:dim8ops1} }
	\end{table}
	
	\clearpage
	\begin{table}[h!]
		\begin{center}
			\resizebox{\textwidth}{!}{
				\begin{tabular}{cclcl}
					\multicolumn{5}{c}{\rclass \large{\textbf{Dimension 8}}} \Tstrut\\[0.2cm]
					%
& \rlname $\mathcal{O}_{W^2 H^2 D^2}^{(1)}$ & \rlop $(D^\mu H^\dagger D^\nu H) W_{\mu\rho}^I W_\nu^{I\rho}$ &%
					\rlname $\mathcal{O}_{W^2 H^2 D^2}^{(2)}$ & \rlop $(D^\mu H^\dagger D_\mu H) W_{\nu\rho}^I W^{I\nu\rho}$ \Tstrut\\[0.2cm]
& \rlname $\mathcal{O}_{W^2 H^2 D^2}^{(3)}$ & \rlop $(D^\mu H^\dagger D_\mu H) W_{\nu\rho}^I \widetilde{W}^{I\nu\rho}$ &%
					\rlname $\mathcal{O}_{W^2 H^2 D^2}^{(4)}$ & \rlop $\ii\epsilon^{IJK}(D^\mu H^\dagger\sigma^I D^\nu H) W_{\mu\rho}^J W_\nu^{K\rho}$ \Tstrut\\[0.2cm]
& \rlname $\mathcal{O}_{W^2 H^2 D^2}^{(5)}$ & \rlop $\epsilon^{IJK}(D^\mu H^\dagger\sigma^I D^\nu H) (W_{\mu\rho}^J \widetilde{W}_\nu^{K\rho}-\widetilde{W}^J_{\mu\rho} W_\nu^{K\rho})$ &%
					\rlname $\mathcal{O}_{W^2 H^2 D^2}^{(6)}$ & \rlop $\ii\epsilon^{IJK}(D^\mu H^\dagger\sigma^I D^\nu H) (W_{\mu\rho}^J \widetilde{W}_\nu^{K\rho}+\widetilde{W}^J_{\mu\rho} W_\nu^{K\rho})$ \Tstrut\\[0.2cm]
					& \rlname \textcolor{gray}{$\mathcal{O}_{W^2 H^2 D^2}^{(7)}$} & \rlop \textcolor{gray}{$\ii\epsilon^{IJK}( H^\dagger \sigma^I D^\nu H-D^\nu H^\dagger \sigma^I  H) D_\mu W^{J\mu\rho}\widetilde{W}^{K}_{\nu\rho}$} &%
					\rlname \textcolor{gray}{$\mathcal{O}_{W^2 H^2 D^2}^{(8)}$} & \rlop \textcolor{gray}{$\epsilon^{IJK}  H^\dagger\sigma^I H D_\nu D_\mu W^{J\mu\rho}\widetilde{W}^{K\nu}_{\,\,\,\,\rho}$} \Tstrut\\[0.2cm]
					& \rlname \textcolor{gray}{$\mathcal{O}_{W^2 H^2 D^2}^{(9)}$} & \rlop \textcolor{gray}{$\ii ( H^\dagger D_\nu H-D_\nu H^\dagger  H) D_\mu W^{I\mu\rho} \widetilde{W}^{I\nu}_{\,\,\,\,\rho}$} &%
					\rlname \textcolor{gray}{$\mathcal{O}_{W^2 H^2 D^2}^{(10)}$} & \rlop \textcolor{gray}{$( H^\dagger D_\nu H+D_\nu H^\dagger  H) D_\mu W^{I\mu\rho} \widetilde{W}^{I\nu}_{\,\,\,\,\rho}$} \Tstrut\\
\end{tabular}}
		\end{center}
		\caption{Green's basis of dimension-eight bosonic operators. Operators in gray are redundant. Original work, extracted from \cite{Chala:2021cgt}.
		}
	\end{table}
	
	\clearpage	
%
%
%
	\begin{table}[h!]
		\begin{center}
			\resizebox{\textwidth}{!}{
				\begin{tabular}{cclcl}
					\multicolumn{5}{c}{\rclass \large{\textbf{Dimension 8}}} \Tstrut\\[0.2cm]
					%
					& \rlname $\mathcal{O}_{\ell^2H^4D}^{(1)}$ & \rlop $\ii(\overline{\ell} \gamma^\mu \ell)(H^{\dagger} \overleftrightarrow{D}_\mu H)(H^{\dagger} H)$ &%
					\rlname $\mathcal{O}_{\ell^2H^4D}^{(2)}$ & \rlop $\ii(\overline{\ell} \gamma^\mu \sigma^I \ell)[(H^{\dagger} \overleftrightarrow{D}_\mu^I H)(H^{\dagger} H)+(H^{\dagger} \overleftrightarrow{D}_\mu H)(H^{\dagger} \sigma^I H)]$ \Tstrut\\[0.2cm]
					& \rlname $\mathcal{O}_{\ell^2H^4D}^{(3)}$ & \rlop $\ii \epsilon^{I J K}(\overline{\ell} \gamma^\mu \sigma^I \ell)(H^{\dagger} \overleftrightarrow{D}_\mu^J H)(H^{\dagger} \sigma^K H)$ &%
					\rlname $\mathcal{O}_{\ell^2H^4D}^{(4)}$ & \rlop $\epsilon^{I J K}(\overline{\ell} \gamma^\mu \sigma^I \ell)(H^{\dagger} \sigma^J H) D_\mu(H^{\dagger} \sigma^K H)$ \Tstrut\\[0.2cm]
					& \rlname \textcolor{gray}{$\mathcal{O}_{\ell^2H^4D}^{(5)}$} & \rlop \textcolor{gray}{$(\overline{\ell} \gamma^\mu \ell)D_\mu(H^{\dagger} H)(H^{\dagger} H) $} &%
					\rlname \textcolor{gray}{$\mathcal{O}_{\ell^2H^4D}^{(6)}$} & \rlop \textcolor{gray}{$\ii(\overline{\ell} \gamma^\mu \overleftrightarrow{D}_{\mu} \ell)(H^{\dagger} H)(H^{\dagger} H) $} \Tstrut\\[0.2cm]
					& \rlname \textcolor{gray}{$\mathcal{O}_{\ell^2H^4D}^{(7)}$} & \rlop \textcolor{gray}{$\ii(\overline{\ell} \gamma^\mu \overleftrightarrow{D}^I_{\mu} \ell)(H^{\dagger} \sigma^I H)(H^{\dagger} H) $} &%
					\rlname \textcolor{gray}{$\mathcal{O}_{\ell^2H^4D}^{(8)}$} & \rlop \textcolor{gray}{$D_{\mu}(\overline{\ell} \gamma^\mu \sigma^I \ell)(H^{\dagger} \sigma^I H)(H^{\dagger} H)$} \Tstrut\\
\end{tabular}}
\end{center}
\caption{Green's Basis of dimension-eight fermionic operators. Operators in grey are redundant. Adapted from Refs.~\cite{Murphy:2020rsh,Ren:2022tvi}. $(*)$ represents a complex class of operators.}
\end{table}
					%
	\begin{table}[h!]
		\begin{center}
			\resizebox{\textwidth}{!}{
				\begin{tabular}{cclcl}
					\multicolumn{5}{c}{\rclass \large{\textbf{Dimension 8}}} \Tstrut\\[0.2cm]
					%
					& \rlname $\mathcal{O}_{\ell eH^3D^2}^{(1)}$ & \rlop $(\overline{\ell} e H)(D_\mu H^{\dagger} D^\mu H)$ &%
					\rlname $\mathcal{O}_{\ell eH^3D^2}^{(2)}$ & \rlop $(\overline{\ell} e \sigma^I H)(D_\mu H^{\dagger} \sigma^I D^\mu H)$ \Tstrut\\[0.2cm]
		& \rlname $\mathcal{O}_{\ell eH^3D^2}^{(3)}$ & \rlop $(\overline{\ell} \sigma^{\mu \nu} e H)(D_\mu H^{\dagger} D_\nu H)$ &%
					\rlname $\mathcal{O}_{\ell eH^3D^2}^{(4)}$ & \rlop $(\overline{\ell} \sigma^{\mu \nu} e \sigma^I H)(D_\mu H^{\dagger} \sigma^I D_\nu H)$ \Tstrut\\[0.2cm]
		& \rlname $\mathcal{O}_{\ell eH^3D^2}^{(5)}$ & \rlop $(\overline{\ell} e D^\mu H)(H^{\dagger} D_\mu H)$ &%
					\rlname $\mathcal{O}_{\ell eH^3D^2}^{(6)}$ & \rlop $(\overline{\ell} \sigma^{\mu \nu} e D_\nu H)(H^{\dagger} D_\mu H)$ \Tstrut\\[0.2cm]
					& \rlname \textcolor{gray}{$\mathcal{O}_{\ell eH^3D^2}^{(7)}$} & \rlop \textcolor{gray}{$(\overline{\ell} D^{\mu}e H)D_{\mu}(H^\dagger H)$} &%
					\rlname \textcolor{gray}{$\mathcal{O}_{\ell eH^3D^2}^{(8)}$} & \rlop \textcolor{gray}{$(\overline{\ell} D^{\mu}e H)(H^\dagger \ii \overleftrightarrow{D}_{\mu} H)$} \Tstrut\\[0.2cm]
					& \rlname \textcolor{gray}{$\mathcal{O}_{\ell eH^3D^2}^{(9)}$} & \rlop \textcolor{gray}{$(\overline{\ell} \sigma^{\mu\nu} D_{\mu}e H)D_{\nu}(H^\dagger H) $} &%
					\rlname \textcolor{gray}{$\mathcal{O}_{\ell eH^3D^2}^{(10)}$} & \rlop \textcolor{gray}{$(\overline{\ell} \sigma^{\mu\nu} D_{\mu}e H)(H^\dagger \ii \overleftrightarrow{D}_{\nu} H) $} \Tstrut\\[0.2cm]
					& \rlname \textcolor{gray}{$\mathcal{O}_{\ell eH^3D^2}^{(11)}$} & \rlop \textcolor{gray}{$(D^{\mu}\overline{\ell} e H)D_{\mu}(H^\dagger H)$} &%
					\rlname \textcolor{gray}{$\mathcal{O}_{\ell eH^3D^2}^{(12)}$} & \rlop \textcolor{gray}{$(D^{\mu}\overline{\ell} e H)(H^\dagger \ii \overleftrightarrow{D}_{\mu} H)$} \Tstrut\\
\end{tabular}}
\end{center}
\caption{Green's Basis of dimension-eight fermionic operators. Operators in grey are redundant. Adapted from Refs.~\cite{Murphy:2020rsh,Ren:2022tvi}. $(*)$ represents a complex class of operators.}
\end{table}

	\begin{table}[h!]
		\begin{center}
			\resizebox{\textwidth}{!}{
				\begin{tabular}{cclcl}
					\multicolumn{5}{c}{\rclass \large{\textbf{Dimension 8}}} \Tstrut\\[0.2cm]
					%
					& \rlname $\mathcal{O}_{W e^2H^2D}^{(1)}$ & \rlop $(\overline{e} \gamma^\nu e) D^\mu(H^{\dagger} \sigma^I H) W_{\mu \nu}^I$ &%
					\rlname $\mathcal{O}_{W e^2H^2D}^{(2)}$ & \rlop $(\overline{e} \gamma^\nu e) D^\mu(H^{\dagger} \sigma^I H) \widetilde{W}_{\mu \nu}^I $ \Tstrut\\[0.2cm]
					& \rlname $\mathcal{O}_{W e^2H^2D}^{(3)}$ & \rlop $\ii(\overline{e} \gamma^\nu e)(H^{\dagger} \overleftrightarrow{D}^I{}^\mu H) W_{\mu \nu}^I$ &%
					\rlname $\mathcal{O}_{W e^2H^2D}^{(4)}$ & \rlop $\ii(\overline{e} \gamma^\nu e)(H^{\dagger} \overleftrightarrow{D}^I{}^\mu H) \widetilde{W}_{\mu \nu}^I$ \Tstrut\\[0.2cm]
					& \rlname \textcolor{gray}{$\mathcal{O}_{W e^2H^2D}^{(5)}$} & \rlop \textcolor{gray}{$D^\mu (\overline{e} \gamma^\nu e) (H^{\dagger} \sigma^I H) W_{\mu \nu}^I$} &%
					\rlname \textcolor{gray}{$\mathcal{O}_{W e^2H^2D}^{(6)}$} & \rlop \textcolor{gray}{$\ii(\overline{e} \gamma^\nu \overleftrightarrow{D}^\mu e) (H^{\dagger} \sigma^I H) W_{\mu \nu}^I $} \Tstrut\\[0.2cm]
					& \rlname \textcolor{gray}{$\mathcal{O}_{W e^2H^2D}^{(7)}$} & \rlop \textcolor{gray}{$\ii(\overline{e} \gamma^\nu \overleftrightarrow{D}^\mu e) (H^{\dagger} \sigma^I H) \widetilde{W}_{\mu \nu}^I$} &%
					\rlname  & \rlop \Tstrut\\[0.2cm]
					& \rlname $\mathcal{O}_{Be^2H^2D}^{(1)}$ & \rlop $(\overline{e} \gamma^\nu e) D^\mu(H^{\dagger} H) B_{\mu \nu}$ &%
					\rlname $\mathcal{O}_{Be^2H^2D}^{(2)}$ & \rlop $(\overline{e} \gamma^\nu e) D^\mu(H^{\dagger} H) \widetilde{B}_{\mu \nu}$ \Tstrut\\[0.2cm]
					& \rlname $\mathcal{O}_{Be^2H^2D}^{(3)}$ & \rlop $\ii(\overline{e} \gamma^\nu e)(H^{\dagger} \overleftrightarrow{D}^\mu H) B_{\mu \nu} $ &%
					\rlname $\mathcal{O}_{Be^2H^2D}^{(4)}$ & \rlop $\ii(\overline{e} \gamma^\nu e)(H^{\dagger} \overleftrightarrow{D}^\mu H) \widetilde{B}_{\mu \nu}$ \Tstrut\\
\end{tabular}}
\end{center}
\caption{Green's Basis of dimension-eight $X\psi^2H^2D$ operators (Part I). Operators in grey are redundant. Adapted from Refs.~\cite{Murphy:2020rsh,Ren:2022tvi}.}
\end{table}

	\begin{table}[h!]
		\begin{center}
			\resizebox{\textwidth}{!}{
				\begin{tabular}{cclcl}
					\multicolumn{5}{c}{\rclass \large{\textbf{Dimension 8}}} \Tstrut\\[0.2cm]
					%
					& \rlname $\mathcal{O}_{Wd^2H^2D}^{(1)}$ & \rlop $(\overline{d} \gamma^\nu d) D^\mu(H^{\dagger} \sigma^I H) W_{\mu \nu}^I$ &%
					\rlname $\mathcal{O}_{Wd^2H^2D}^{(2)}$ & \rlop $(\overline{d} \gamma^\nu d) D^\mu(H^{\dagger} \sigma^I H) \widetilde{W}_{\mu \nu}^I $ \Tstrut\\[0.2cm]
					& \rlname $\mathcal{O}_{Wd^2H^2D}^{(3)}$ & \rlop $\ii(\overline{d} \gamma^\nu d)(H^{\dagger} \overleftrightarrow{D}^I{}^\mu H) W_{\mu \nu}^I$ &%
					\rlname $\mathcal{O}_{Wd^2H^2D}^{(4)}$ & \rlop $\ii(\overline{d} \gamma^\nu d)(H^{\dagger} \overleftrightarrow{D}^I{}^\mu H) \widetilde{W}_{\mu \nu}^I$ \Tstrut\\[0.2cm]
					& \rlname \textcolor{gray}{$\mathcal{O}_{Wd^2H^2D}^{(5)}$} & \rlop \textcolor{gray}{$D^\mu (\overline{d} \gamma^\nu d) (H^{\dagger} \sigma^I H) W_{\mu \nu}^I $} &%
					\rlname \textcolor{gray}{$\mathcal{O}_{Wd^2H^2D}^{(6)}$} & \rlop \textcolor{gray}{$\ii(\overline{d} \gamma^\nu \overleftrightarrow{D}^\mu d) (H^{\dagger} \sigma^I H) W_{\mu \nu}^I $} \Tstrut\\[0.2cm]
					& \rlname \textcolor{gray}{$\mathcal{O}_{Wd^2H^2D}^{(7)}$} & \rlop \textcolor{gray}{$\ii(\overline{d} \gamma^\nu \overleftrightarrow{D}^\mu d) (H^{\dagger} \sigma^I H) \widetilde{W}_{\mu \nu}^I$} &%
					\rlname  & \rlop \Tstrut\\[0.2cm]
					& \rlname $\mathcal{O}_{Bd^2H^2D}^{(1)}$ & \rlop $(\overline{d} \gamma^\nu d) D^\mu(H^{\dagger} H) B_{\mu \nu} $ &%
					\rlname $\mathcal{O}_{Bd^2H^2D}^{(2)}$ & \rlop $(\overline{d} \gamma^\nu d) D^\mu(H^{\dagger} H) \widetilde{B}_{\mu \nu}  $ \Tstrut\\[0.2cm]
					& \rlname $\mathcal{O}_{Bd^2H^2D}^{(3)}$ & \rlop $\ii(\overline{d} \gamma^\nu d)(H^{\dagger} \overleftrightarrow{D}^\mu H) B_{\mu \nu}$ &%
					\rlname $\mathcal{O}_{Bd^2H^2D}^{(4)}$ & \rlop $\ii(\overline{d} \gamma^\nu d)(H^{\dagger} \overleftrightarrow{D}^\mu H) \widetilde{B}_{\mu \nu}$ \Tstrut\\
\end{tabular}}
\end{center}
\caption{Green's Basis of dimension-eight $X\psi^2H^2D$ operators (Part II). Operators in grey are redundant. Adapted from Refs.~\cite{Murphy:2020rsh,Ren:2022tvi}.}
\end{table}

	\begin{table}[h!]\vspace{-2cm}
		\begin{center}
			\resizebox{\textwidth}{!}{
				\begin{tabular}{cclcl}
					\multicolumn{5}{c}{\rclass \large{\textbf{Dimension 8}}} \Tstrut\\[0.2cm]
					%
					& \rlname $\mathcal{O}_{B\ell^2 H^2D}^{(1)}$ & \rlop $(\overline{\ell} \gamma^\nu \sigma^I \ell) D^\mu(H^{\dagger} \sigma^I H) B_{\mu \nu} $ &%
					\rlname $\mathcal{O}_{B\ell^2 H^2D}^{(2)}$ & \rlop $(\overline{\ell} \gamma^\nu \sigma^I \ell) D^\mu(H^{\dagger} \sigma^I H) \widetilde{B}_{\mu \nu}$ \Tstrut\\[0.2cm]
					& \rlname $\mathcal{O}_{B\ell^2 H^2D}^{(3)}$ & \rlop $\ii(\overline{\ell} \gamma^\nu \sigma^I \ell)(H^{\dagger} \overleftrightarrow{D}^I{}^\mu H) B_{\mu \nu}$ &%
					\rlname $\mathcal{O}_{B\ell^2 H^2D}^{(4)}$ & \rlop $\ii(\overline{\ell} \gamma^\nu \sigma^I \ell)(H^{\dagger} \overleftrightarrow{D}^I{}^\mu H) \widetilde{B}_{\mu \nu}$ \Tstrut\\[0.2cm]
					& \rlname $\mathcal{O}_{B\ell^2 H^2D}^{(5)}$ & \rlop $(\overline{\ell} \gamma^\nu \ell) D^\mu(H^{\dagger} H) B_{\mu \nu} $ &%
					\rlname $\mathcal{O}_{B\ell^2 H^2D}^{(6)}$ & \rlop $(\overline{\ell} \gamma^\nu \ell) D^\mu(H^{\dagger} H) \widetilde{B}_{\mu \nu} $ \Tstrut\\[0.2cm]
					& \rlname $\mathcal{O}_{B\ell^2 H^2D}^{(7)}$ & \rlop $\ii(\overline{\ell} \gamma^\nu \ell)(H^{\dagger} \overleftrightarrow{D}^\mu H) B_{\mu \nu}$ &%
					\rlname $\mathcal{O}_{B\ell^2 H^2D}^{(8)}$ & \rlop $\ii(\overline{\ell} \gamma^\nu \ell)(H^{\dagger} \overleftrightarrow{D}^\mu H) \widetilde{B}_{\mu \nu}$ \Tstrut\\[0.2cm]
					& \rlname \textcolor{gray}{$\mathcal{O}_{B\ell^2 H^2D}^{(9)}$} & \rlop \textcolor{gray}{$D^\mu (\overline{\ell} \gamma^\nu \sigma^I \ell) (H^{\dagger} \sigma^I H) B_{\mu \nu}$} &%
					\rlname \textcolor{gray}{$\mathcal{O}_{B\ell^2 H^2D}^{(10)}$} & \rlop \textcolor{gray}{$\ii(\overline{\ell} \gamma^\nu \overleftrightarrow{D}^I{}^\mu \ell) (H^{\dagger} \sigma^I H) B_{\mu \nu} $} \Tstrut\\
\end{tabular}}
\end{center}
\caption{Green's Basis of dimension-eight $X\psi^2H^2D$ operators (Part III). Operators in grey are redundant. Adapted from Refs.~\cite{Murphy:2020rsh,Ren:2022tvi}.}
\end{table}

	\begin{table}[h!]\vspace{-2cm}
		\begin{center}
			\resizebox{\textwidth}{!}{
				\begin{tabular}{cclcl}
					\multicolumn{5}{c}{\bclass \large{\textbf{Dimension 7}}} \Tstrut\\[0.2cm]
					%
					\boldmath{$\psi^2  H^4(*)$} & \blname $ \mathcal{O}_{\ell  H}^{(7)}$  & \blop $\epsilon_{ij}\epsilon_{mn}(\ell^iC\ell^m)( H^j  H^n)( H^\dagger  H)$ & %
					\blop & \blop \Tstrut\\[0.2cm]
					%
					\boldmath{$\psi^2  H^2 D^2(*)$} & \bhname $  \mathcal{O}_{\ell  H D}^{(1)} $  & \bhop$\epsilon_{ij}\epsilon_{mn}\ell^iC(D^\mu \ell^j) H^m (D_\mu H^n)$ & %
					\bhname $  \mathcal{O}_{\ell H D}^{(2)} $ & \bhop $\epsilon_{im}\epsilon_{jn}\ell^iC(D^\mu \ell^j) H^m (D_\mu H^n)$ \Tstrut\\[0.2cm]
					%
					\boldmath{$\psi^2  H^3 D(*)$} &  \blname$ \mathcal{O}_{\ell H De}$ & \blop $\epsilon_{ij}\epsilon_{mn}(\ell^iC\gamma_\mu e)  H^j H^m D^\mu  H^n$ & %
					\blop & \blop \Tstrut\\[0.2cm]
					%
					\boldmath{$\psi^2  H^2 X(*)$} & \bhname $\mathcal{O}_{\ell H B} $ & \bhop $\epsilon_{ij}\epsilon_{mn}\ell^iC(\sigma_{\mu\nu}\ell^m) H^j H^n B^{\mu\nu}$ & 
					\bhname $ \mathcal{O}_{\ell H W} $ & \bhop $\epsilon_{ij}(\sigma^I\epsilon)_{mn}\ell^iC(\sigma_{\mu\nu}\ell^m) H^j H^n W^{I\mu\nu}$ \Tstrut\\[0.2cm]
					%
					\multicolumn{5}{c}{\gclass \large{\textbf{Dimension 6}}} \Tstrut\\[0.2cm]
					\boldmath{$ H^6$} & \glname $\mathcal{O}_{ H}$ & \glop $ ( H^\dagger  H)^3 $ &%
					\glop & \glop \Tstrut\\[0.2cm]
					%
					& \ghname $\mathcal{O}_{ H \square} $ & \ghop $ ( H^\dagger  H)\Square ( H^\dagger  H) $ &%
					\ghname $\mathcal{O}_{ H D} $ & \ghop $ ( H^\dagger D^\mu  H)^\dagger ( H^\dagger D_\mu  H) $ \Tstrut\\
\end{tabular}}
		\end{center}
		\caption{\label{tab:op67} Basis of dimension-five, -six and -seven operators needed for the renormalisation of dimension-eight bosonic operators. Operators in grey are redundant. $(*)$ represents a complex class of operators. Adapted from Refs.~\cite{Grzadkowski:2010es,Gherardi:2020det,Liao:2016hru,Lehman:2014jma}.
		}
	\end{table}

\thispagestyle{empty}

\clearpage\hbox{}\thispagestyle{empty}\newpage

\AtEndDocument{\includepdf[pages=1]{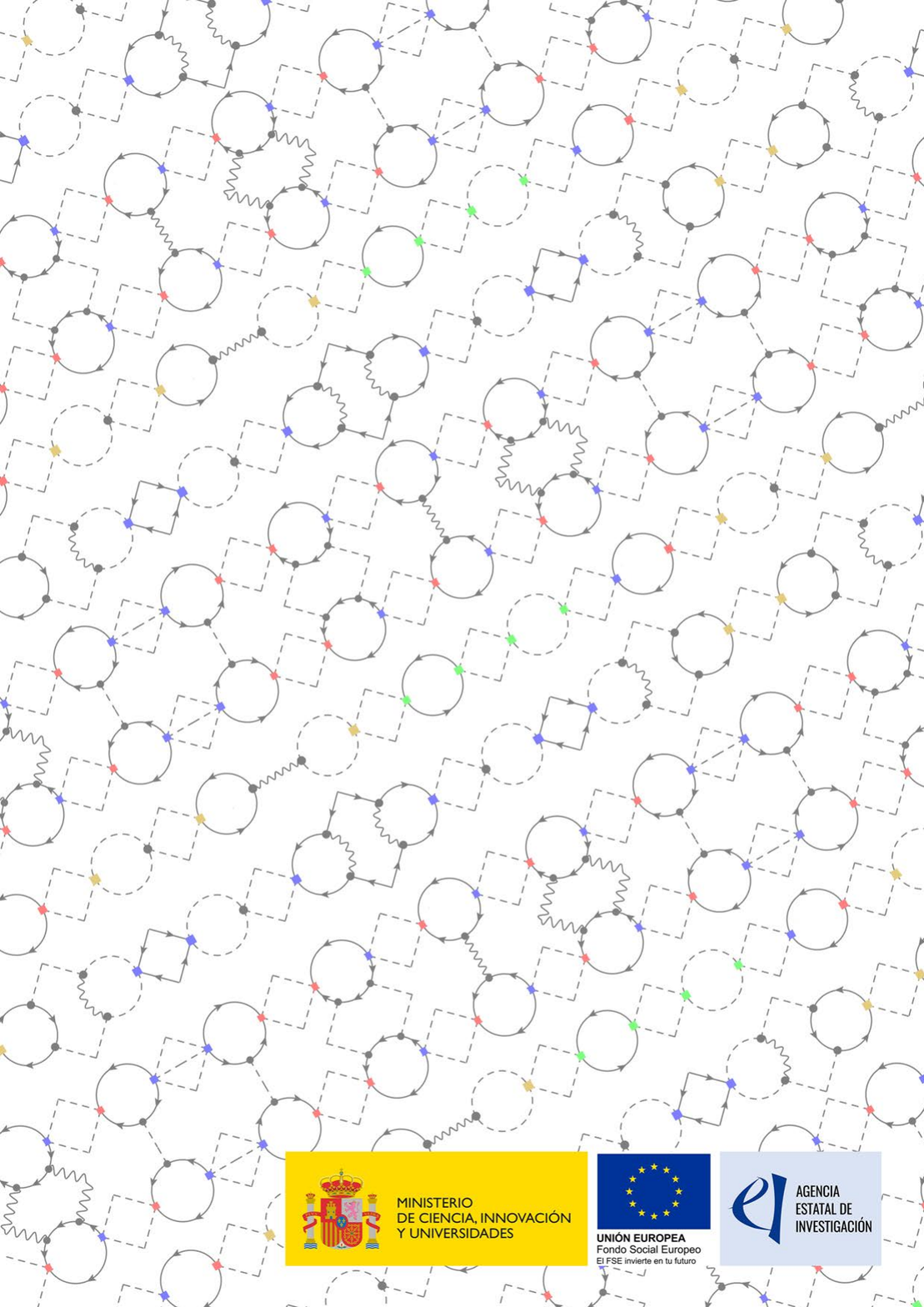}}

\end{document}